\documentclass[aps,pra,groupedaddress,reprint]{revtex4-1}

\usepackage[german,english]{babel}
\usepackage{amsmath}
\usepackage[colorlinks=true,citecolor=black,linkcolor=black]{hyperref}   
\usepackage{amsfonts}
\usepackage{amssymb}

\usepackage[T1]{fontenc} 
\usepackage{palatino}
\usepackage{fix-cm}

\usepackage{natbib}
\usepackage{graphicx}
\usepackage[labelfont=footnotesize,font=footnotesize]{caption} 
\usepackage[labelfont=scriptsize,font=scriptsize,singlelinecheck=off,justification=centering]{subcaption}
\usepackage{booktabs} 
\usepackage{tabularx}
\usepackage{tabulary}
\newcolumntype{Z}{>{\centering\arraybackslash}X} 

\newcommand{\bra}[1]{\mathop{\left\langle#1\right|}}
\newcommand{\ket}[1]{\mathop{\left|#1\right\rangle}}
\newcommand{\pouter}[2]{\ensuremath{\mathopen{}\left|{#1}\right\rangle\!\left\langle{#2}\right|}} 

\newcommand{\idd}{\mathbf{1}}

\newcommand{\bpm}{\begin{pmatrix}}
\newcommand{\epm}{\end{pmatrix}}

\newcommand{\E}{\mathrm{e}}

\newcommand{\II}{\mathrm{i}}

\newcommand{\MTR}{\text{Tr}}
\newcommand{\mtr}{\MTR}
\newcommand{\hilb}{\mathcal{H}}
\newcommand{\sun}{\ensuremath{\mathrm{SU}(N)}}

\newcommand{\dddd}{\mathrm{d}}

\newcommand{\ppp}{\mathbf{p}}
\newcommand{\qqq}{\mathbf{q}}

\newcommand{\eee}{\mathbf{e}}
\newcommand{\vvv}{\mathbf{v}}
\newcommand{\nnn}{\mathbf{n}}

\newcommand{\degree}{\ensuremath{^\circ}}
\newcommand{\comment}[1]{}

\begin{document}

\title{Entanglement of two relativistic particles with discrete momenta}
\author{Veiko Palge}
\email[Email: ]{veiko.palge@gmail.com}
\affiliation{Graduate School of Information Science, Nagoya University, Nagoya 464-8601, Japan}
\author{Jacob Dunningham}
\email[Email: ]{J.Dunningham@sussex.ac.uk}
\affiliation{Department of Physics and Astronomy, University of Sussex, Brighton BN1 9QH, United Kingdom}

\begin{abstract}
We study the structure of maps that Lorentz boosts induce on the spin degree of freedom of a system consisting of two massive spin-$1/2$ particles. We consider the case where the spin state is described by the Werner state and the momenta are discrete. Transformations on the spins are systematically investigated in various boost scenarios by calculating the orbit and concurrence of the bipartite spin state with different kinds of product and entangled momenta. We confirm the general conclusion that Lorentz boosts cause non-trivial behavior of bipartite spin entanglement. Visualization of the evolution of the spin state is shown to be valuable in explaining the pattern of concurrence. The idealized model provides a basis of explanation in terms of which phenomena in systems involving continuous momenta can be understood.
\end{abstract}

\maketitle

\section{Introduction}

Entanglement is widely regarded as one of the central features that sets the quantum and classical worlds apart. Motivated by its fundamental importance as well as promises of application, the theory of entanglement has made vast progress over the last decades. Recently there has been growing interest in relativistic quantum information. This takes seriously the notion that the ultimate description of physical reality is relativistic and seeks to provide an account of how the quantum information theoretic notions like entanglement behave in the relativistic regime.

Extensive research on both single and two particle systems has uncovered a wealth of results about how relativity affects entanglement \cite{czachor_einstein-podolsky-rosen-bohm_1997, peres_quantum_2002, gingrich_quantum_2002, ahn_relativistic_2002, ahn_lorentz-covariant_2003, alsing_lorentz_2002, ahn_relativistic_2003, czachor_relativistic_2003, moon_relativistic_2004, lee_quantum_2004, czachor_comment_2005, lamata_relativity_2006, alsing_entanglement_2006, chakrabarti_entangled_2009, friis_relativistic_2010, fuentes_entanglement_2010, palge_generation_2012}. Early work found that spin entanglement of a bipartite system does not remain invariant under Lorentz boosts \cite{gingrich_quantum_2002}. This was confirmed by \cite{peres_quantum_2002} who reported that spin entropy of a single particle is not a relativistic scalar. On the other hand, \cite{alsing_lorentz_2002} argued that the entanglement fidelity of a Bell state remains invariant for a Lorentz boosted observer. Further research found that no sum of entanglements remains invariant under boosts \cite{jordan_lorentz_2007}.

A key aspect one notices is that many of these sometimes seemingly conflicting results involve systems containing different momentum states and boost geometries. This confirms what has been observed in single particle systems: entanglement under Lorentz boosts is highly dependent on the boost scenario in question \cite{palge_generation_2012}. In the same vein, the literature on the Wigner rotation is quite clear about the fact that its nature is highly geometric, yet aside from a few cases \cite{palge_behavior_2011}, there is little work in relativistic quantum information that systematically takes this into account.

In this paper, we set out to address this by exploring how entanglement behaves in a number of boost scenarios with different momenta as well as geometries. We focus on massive two particle spin-$1/2$ systems with discrete momenta in product and entangled states. While it is common to assume that the spin state is given by a maximally entangled Bell state, we extend the treatment to mixed states by considering spins in the Werner state. We also discuss how one can visualize the orbit of the spin state in a 3D manner in order to gain deeper insight into how entanglement changes under Lorentz boosts. The aim is to provide a simple discrete framework which can be used to explain the results involving both discrete as well as continuous momenta. Surveying a range of different momenta and geometries will also contribute to an overview of the kinds of systems that could be of interest for relativistic quantum information.

The paper is organized as follows. We begin by setting the stage in section \ref{sec:theGeneralSetting}, followed by a characterization of the Thomas--Wigner rotation. The next three sections describe the discrete model used throughout the paper, focussing on the momentum and spin states in sections \ref{sec:boostScenariosAndSpinRotations} and \ref{sec:spinStateAndItsVisualization}, respectively. Thereafter we turn to studying the behavior of mixed spin states in boost scenarios which contain different kinds of product and entangled momenta. We conclude with a discussion of the results obtained.

\section{The general setting}\label{sec:theGeneralSetting}

We will focus on a system consisting of two massive spin-$1/2$ particles with spin and momentum, and ask how the spin state changes when viewed from a different inertial frame. This question has a trivial answer in non-relativistic quantum theory: the state will remain unchanged. But the relativistic world is different. A Lorentz boosted observer will see in general a transformed spin state and the reason is the so-called Wigner rotation, or Thomas--Wigner rotation (TWR), where the latter form is commonly used in honor of Thomas's contribution by discovering the Thomas precession \cite{thomas_motion_1926,thomas_kinematics_1927}. By way of illustration, consider a simple, one particle system which forms the smallest entity---the `qubit'---of relativistic quantum information in inertial frames. Suppose the particle is moving relative to observer $O$ who describes its state by
\begin{align}\label{eq:genericDiscreteState}
\ket{\Psi} = \sum_{\ppp, \lambda} \psi_{\lambda}(\ppp) \ket{\ppp, \lambda},
\end{align}
where $\ket{\ppp, \lambda}$ is a basis vector with $\ppp$ labeling the momentum and $\lambda$ the spin of the particle. For the sake of simplicity we restrict our attention to discrete momentum states. Observer $O^{\Lambda}$ who is Lorentz boosted by $\Lambda^{-1}$ relative to $O$ assigns in general a different state $\ket{\Psi^{\Lambda}} = U(\Lambda)\ket{\Psi}$ to the same system, where $U(\Lambda)$ is the unitary representation of $\Lambda$. In order to calculate $\ket{\Psi^{\Lambda}}$, we need the action of $U(\Lambda)$ on a basis vector,  
\begin{align}\label{eq:LorentzBoostActionOnBasis}
U(\Lambda) \ket{\ppp, \lambda} = \ket{\Lambda\ppp} U\!\left[W(\Lambda, \ppp)\right] \ket{\lambda}, 
\end{align}
where $U\!\left[W(\Lambda, \ppp)\right]$ is the unitary representation of the TWR. This means that to observer $O^{\Lambda}$ the spin appears rotated by $U\!\left[W(\Lambda, \ppp)\right]$. The rotation depends on the geometry, i.e.\ the angle between the two boosts, and the momenta of both the system and the observer. The state (\ref{eq:genericDiscreteState}) then transforms as follows, 
\begin{align}
\ket{\Psi^{\Lambda}} = \sum_{\ppp, \lambda} \psi_{\lambda}(\ppp) \ket{\Lambda\ppp} U\!\left[W(\Lambda, \ppp)\right] \ket{\lambda}. 
\end{align}
An interesting implication is that states whose spin and momentum are separable for the rest observer $O$ may display spin--momentum entanglement to the moving observer $O^{\Lambda}$, see \cite{peres_quantum_2002, dunningham_entanglement_2009, palge_generation_2012} for details.

The curious dependency of spin on momentum can be conceptualized using an analogy from quantum information theory \cite{peres_quantum_2002, gingrich_quantum_2002}. Consider a quantum gate with two input qubits, the control qubit and the target qubit. Suppose the action of the gate is to change the target qubit depending on the value of the control qubit. If the target qubit is transformed by a unitary transformation $U$, then such a gate is called a controlled-$U$ gate. If we think of momenta as control qubits, the Lorentz boost in (\ref{eq:LorentzBoostActionOnBasis}) can be conceived of as a controlled unitary: when the boost angle and rapidity are fixed, then the transform on the spin state depends solely on the momentum state \footnote{Note that there is a slight discrepancy between the notions of a controlled unitary and a Lorentz boost. The former does not change the control qubit whereas the latter does: a Lorentz boost alters the momentum state.}. For a discrete system this can be formally written as
\begin{align}
\sum_{\ppp} \pouter{\Lambda\ppp}{\ppp} \otimes U[W(\Lambda, \ppp)].
\end{align}
Although we will not make explicit use of this expression, the notion that Lorentz boosts are controlled unitaries where momentum qubits govern the behavior of spin qubits is central to our thinking of the relativistic spin--momentum systems studied in this paper.

Let us next consider two particles, the system we are primarily interested in, and assume that observer $O$ describes it by the pure state
\begin{align}
\ket{\Psi} = \sum_{\ppp, \qqq, \lambda, \kappa} \psi_{\lambda\kappa}(\ppp, \qqq) \ket{\ppp, \qqq}\ket{\lambda, \kappa}. 
\end{align}
where the first label of either momentum or spin refers to the first particle and the second to the second particle. The boosted observer $O^\Lambda$ sees the state transformed by the tensor product of single particle unitaries, $\ket{\Psi^{\Lambda}} = U(\Lambda) \otimes U(\Lambda)\ket{\Psi}$, 
\begin{align}\label{eq:genericBoostedPureState}
\ket{\Psi^\Lambda} &= \sum_{\ppp, \qqq, \lambda, \kappa} 
\psi_{\lambda\kappa}(\ppp, \qqq) \ket{\Lambda\ppp, \Lambda\qqq} U\!\left[W(\Lambda, \ppp)\right]\ket{\lambda} \nonumber\\
&\phantom{AAAAAAAAAAAAAAAA}\otimes U\!\left[W(\Lambda, \qqq)\right]\ket{\kappa}. 
\end{align} 
with each spin undergoing a momentum dependent rotation and for a generic state this induces a non-trivial transformation on the spin degree of freedom. Our focus will now shift importantly. Whereas in the case of the single particle we were interested in how boosts entangle spin and momentum, in the case of two particle systems we are concerned with how boosts change the entanglement between the spins of the particles. It is in this sense that single particle systems provide a foundation: the physical mechanism which leads to nontrivial transformations of two spins is precisely the one that causes entanglement between the momentum and spin of a single particle.

Although the characterization of composite spin behavior is considerably less straightforward because the structure of the two particle state space is more complicated, it will be our aim to determine this behavior by surveying the landscape of maps that momenta induce on the spin degree of freedom. The motivation to do so comes from the single particle. In \cite{palge_generation_2012} we learned that single particle entanglement is highly sensitive to the boost geometry in question. This naturally leads to the question of how different momentum states and boost geometries affect the entanglement of a bipartite spin state under Lorentz boosts. We will analyze the situation by studying different kinds of product and entangled momenta, and by exploring the geometry of the TWR. Conceiving of momenta in a discrete manner as qubits provides a simple yet powerful model to probe the structure of maps that boosts induce on spins. In analogy to quantum information theory, we view momentum qubits (either product or entangled) as a relativistic resource that enables the manipulation of the spin qubits.

\section{Thomas--Wigner rotation}\label{sec:WignerRotation}

\begin{figure}[htb]
\includegraphics[width=0.45\textwidth]{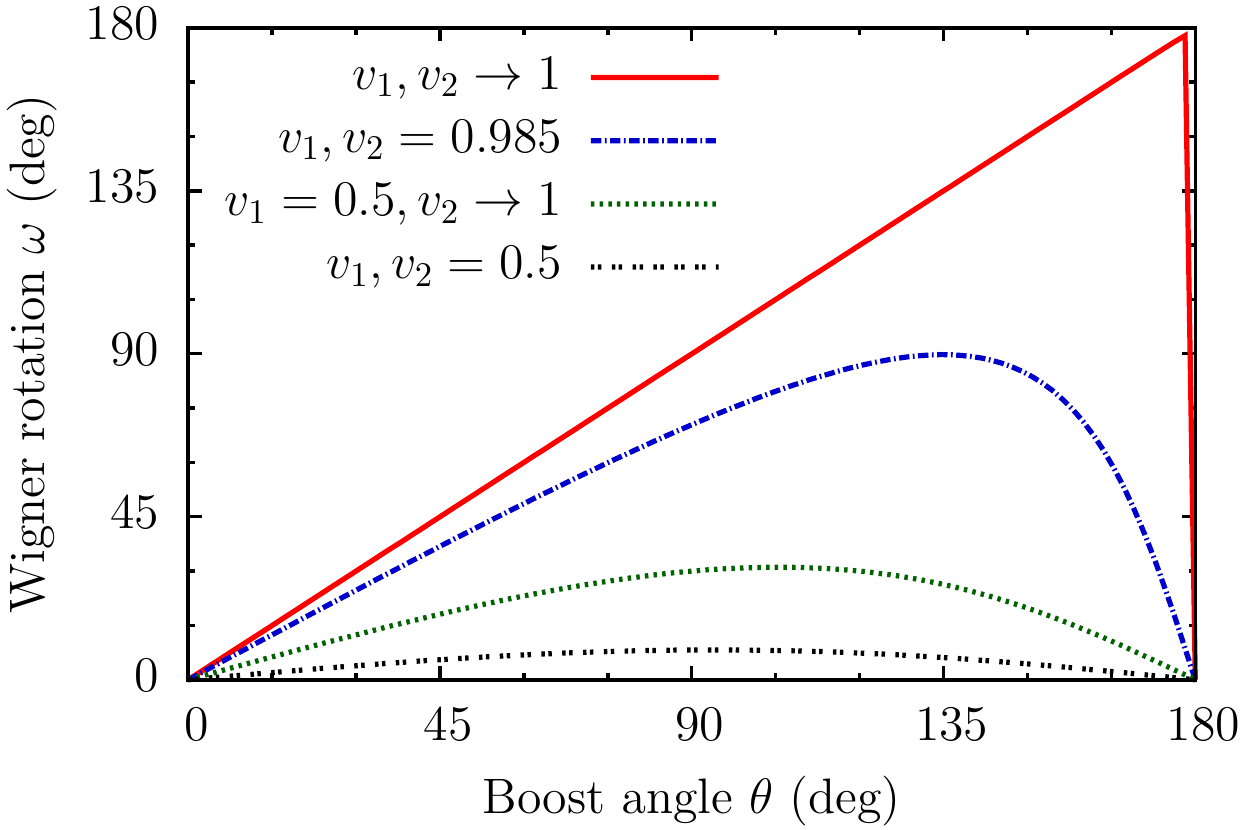}
\caption{\label{fig:fig_1_wigner_rotation}Dependence of TWR on the angle $\theta$ between two boosts.}
\end{figure}

TWR arises from the fact that the subset of Lorentz boosts does not form a subgroup of the Lorentz group. Consider three inertial observers $O$, $O'$ and $O''$ where $O'$ has velocity $\vvv_1$ relative to $O$ and $O''$ has $\vvv_2$ relative to $O'$. Then the combination of two canonical boosts $\Lambda(\vvv_1)$ and $\Lambda(\vvv_2)$ that relates $O$ to $O''$ is in general a boost \emph{and} a rotation,
\begin{align}
\Lambda(\vvv_2) \Lambda(\vvv_1) = R(\omega) \Lambda(\vvv_3),
\end{align}
where $R(\omega)$ is the TWR with 
angle $\omega$. To an observer $O$, the frame of $O''$ appears to be rotated by $\omega$. We will immediately specialize to massive systems, then 
$R(\omega) \in \mathrm{SO(3)}$ and $\omega$ is given by \citep{rhodes_relativistic_2004,halpern_special_1968},
\begin{align}\label{eq:TWRHalfTanFormula}
\tan \frac{\omega}{2} = \frac{\sin \theta}{\cos \theta + D},
\end{align}
where $\theta$ is the angle between two boosts or, equivalently,
$\vvv_1$ and $\vvv_2$, and
\begin{align}\label{eq:theDFactor}
D = \sqrt{\left( \frac{\gamma_1 + 1}{\gamma_1 - 1} \right) \left( \frac{\gamma_2 + 1}{\gamma_2 - 1} \right)},
\end{align}
with 
$\gamma_{1,2} = (1 - v_{1,2}^2)^{-1/2}$ and $v_{1,2} = |\vvv_{1,2}|$.
We assume natural units throughout, $\hbar = c = 1$. The axis of rotation specified by $\nnn = \vvv_2 \times \vvv_1 / |\vvv_2 \times \vvv_1|$
is orthogonal to the plane defined by $\vvv_1$ and $\vvv_2$. The dependence of TWR on the angle between two boosts is shown in Fig.~\ref{fig:fig_1_wigner_rotation}.

Several interesting characteristics are immediately noticeable. First, for any two boosts with velocities $\vvv_1, \vvv_2$ at an angle $\theta$, the TWR increases with both $v_1, v_2$, approaching the maximum value $180\degree$ as $v_1, v_2$ approach the speed of light. Second, the maximum value of $\omega$ is bounded by the smaller boost. If $v_1 = 0.5$, then even if $v_2$ becomes arbitrarily close to the speed of light, $\omega$ will be considerably lower than in the case when both boosts approach the speed of light. Third, the angle $\theta$ at which the maximum TWR occurs depends on the magnitudes of both $v_1$ and $v_2$. It is worth noting that $\omega$ approaches the maximum value $180\degree$ when both boosts are almost opposite and both $v_1, v_2 \rightarrow 1$. At lower velocities, maximum rotation occurs earlier.

In the subsequent discussion, we will often make use of the fact that TWR ranges from $0$ to $\pi$, assuming that the rotation by a given angle $\omega$ is realized by some boost configuration. The latter can be specified in terms of velocities and boost angle, $(\vvv_1, \vvv_2, \theta)$, which need not be unique as the foregoing discussion shows because a given rotation can be realized by several different configurations. However, below we will follow common practice in quantum theory and write the TWR operators using a parameterization in terms of momentum $\ppp = \vvv_1 E(\ppp)$ and boost $\Lambda \equiv \Lambda(\xi, \eee)$, where $\xi = \mathrm{arctanh}\, v_2$ is rapidity of the boost in the direction of $\eee = \vvv_2 / |\vvv_2|$.

\section{The discrete model}\label{sec:theDiscreteModel}

In this section we will describe the model of the two particle particle system to be studied in detail below. The next two sections elaborate on the momentum and the spin subsystems respectively.

To ease the investigation, we begin by assuming that momenta can be treated as discrete variables \cite{jordan_lorentz_2007}. This is justified when they are given by narrow distributions centered around different momentum values such that we can represent them by orthogonal state vectors which formally satisfy the relationship $\langle \ppp | \ppp' \rangle = \delta_{\ppp\ppp'}$. For each $\ket{\ppp}$ we use a single TWR, $U\!\left[W(\Lambda, \ppp)\right]$. Although narrow momenta are an idealization, they constitute a system worth studying. Discrete momenta are computationally easier to deal with than continuous ones but display qualitative features that carry over to systems with continuous momenta. Examples can be found in \cite{palge_surveying_2014} where we are concerned with continuous systems.

Throughout we will assume that spin and momentum are initially, i.e.\ in the rest frame, in a product state, so the total state of the system is given by 
\begin{align}\label{eq:generalStateFiniteDimMomentaSpinMomFact}
\rho = \rho_M \otimes \rho_S, 
\end{align}
where $\rho_S$ is the spin state of the two particle system, and $\rho_M = \pouter{M}{M}$ is a projector on the pure momentum 
\begin{align}\label{eq:generalFiniteDimMomentum}
\ket{M} = \sum_{\ppp, \qqq} \psi_M(\ppp, \qqq) \ket{\ppp, \qqq}. 
\end{align}
To the boosted observer $O^{\Lambda}$, the state of the composite system is transformed by the tensor product of single particle transforms $U(\Lambda) \otimes U(\Lambda)$,
\begin{align}\label{eq:boostedTotalDensityOperator}
\rho \longmapsto \rho^{\Lambda} = U(\Lambda) \otimes U(\Lambda) \, \rho \, U^{\dagger}(\Lambda) \otimes U^{\dagger}(\Lambda).
\end{align}
By combining (\ref{eq:generalStateFiniteDimMomentaSpinMomFact}), (\ref{eq:generalFiniteDimMomentum}) and (\ref{eq:boostedTotalDensityOperator})
we write the boosted state as
\begin{align}\label{eq:LorentzBoostedFiniteDimMomentaSpinMomFact}
\rho^\Lambda &= \sum_{\ppp, \qqq, \ppp', \qqq'} \psi_M(\ppp, \qqq) \psi^*_M(\ppp', \qqq') \ket{\Lambda\ppp, \Lambda\qqq} \bra{\Lambda\ppp', \Lambda\qqq'} \nonumber\\
&\;\;\otimes U\!\left[W(\Lambda, \ppp, \qqq)\right] \, \rho_S \, U^{\dagger}\!\left[W(\Lambda, \ppp', \qqq')\right],
\end{align}
where $U\!\left[W(\Lambda, \ppp, \qqq)\right]$ stands for the product of single particle unitaries,
\begin{align}
U\!\left[W(\Lambda, \ppp, \qqq)\right] \equiv U\!\left[W(\Lambda, \ppp)\right] \otimes U\!\left[W(\Lambda, \qqq)\right].
\end{align}

Note that whereas the rest frame state (\ref{eq:generalStateFiniteDimMomentaSpinMomFact}) factorizes between spin and momentum, the boosted state (\ref{eq:LorentzBoostedFiniteDimMomentaSpinMomFact}) does not. This means that the assumption made at the beginning, namely that spin and momentum factorize, is less restricting than it seems at first sight. By studying how spin--momentum product states are transformed to entangled states, we are also investigating the dual situation where entangled states are mapped to product states. This is simply because we can regard either frame as a rest frame and the other a moving frame since all inertial frames are on equal footing. Neither can be singled out as \emph{the} rest frame or \emph{the} moving frame. Also, we are always guaranteed to have inverses of maps since Lorentz boosts form a group. In this paper, however, we will be concerned mostly with the analysis of spin--momentum product states, proper analysis of spin--momentum entanglement is beyond the scope of this paper and will be left for another occasion.

Since we are interested in how the spin state changes under boosts, we trace out momenta in Eq.~(\ref{eq:LorentzBoostedFiniteDimMomentaSpinMomFact}), obtaining the spin state $\rho^{\Lambda}_S = \MTR_{\ppp,\qqq} \left( \rho^\Lambda \right)$, 
\begin{align}\label{eq:LorentzBoostedSpinStateSpinMomFact} 
\rho^{\Lambda}_S = \sum_{\ppp, \qqq} |\psi_M(\ppp, \qqq)|^2 U\!\left[W(\Lambda, \ppp, \qqq)\right] \, \rho_S \, U^{\dagger}\!\left[W(\Lambda, \ppp, \qqq)\right].
\end{align}

Because the spins will be generally in a mixed state, we will use concurrence to quantify the degree of entanglement. Concurrence $C$ of a bipartite state $\rho$ of two qubits is defined as
\begin{align} 
C(\rho) = \mathrm{max} \{0, \lambda_1 - \lambda_2 -\lambda_3 - \lambda_4 \},
\end{align}
where the $\lambda_i$ are square roots of eigenvalues of a non-Hermitian matrix $\rho \widetilde{\rho}$ in decreasing order and 
\begin{align} 
\widetilde{\rho} = \left( \sigma_y \otimes \sigma_y \right) \rho^* \left( \sigma_y \otimes \sigma_y \right),
\end{align}
with $\sigma_y$ a Pauli matrix, is the spin-flipped state with the complex conjugate $^*$ taken in the standard basis~\cite{wootters_entanglement_1998}.

\section{Momenta and spin rotations}\label{sec:boostScenariosAndSpinRotations}

Since the behavior of spin entanglement depends on the map generated by the momenta, it will be of interest to study momentum states of various forms. We will explore both product and entangled momenta, and begin with the simplest case of the former, 
\begin{align}
\ket{M_{\text{EPRB}}} = \ket{\ppp, \qqq}.
\end{align}
States of this form represent an oft discussed scenario in the literature~\cite{czachor_einstein-podolsky-rosen-bohm_1997, alsing_lorentz_2002, terashima_relativistic_2003, caban_lorentz-covariant_2005, caban_einstein-podolsky-rosen_2006}. Setting $\qqq = -\ppp$ corresponds to the familiar EPR--Bohm setup where the spins are in a Bell state, and the first particle moves in the $\ppp$-direction while the other particle moves in the opposite direction \cite{einstein_can_1935, bohm_quantum_1951}.

In the case of single particle in \cite{dunningham_entanglement_2009} the momentum state was of the form of symmetrically displaced terms $\left( \ket{\ppp} + \ket{-\ppp} \right) / \sqrt{2}$, and we saw that such a state generated maximal entanglement between spin and momentum. This suggests that similar behavior for two particles might be observed when momenta contain analogous terms for both particles,   
\begin{align}\label{eq:twoParticleGeneralSymmetricMomentum}
\ket{M_\Sigma} = \frac{1}{2} \left( \ket{\ppp} + \ket{-\ppp} \right) \left( \ket{\qqq} + \ket{-\qqq} \right).
\end{align}
where $\Sigma$ signifies the fact momenta take symmetric values.

Generalizing further, we get a momentum state where both particles are in a superposition of momenta along a given direction $\pm\ppp$ and a direction perpendicular to this,   
$\pm \ppp_{\perp}$, 
\begin{align}
\ket{M_{\times}} = &\frac{1}{4}
\left( \ket{\ppp} + \ket{-\ppp} + \ket{\ppp_{\perp}} + \ket{-\ppp_{\perp}} \right) \nonumber\\
&\otimes \left( \ket{\qqq} + \ket{-\qqq} + \ket{\qqq_{\perp}} + \ket{-\qqq_{\perp}} \right).
\end{align}   
We will see below that momenta of such a form provide a good approximation to the two particle model considered in the seminal paper \cite{gingrich_quantum_2002}.

We would also like to study entangled momenta since they give rise to interesting behavior in the quantum domain. We assume the generic form of such momenta is given by 
\begin{align}\label{eq:twoParticle_ch1:genericEntangledMomenta} 
\ket{M_{\pm}} &= \frac{1}{\sqrt{2}} \left( \ket{\ppp_1, \qqq_1} \pm \ket{\ppp_2, \qqq_2} \right).
\end{align} 
Since we are surveying the logical structure of spin rotations and would like to study the maximal changes that momenta might generate, we will choose momenta to be maximally entangled. For instance, by setting $\ppp_1 = \qqq_1 = -\ppp_2 = -\qqq_2 = \ppp$, we get 
\begin{align}\label{eq:twoParticle_ch1:BellMomenta1}
\ket{M_{\Phi\pm}} &= \frac{1}{\sqrt{2}} \left( \ket{\ppp, \ppp} \pm \ket{-\ppp, -\ppp} \right), 
\end{align}
which correspond to the Bell states $\ket{\Phi^{\pm}}.$ Likewise, by choosing $\ppp_1 = -\qqq_1 = -\ppp_2 = \qqq_2 = \ppp$, we obtain counterparts of the Bell states $\ket{\Psi^{\pm}},$
\begin{align}\label{eq:twoParticle_ch1:BellMomenta2}
\ket{M_{\Psi\pm}} &= \frac{1}{\sqrt{2}} \left( \ket{\ppp, -\ppp} \pm \ket{-\ppp, \ppp} \right). 
\end{align}
This state has been studied to some extent in~\cite{jordan_lorentz_2007, friis_relativistic_2010}.

In general momenta may lie along different axes. For example, if we specify that the momenta of the first particle are given by 
$\ppp_1 = -\ppp_2 = \ppp$, 
whereas the second particle has 
$\qqq_1 = -\qqq_2 = \ppp_{\perp}$, 
then we get states that resemble $\ket{\Phi_+}$. We signify
\begin{align}\label{eq:twoParticle_ch1:BellMomenta1}
\ket{M_{[\Phi\pm]}} &= \frac{1}{\sqrt{2}} \left( \ket{\ppp, \ppp_{\perp}} \pm \ket{-\ppp, -\ppp_{\perp}} \right). 
\end{align}
For $\ket{\Psi_+}$ we obtain a similar state when we choose 
$\ppp_1 = -\ppp_2 = \ppp$ and $\qqq_1 = -\qqq_2 = -\ppp_{\perp}$,  
\begin{align}\label{eq:twoParticle_ch1:BellMomenta1}
\ket{M_{[\Psi\pm]}} &= \frac{1}{\sqrt{2}} \left( \ket{\ppp, -\ppp_{\perp}} \pm \ket{-\ppp, \ppp_{\perp}} \right). 
\end{align}

Note that as long as we are interested in the behavior of spins, the relative phases of momenta do not matter. This is because the expression for the boosted spin state, Eq.~(\ref{eq:LorentzBoostedSpinStateSpinMomFact}), contains only the squared modulus of the momentum wave function, entailing that two momenta $\psi(\ppp)$ and $\psi'(\ppp)$ that are related by a local gauge transformation 
\begin{align}
\psi(\ppp) \mapsto \psi'(\ppp) = \E^{\phi(\ppp)} \psi(\ppp)
\end{align}
induce the same spin orbits~\cite{jordan_lorentz_2007}. Thus it suffices to consider only $\ket{M_{\Phi+}}$, $\ket{M_{\Psi+}}$, $\ket{M_{[\Phi+]}}$ and $\ket{M_{[\Psi+]}}$ the other Bell states will produce exactly the same spin behavior.

Eq.~(\ref{eq:LorentzBoostedSpinStateSpinMomFact}) also leads to another simplification. As long as we are only interested in the boosted spin state, we can also take momenta to be the mixed states that consist of the diagonal elements of the projector on the corresponding pure momenta,
\begin{align}\label{eq:mixedDiagonalMomentumStates}
\rho_{M \dddd} = \mathrm{diag}\,\pouter{M}{M}. 
\end{align} 
This is because if one assumes that the initial momenta are given by a mixed state that consists of the diagonal elements of the corresponding pure momenta, 
\begin{align}
\rho &= \sum_{\ppp, \qqq} |\psi_M(\ppp, \qqq)|^2 \pouter{\ppp, \qqq}{\ppp, \qqq} \otimes \rho_{S},
\end{align}
then a Lorentz boost $\Lambda$ transforms this to 
\begin{align}
\rho^{\Lambda} &= \sum_{\ppp, \qqq} |\psi_M(\ppp, \qqq)|^2
\pouter{\Lambda\ppp, \Lambda \qqq}{\Lambda\ppp, \Lambda\qqq} \nonumber\\
&\phantom{\sum\sum}\otimes U\!\left[W(\Lambda, \ppp, \qqq)\right] \rho_{S} \, U^{\dagger}\!\left[W(\Lambda, \ppp, \qqq)\right].
\end{align}
By tracing out momenta we obtain the spin state
\begin{align} 
\rho_S^{\Lambda} 
&= \sum_{\ppp, \qqq} |\psi_M(\ppp, \qqq)|^2 
U\!\left[W(\Lambda, \ppp, \qqq)\right] \rho_{S} \, U^{\dagger}\!\left[W(\Lambda, \ppp, \qqq)\right],
\end{align}
which is identical to the expression (\ref{eq:LorentzBoostedSpinStateSpinMomFact}) that describes the boosted spin generated by pure momentum states~\cite{jordan_lorentz_2007}. In other words, only the diagonal elements of the momentum matrix contribute to the final spin state. In the following calculations we will use the simpler form given by the mixed momenta (\ref{eq:mixedDiagonalMomentumStates}) since we will be only interested in the spin state \footnote{However, if we were interested in the total state, then we would need to distinguish between pure and mixed momenta since they generate different total spin--momentum states.}.

\subsection{From momenta to rotations}

Although we have specified the general forms that momenta will take, the geometry they might realize is still undetermined. We will now turn to the discussion of how the generic states are implemented by particular momenta and relate them to different types of rotations generated on spins.

Momenta of both particles may be aligned along the same axes, for instance two particles can be in a superposition of momenta along the $x$-axis, yielding the state,
\begin{align}\label{eq:momentaAlongTheSameAxis}
\ket{M_\Sigma^{XX}} = \frac{1}{2} \left( \ket{\ppp_x} + \ket{-\ppp_x} \right) \left( \ket{\qqq_x} + \ket{-\qqq_x} \right).
\end{align}
Or momenta of both particles may be aligned along different axes, for instance the first particle might be in a superposition of momenta along the $x$-axis and the second particle in a superposition along the $y$-axis, 
\begin{align}\label{eq:momentaAlongDifferentAxis}
\ket{M_\Sigma^{XY}} = \frac{1}{2} \left( \ket{\ppp_x} + \ket{-\ppp_x} \right) \left( \ket{\qqq_y} + \ket{-\qqq_y} \right).
\end{align}
Assuming for simplicity that initially the system is in a pure state 
\begin{align}
\ket{\Psi} = \ket{M} \otimes \ket{S}, 
\end{align}
and substituting momentum $\ket{M_\Sigma^{XX}}$ into (\ref{eq:genericBoostedPureState}) we obtain the boosted state 
\begin{align}\label{eq:momentumExplicitlyWithURotations}
\ket{\Psi^{\Lambda}} = &\frac{1}{2} \biggl\{
	\ket{\Lambda_z \ppp_x, \Lambda_z \qqq_x} \, U\!\left[W(\Lambda_z, \ppp_x)\right] \otimes U\!\left[W(\Lambda_z, \qqq_x)\right] \biggr. \nonumber\\
	&+ \ket{\Lambda_z \ppp_x, -\Lambda_z \qqq_x} \, U\!\left[W(\Lambda_z, \ppp_x)\right] \otimes U\!\left[W(\Lambda_z, -\qqq_x)\right] \nonumber\\
	&+ \ket{-\Lambda_z \ppp_x, \Lambda_z \qqq_x} \, U\!\left[W(\Lambda_z, -\ppp_x)\right] \otimes U\!\left[W(\Lambda_z, \qqq_x)\right] \nonumber\\
	&+ \biggl.  \ket{-\Lambda_z \ppp_x, -\Lambda_z \qqq_x} \, U\!\left[W(\Lambda_z, -\ppp_x)\right] \nonumber\\
	&\otimes U\!\left[W(\Lambda_z, -\qqq_x)\right] \biggr\} \ket{S},
\end{align} 
where we have taken the boost in the $z$-direction. Now the operators $U\!\left[W(\Lambda, \ppp)\right]$ for the unitary representation of the TWR in this expression are given in terms of momenta, the direction of boost and rapidity, that is, variables which specify the configuration of the boost in the physical three space. Formally they are $\mathrm{SU}(2)$ operators parameterized by the latter three quantities. However, as long as our main interest lies in how boosts affect spins, we can simplify the calculations by hiding away the concrete physical situation and using a well known parameterization of $\mathrm{SU}(2)$ in terms of the angle of rotation $\omega$, 
\begin{align}\label{eq:theSpinHalfRotationOperator}
R_{\nnn}(\omega) = \exp \left( -\II \frac{\omega}{2} \boldsymbol\sigma \nnn \right),
\end{align} 
where $\nnn = (n_x, n_y, n_z)$ is the axis of rotation given by a real unit vector in three dimensions and $\boldsymbol\sigma = (\sigma_x, \sigma_y, \sigma_z)$ denotes the three component vector of Pauli matrices.

\begin{figure}[htb]
\centering
\includegraphics[width=0.35\textwidth]{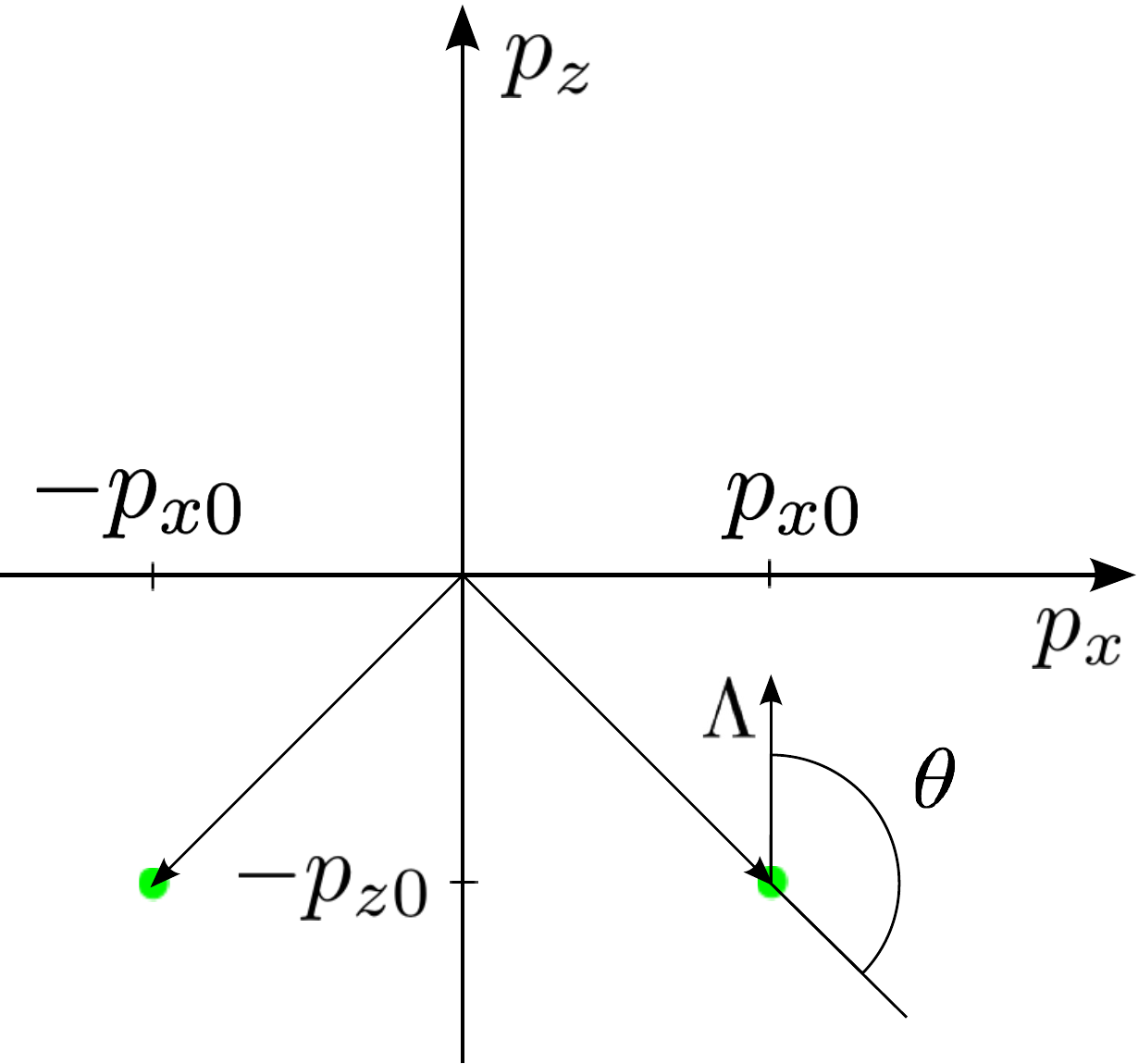}
\caption{Schematic illustration of a boost configuration at a large angle $\theta$. Momenta (green circles) are located at $(\pm p_{x0}, 0, -p_{z0})$. The $z$-projection of the spin field is indicated by an arrow at the momentum. Boost $\Lambda$ is in the positive $z$-direction.}
\label{fig:D3}
\end{figure}

Indeed, this is how we will proceed. In the following calculations we will represent TWR by operators of the form (\ref{eq:theSpinHalfRotationOperator}) where $\omega$ is a rotation angle that ranges from $0$ to $\pi$. Accordingly, we will write $R(\omega)$ instead of $U\!\left[W(\Lambda, \ppp)\right]$ for single particle rotations, and $R(\omega, \chi)$ instead of $U\!\left[W(\Lambda, \ppp, \qqq)\right]$ for two particles. The abstraction is legitimate since, as we saw in section \ref{sec:WignerRotation}, any Wigner angle between $0$ and $\pi$ can be realized by some actual boost configuration comprising momenta, direction of boost and rapidity. In particular, although we have been speaking as if momenta in Eqs.~(\ref{eq:momentaAlongTheSameAxis}) and (\ref{eq:momentaAlongDifferentAxis}) lie along some axis, it need not be and typically it is not the case in a general boost configuration. To generate maximal spin rotations large boost angles are needed, which are implemented by momentum vectors typically not aligned with an axis. For instance, if the boost is in the positive $z$-direction, then momenta centered at $\ppp_x = (\pm p_{x0}, 0, -p_{z0})$ realize a state not lying along the $x$-axis and making an angle to the boost direction which increases as the $z$-component decreases, see Fig.~\ref{fig:D3}. 
However, it is the $x$-component that determines the boost plane (when the boost is assumed to be in the $z$-direction), and hence the direction of the TWR occurring on the spin. We will therefore adopt the convention that we denote by $\ket{\ppp_i}$, $i \in \{ x, y, z \}$ any state that lies in the boost plane, that is, the plane defined by the $i$-axis and the unit vector $\eee$ in the direction of the boost, but where the boost angle is chosen so as to realize any TWR $\omega \in [0, \pi)$. For instance, for large rotations when $\omega$ approaches $\pi$, the boost angle $\theta$ approaches $\pi$ as well.

\begin{figure*}	
	\centering
	\begin{subfigure}[t]{0.28\textwidth}
		\centering
		\includegraphics[width=\textwidth]{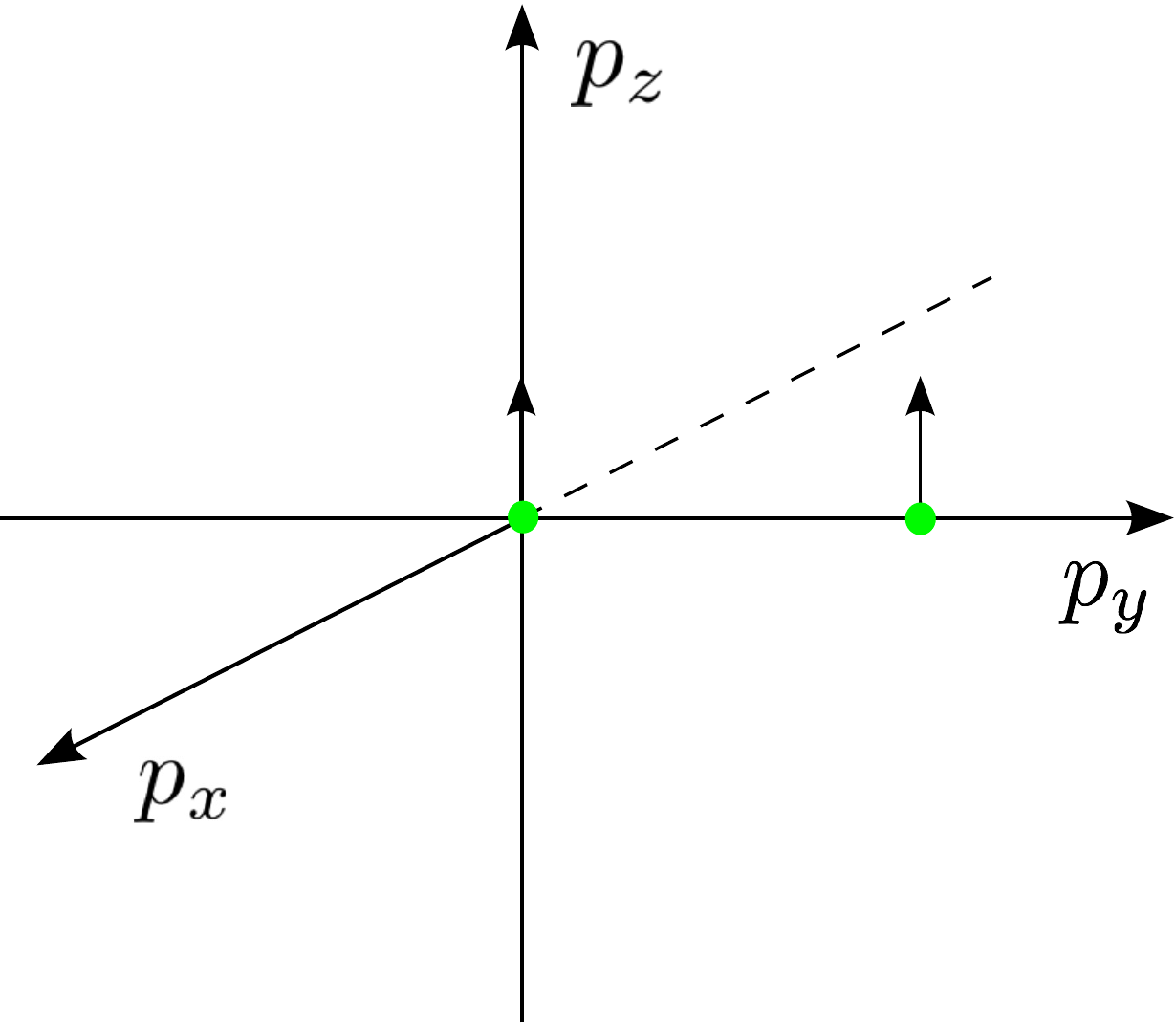}
		\caption{}
		\label{fig:D4a}
	\end{subfigure}
	\hspace{2em}
	\begin{subfigure}[t]{0.28\textwidth}
		\centering
		\includegraphics[width=\textwidth]{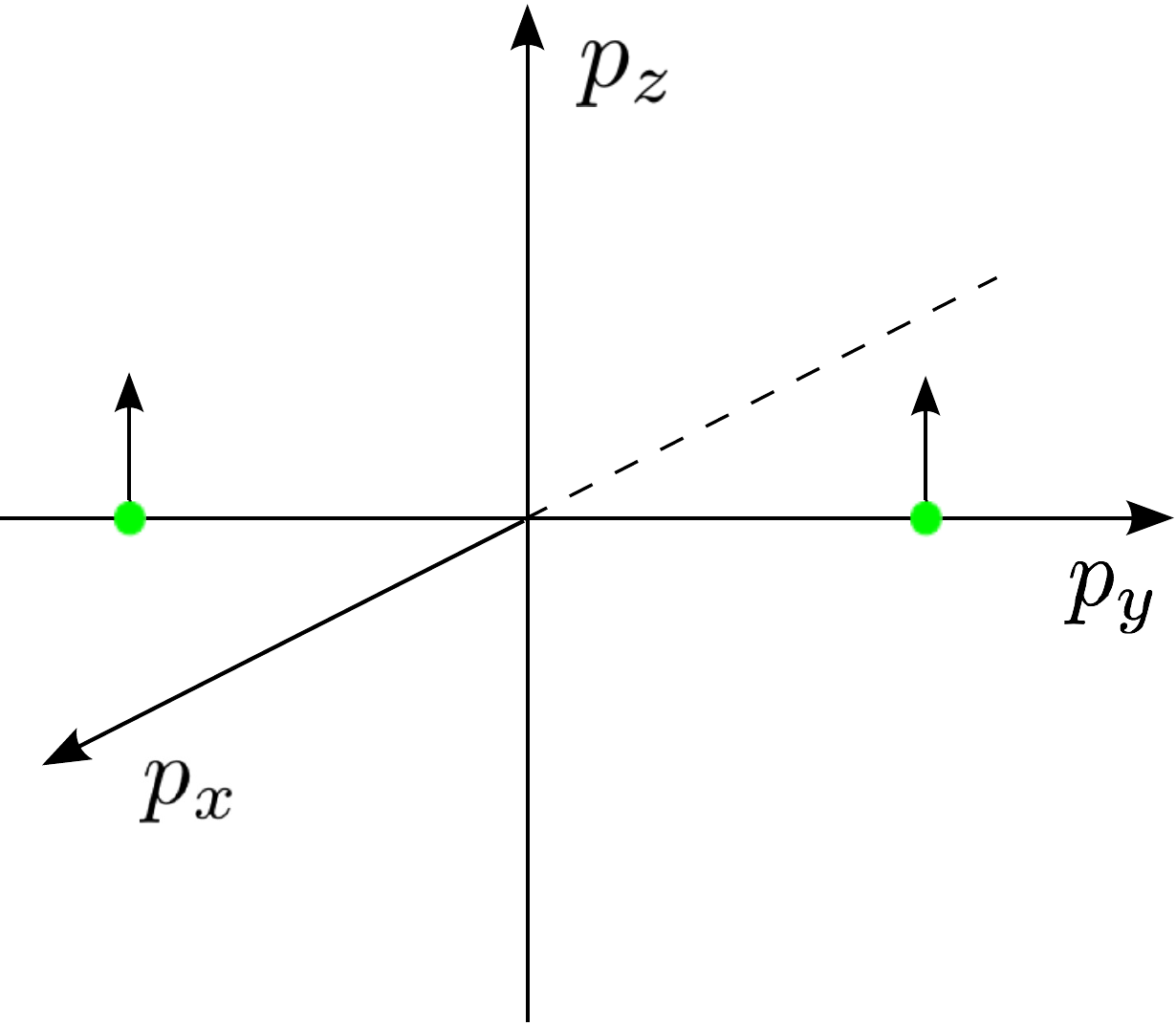}
		\caption{}
		\label{fig:D4b}
	\end{subfigure}
	\hspace{2em}
	\begin{subfigure}[t]{0.28\textwidth}
		\centering
		\includegraphics[width=\textwidth]{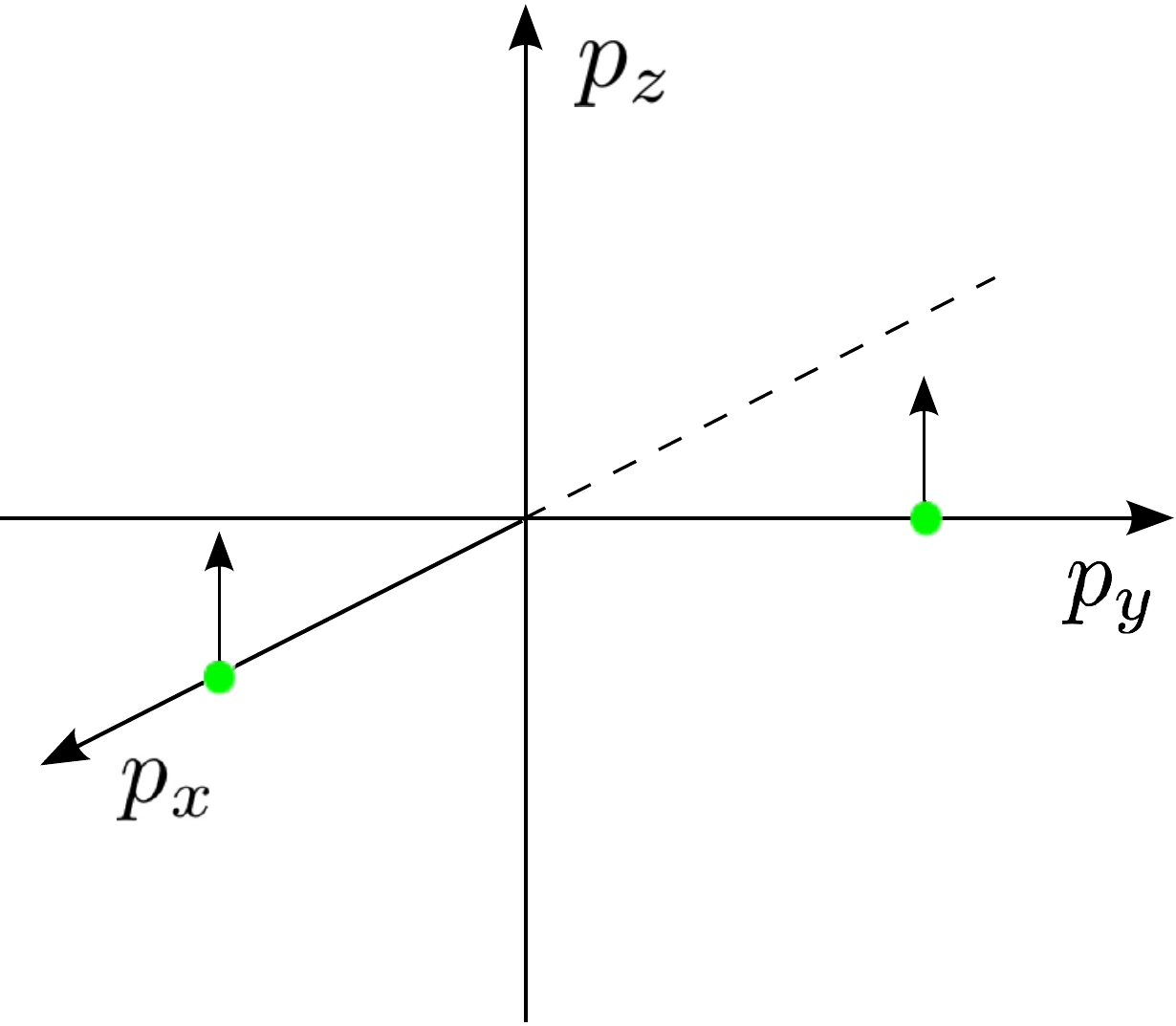}
		\caption{}
		\label{fig:D4c}
	\end{subfigure}
	\caption{Schematic illustration. Examples of geometric configurations of momenta (green circles) for realizations of different types of rotations on spins, with (a) $R_i \otimes \idd$, (b) $R_i \otimes R_i$, (c) $R_i \otimes R_j, i \ne j$. The $z$-projection of the spin field is indicated by an arrow at the momentum.}
	\label{fig:D4}
\end{figure*}

Using parameterization with angles, Eq.~(\ref{eq:momentumExplicitlyWithURotations}) can be written as
\begin{align}
\ket{\Psi^{\Lambda}} = &\frac{1}{2} \bigl[  
	\ket{\Lambda_z \ppp_x, \Lambda_z \qqq_x} \, R_Y(\omega) \otimes R_Y(\chi) \big. \nonumber\\
	&+ \ket{\Lambda_z \ppp_x, -\Lambda_z \qqq_x} \, R_Y(\omega) \otimes R_Y(-\chi) \nonumber\\
	&+ \ket{-\Lambda_z \ppp_x, \Lambda_z \qqq_x} \, R_Y(-\omega) \otimes R_Y(\chi) \nonumber\\
	&+ \big.  \ket{-\Lambda_z \ppp_x, -\Lambda_z \qqq_x} \, R_Y(-\omega) \otimes R_Y(-\chi) 
	\bigr] \ket{S},
\end{align}
where $R_Y(\omega)$ signifies a rotation around the $y$-axis given by~(\ref{eq:theSpinHalfRotationOperator}). Thus we see that the momenta $\ket{M_\Sigma^{XX}}$ generate rotation terms of the form
\begin{align}
R_Y(\pm\omega) \otimes R_Y(\pm\chi), \quad R_Y(\pm\omega) \otimes R_Y(\mp\chi) 
\end{align}
on the spin state. In the same vein, if the momenta are given by $\ket{M_\Sigma^{XY}}$ the $z$-boosted state will have terms that generate rotations  
\begin{align}
R_Y(\pm\omega) \otimes R_X(\pm\chi), \quad R_Y(\pm\omega) \otimes R_X(\mp\chi)
\end{align}
on the spin state. Following considerations along these lines we see that by taking momenta along different combinations of axes for both product and entangled momenta, one obtains three different types of rotations that can occur on the spin state,
\begin{align}\label{eq:rotationTypesGeneral}
\mathrm{(i) }\;\;  &R_i \otimes \idd,  \nonumber\\
\mathrm{(ii) }\;\; &R_i \otimes R_i,  \\ 
\mathrm{(iii) }\;\; &R_i \otimes R_j, \quad i \ne j, \nonumber
\end{align}
where $i, j \in \{ X, Y, Z \}$ and each type of rotation can be realized by some set of suitably chosen momenta, see Fig.~\ref{fig:D4}. 
For instance, we saw that $R_i \otimes R_i$ is instantiated by $R_Y \otimes R_Y$ when momenta are given by the product state $\ket{M_\Sigma^{XX}}$ and the boost is in the $z$-direction. Another implementation of the same type is $R_X \otimes R_X$ when momenta are again product but located along the $y$-axis, $\ket{M_\Sigma^{YY}}$, and the boost is in the $z$-direction.

We will next give a few examples of momenta and boost geometries that implement the different types of rotations listed in~(\ref{eq:rotationTypesGeneral}).

\paragraph{Type $R_i \otimes \idd$.} In this scenario, only the first particle undergoes rotation. The momentum of the second particle is chosen so that it leaves the spin alone. Denoting such a momentum by $\ket{0}$, the following pairs of boosts and momenta listed on the left hand side generate rotations given on the right hand side, 
\begin{align}\label{eq:realizationOfR_i_1_}
\Lambda_z \, , \ket{\ppp_y, 0}  \longmapsto  R_X \otimes \idd, \nonumber\\
\Lambda_z \, , \ket{\ppp_x, 0}  \longmapsto  R_Y \otimes \idd, \\
\Lambda_y \, , \ket{\ppp_x, 0}  \longmapsto  R_Z \otimes \idd. \nonumber
\end{align}

\paragraph{Type $R_i \otimes R_i$.} For scenarios in which both particles are rotated around the same axis but not necessarily in the same direction, we obtain the following boosts and momenta,  
\begin{align}\label{eq:realizationOfR_i_R_i_}
\Lambda_z \, , \ket{\ppp_y, \qqq_y}   \longmapsto   R_X \otimes R_X, \nonumber\\
\Lambda_z \, , \ket{\ppp_x, \qqq_x}   \longmapsto   R_Y \otimes R_Y, \\
\Lambda_y \, , \ket{\ppp_x, \qqq_x}   \longmapsto   R_Z \otimes R_Z. \nonumber
\end{align}

\paragraph{Type $R_i \otimes R_j$, $i \ne j$.} Scenarios where particles undergo rotations around different axes can be realized by 
\begin{align}\label{eq:realizationOfR_i_R_j_}
\Lambda_y \, , \ket{\ppp_z, \qqq_x}   \longmapsto   R_X \otimes R_Z, \nonumber\\
\Lambda_z \, , \ket{\ppp_y, \qqq_x}   \longmapsto   R_X \otimes R_Y, \\
\Lambda_x \, , \ket{\ppp_z, \qqq_y}   \longmapsto   R_Y \otimes R_Z. \nonumber
\end{align}

\vspace{1em}
These scenarios admit an obvious generalization. By choosing momenta and boosts appropriately, one can consider single particle rotations around an arbitrary axis $\nnn = (n_x, n_y, n_z)$. This leads to combinations of generic rotations $R_{\nnn_1} \otimes R_{\nnn_2}$ for two particle systems, opening up a wide avenue of research. However, when surveying the situation for the first time, we would like to keep the situation tractable by confining attention to the cases listed above and leave the more general approach for another occasion.

\section{Spin state and its visualization}\label{sec:spinStateAndItsVisualization}

We will next characterize the spin state of the system. Most of previous work has focussed on the Bell states,  
\begin{align}
\ket{\Phi_{\pm}} = \frac{1}{\sqrt{2}} \left( \ket{00} \pm \ket{11}\right), \quad
\ket{\Psi_{\pm}} = \frac{1}{\sqrt{2}} \left( \ket{01} \pm \ket{10}\right),  
\end{align}
the maximally entangled bipartite states of two level systems. Understanding their behavior in relativity is very important for quantum information. However, pure states are an idealization and in practical situations one would like to know how mixed states behave as well. From the theoretical perspective we are likewise interested in exploring how boosts affect the properties of states with less than maximal entanglement. We will therefore extend the treatment to include the mixed states.

To find the possibly widest range of behavior we would like to study mixed states whose entanglement ranges from maximal to no entanglement at all. These considerations naturally lead to the so-called Werner states. The family of generalized Werner states are the states that interpolate between the maximally mixed and maximally entangled state $P_{+} = \pouter{\Phi_+}{\Phi_+}$, 
\begin{align}
\rho_W(\lambda) = \lambda \pouter{\Phi_+}{\Phi_+} + (1 - \lambda) \frac{1}{N} \idd \quad \mathrm{with}\;\; \lambda \in [0,1] ,
\end{align}
where in the present case $N = 4$ for the bipartite two level systems. For $\lambda = 1$ we recover the Bell state $\ket{\Phi_+}$ and for $\lambda = 0$ we obtain the maximally mixed state $\tfrac{1}{4} \idd$. Values between these two extreme cases correspond to mixed states which range from entangled to separable systems with interesting properties: 
the states with $\lambda > 1/3$ are entangled, but they do not violate the Bell inequality until $\lambda$ becomes larger than $1/\sqrt{2}$, see \cite{werner_quantum_1989, horodecki_separability_1996}. These features of Werner states make them particularly suitable for the purpose of probing the behavior of a wide range of mixed states with different degrees of entanglement.

As regards the geometric configuration, we will assume throughout that the spins are aligned with the $z$-axis irrespective of the direction of the boost. We adopt the convention that $\ket{0}$ signifies that `up' spin and $\ket{1}$ the `down' spin.

\subsection{Visualization}\label{sec:VisualizationStateChange}

In order to gain a better understanding of the state change of a single qubit, one commonly uses visualization on the Bloch sphere. Visualization of two qubits, however, is in general impossible since one needs $15$ real parameters to characterize the density matrix. However, some cases still allow for a representation in three space, for instance when the state is restricted to evolve in a subspace of few dimensions. Fortunately this turns out to be the case for our system.

To characterize mixed states, it is useful to work in the Hilbert-Schmidt space of operators $B(\hilb)$, defined on the Hilbert space $\hilb$ with \mbox{$\text{dim} = N$} \cite{bengtsson_geometry_2006}. $B(\hilb)$ becomes a Hilbert space of $N^2$ complex dimensions when equipped with a scalar product defined as \mbox{$\langle A | B \rangle = \mtr(A^{\dagger}B)$}, with $A, B \in B(\hilb)$, where the squared norm is \mbox{$\| A \|^2 = \mtr(A^{\dagger}A)$}. The vector space of Hermitian operators is an $N^2$ real-dimensional subspace of Hilbert-Schmidt space and can be coordinatized using a basis that consists of the identity operator and the generators of $\sun$. For a qubit $N = 2$ and we obtain the familiar Bloch ball. For a bipartite qubit system $N = 4$, $B(\hilb) = B(\hilb_A) \otimes B(\hilb_B)$ where $\hilb_i$ is the single particle space, and we can use a basis whose elements are tensor products $\{ \idd \otimes \idd, \idd \otimes \boldsymbol\sigma, \boldsymbol\sigma \otimes \idd, \boldsymbol\sigma \otimes \boldsymbol\sigma \}$, where $\boldsymbol\sigma = \left( \sigma_x, \sigma_y, \sigma_z \right)$ is the vector of Pauli operators. The density operator for a $2 \times 2$ dimensional system 
can be written in the general form, 
\begin{align} 
\rho = \frac{1}{4} \left( \idd \otimes \idd + \mathbf{r} \boldsymbol\sigma \otimes \idd + \idd \otimes \mathbf{s} \boldsymbol\sigma + \sum_{i,j} t_{ij} \sigma_i \otimes \sigma_j \right), 
\end{align}
where the coefficients $\mathbf{r} = (r_x, r_y, r_z)$, $\mathbf{s} = (s_x, s_y, s_z)$ and $t_{ij}$, $i, j \in \{ x, y, z \}$ are the expectation values of the operators $\mathbf{r} \boldsymbol\sigma \otimes \idd$, $\idd \otimes \mathbf{s} \boldsymbol\sigma$ and $\sigma_i \otimes \sigma_j$.

\begin{figure}[htb]
\centering
\includegraphics[width=0.4\textwidth]{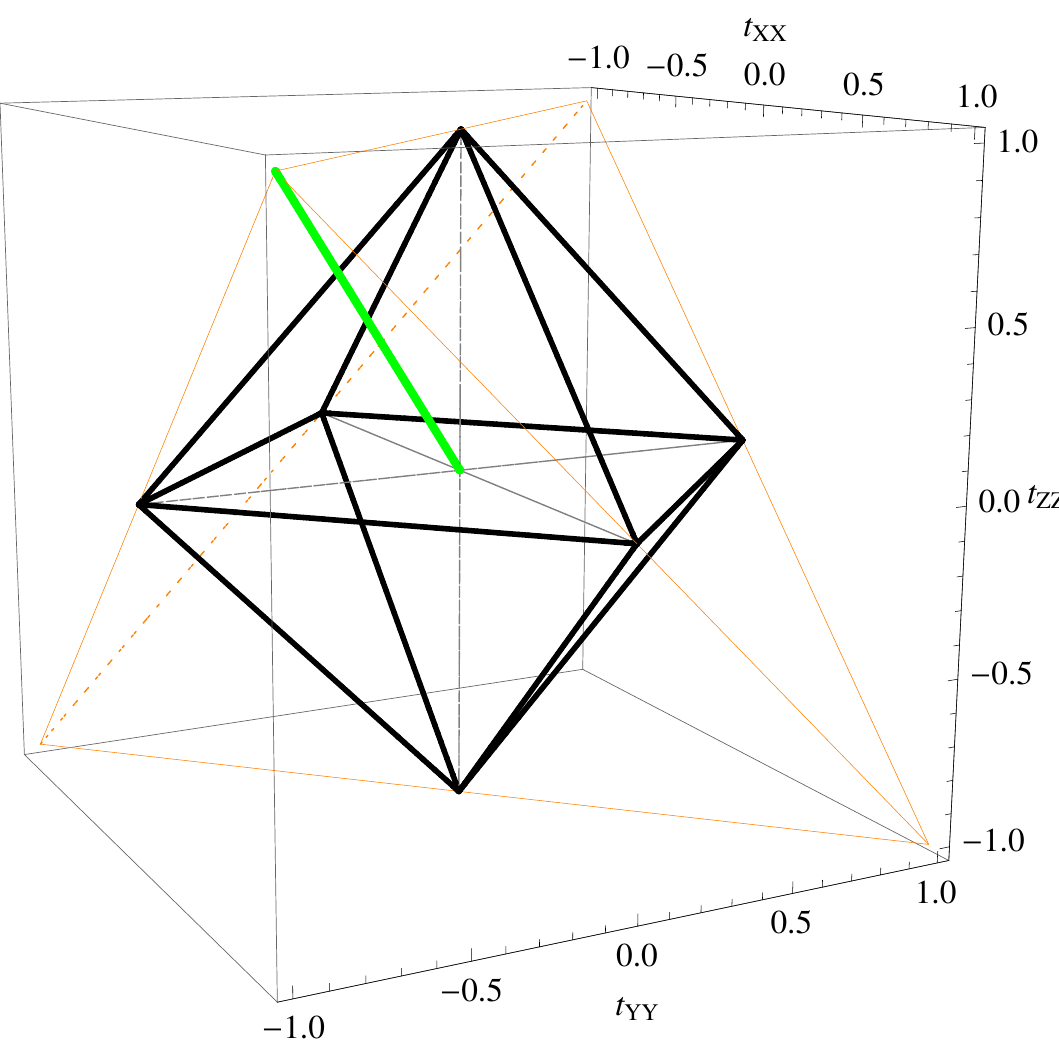}
\caption[The geometry of Bell diagonal states.]{The geometry of Bell diagonal states. The vertices of the tetrahedron $\mathcal{T}$ (yellow) correspond to the four Bell states $\ket{\Phi_+}$, $\ket{\Phi_-}$, $\ket{\Psi_+}$, and $\ket{\Psi_-}$. Convex combinations of projectors on the Bell states, the Bell diagonal states, lie on or in the tetrahedron. A Bell diagonal state is separable iff it lies in the double pyramid formed by the intersection of the tetrahedron $\mathcal{T}$ and its reflection through the origin~$\mathcal{-T}$. Werner states $\rho_W(\lambda)$ (shown green) lie on the line connecting the origin and the vertex at $(1, -1, 1)$.%
}\label{fig:bipartiteTetrahedronGeneral}
\end{figure}

For the projectors on the Bell states $s_i = r_i = 0$ and the matrix $t_{ij}$ is diagonal. This implies we only need to consider the values of diagonal components $t_{ii}$ which constitute a vector in $3$-dimensional space, allowing us to represent the states in Euclidean three space~\cite{bertlmann_geometric_2002}. The Bell states correspond to vectors, 
\begin{align}
t_{\Phi_+} = \left( 1, -1, \phantom{-}1 \right), \quad\quad
t_{\Phi_-} = \left( -1, \phantom{-}1, \phantom{-}1 \right), \nonumber \\
t_{\Psi_+} = \left( 1, \phantom{-}1, -1 \right), \quad\quad
t_{\Psi_-} = \left( -1, -1, -1 \right).
\end{align}
which, in turn, correspond to the vertices of a tetrahedron $\mathcal{T}$ in Fig.~\ref{fig:bipartiteTetrahedronGeneral}. By taking convex combinations of these, one obtains further diagonal states; the set of all such states is called \emph{Bell-diagonal} and is represented by the (yellow) tetrahedron $\mathcal{T}$ in Fig.~\ref{fig:bipartiteTetrahedronGeneral}. 
The set of separable states forms a double pyramid, an octahedron, in the tetrahedron. The octahedron is given by the intersection of $\mathcal{T}$ with its reflection through the origin, $-\mathcal{T}$. The maximally mixed state $\frac{1}{4} \idd_4$ has coordinates $(0, 0, 0)$ and it lies at the origin. The entangled states are located outside the octahedron in the cones of the tetrahedron, see Fig.~\ref{fig:bipartiteTetrahedronGeneral}.

The Werner states lie on the line connecting the origin to the vertex $(1, -1, 1)$ that represents the Bell state $\ket{\Phi_+}$, see Fig.~\ref{fig:bipartiteTetrahedronGeneral}. As the mixture moves from the origin, which represents the maximally mixed state, towards the vertex corresponding to the Bell state, it becomes entangled when crossing the face of the octahedron. This corresponds to the distance $1/\sqrt{3}$ from the origin, or, as mentioned above, to $\lambda_{\text{sep}} = 1/3$. The mixed state state violates the Bell inequality only when the distance from the origin is greater than $\sqrt{3}/ \sqrt{2}$, corresponding to $\lambda = 1 / \sqrt{2}$, see \cite{horodecki_separability_1996}.

We can now visualize the behavior of spin by calculating the coefficients $t_{ii}$ under a given rotation as a function of the Wigner angle $\omega$ and the parameter $\lambda$, 
\begin{align}\label{eq:stateAsThreeVector}
t(\omega, \lambda) = \left(t_{xx}, t_{yy}, t_{zz}\right), 
\end{align}
where
\begin{align}\label{eq:coefficientsOfStateVector}
t_{ii} = \MTR\!\left[\rho^{\Lambda}_S(\omega, \lambda)\, \sigma_i \otimes \sigma_i\right],\quad i \in \{ x, y, z \}.
\end{align}
More precisely, we will choose an initial state $\rho_W(\lambda_1)$ by fixing a particular $\lambda_1$ and then let $\omega$ vary between $0$ and $\pi$. The resulting set of three vectors 
\begin{align}
\Gamma\!\left[\rho^{\Lambda}_W(\omega, \lambda_1)\right] = \{ t(\omega, \lambda_1) \;|\; \omega \in [0, \pi] \}
\end{align}
we call an \emph{orbit} of a given initial state. It can be represented as a curve in three space in the manner described above.

We will use a single parameter $\omega$ to characterize rotations on both particles, making the assumption that momenta of both particles are of equal magnitude, $|\ppp| = |\qqq|$, and both are transformed by boosts with the same rapidity $\xi$. In a more general setting these assumptions may be relaxed, meaning that particles could be subject to different boost geometries, which in turn implies that spins may undergo different rotations. When surveying the topic for the first time, however, we would like to keep the model simple enough in order to gain some insight into how various kinds of momenta affect spin entanglement. In principle, these results can be then later refined by allowing a distinct boost scenario for each particle.

\section{Product momenta}\label{sec:ProductMomenta}

\subsection{Product momenta $\rho_{\text{EPRB}}$}\label{subsec:productSimpleMomentaPQWithWernerState}

We begin by discussing the simplest product state 
\begin{align}
\rho_{\text{EPRB}} = \pouter{\ppp, \qqq}{\ppp, \qqq}. 
\end{align}
Since we have only one momentum term, $\rho_{\text{EPRB}}$ generates a map on the spin state given by a local unitary of the form $U_1 \otimes U_2$,    
\begin{align}
U_1 \otimes U_2 : \rho_{W} \longmapsto  \rho^{\Lambda}_{W} = \lambda \pouter{\Phi_+^{\Lambda}}{\Phi_+^{\Lambda}} + \left( 1 - \lambda \right) \frac{1}{4} \idd_4,
\end{align}
where $\ket{\Phi_{+}^{\Lambda}} = U_1 \otimes U_2 \ket{\Phi_+}$ is the boosted Bell state. This is a maximally entangled state because local unitaries do not change the the degree of entanglement of a Bell state. The final spin state $\rho^{\Lambda}_{W}$ again displays the form of a mixture of a maximally entangled and maximally mixed state parameterized by $\lambda$, thus containing the same amount of entanglement as the initial $\rho_{W}$. In summary, the degree of entanglement of spin Werner states remains invariant under maps generated by simple product momenta $\rho_{\text{EPRB}}$. The conclusion holds for all three types of rotations $R_i \otimes \idd$, $R_i \otimes R_i$ and $R_i \otimes R_j$ because they are all special cases of the form $U_1 \otimes U_2$.

The result concerning the Bell states was first noted in \cite{alsing_lorentz_2002}, where the authors carry out a thorough study of both massive spin-$1/2$ particles and massless photons.

We will see shortly that momenta of such a form represent a special case. In general, the entanglement will not remain invariant in boost scenarios where the momentum part of the state contains more terms since the spins will undergo more complicated transformations.

\subsection{Product momenta $\rho_{\Sigma}$}\label{subsec:productMomentaWithWernerState}

In the following sections \ref{sec:spinWernerStateWithProductMomenta_R_i_1}--\ref{sec:spinWernerStateWithProductMomenta_R_i_R_j} we will focus on mixed momenta of the form
\begin{align}
\rho_{\Sigma} = \frac{1}{4} 
&\left( 
\pouter{\ppp}{\ppp} + \pouter{-\ppp}{-\ppp}
\right) 
\left(
\pouter{\qqq}{\qqq} + \pouter{-\qqq}{-\qqq}
\right),
\end{align}
which are the counterpart of the pure product momenta $\ket{M_{\Sigma}}$.

\subsubsection{Case $R_i \otimes \idd$}\label{sec:spinWernerStateWithProductMomenta_R_i_1}

Rotations of the type $R_i \otimes \idd$, which act only on one particle, can be realized by the various geometries listed in (\ref{eq:realizationOfR_i_1_}). For instance, if we choose the boost to be in the $z$-direction, the rotation $R_X \otimes \idd$ occurs in a scenario where the momenta of the first particle lie in the $y-z$-plane while the second particle's momentum $\ket{\ppp_0}$ is located at the origin. The total momentum state is then of the form 
\begin{align}
\frac{1}{2} \left( \pouter{\ppp_y}{\ppp_y} + \pouter{-\ppp_y}{-\ppp_y} \right) \pouter{\ppp_0}{\ppp_0}
\end{align}
Boosting in the $z$-direction translates $\ket{\ppp_0}$ along the $z$-axis, yielding no rotation on the second particle.

Using Eqs.~(\ref{eq:stateAsThreeVector}) and (\ref{eq:coefficientsOfStateVector}) we calculate that the vector corresponding to the boosted spin state $\rho_S^{\Lambda}$ is given by 
\begin{align}
t_{X \otimes \idd}(\omega, \lambda) = \lambda \left(1, -\cos\omega, \cos\omega\right).  
\end{align}
The concurrence is given by
\begin{align}\label{eq:concurrenceForSpinWernerStateWithProductMomenta_R_i_1}
C(\omega, \lambda) = \left\{
 \begin{array}{ll}
 \frac{1}{2} \left( - 1 + \lambda + 2 \lambda |\! \cos\omega | \right) \quad & \text{if}\quad \lambda \in (\lambda_{\text{sep}}, 1] \\
  0\;\;\,\quad\quad\quad\quad\quad\quad\quad\quad\quad & \text{if}\quad \lambda \in [0, \lambda_{\text{sep}}]\\  
 \end{array} \right.
\end{align}
where $\lambda_{\text{sep}}$ corresponds to the point on the face of the octahedron where the initial state crosses the boundary of entangled and separable states.

Direct calculation shows that other rotations induce similar orbits. For $R_Y \otimes \idd$ and $R_Z \otimes \idd$ we obtain
\begin{align}\label{eq:vectorsForR_Y_1AndR_Z_1WithM_Sigma}
t_{Y \otimes \idd}(\omega, \lambda) &= \lambda \left( \cos\omega, -1, \cos\omega \right) , \nonumber\\
t_{Z \otimes \idd}(\omega, \lambda) &= \lambda \left( \cos\omega, -\cos\omega, 1 \right) ,
\end{align}
with the concurrence given by Eq.~(\ref{eq:concurrenceForSpinWernerStateWithProductMomenta_R_i_1}).

\begin{figure*}	
	\centering
	\begin{subfigure}[t]{0.38\textwidth}
		\centering
		\includegraphics[width=\textwidth]{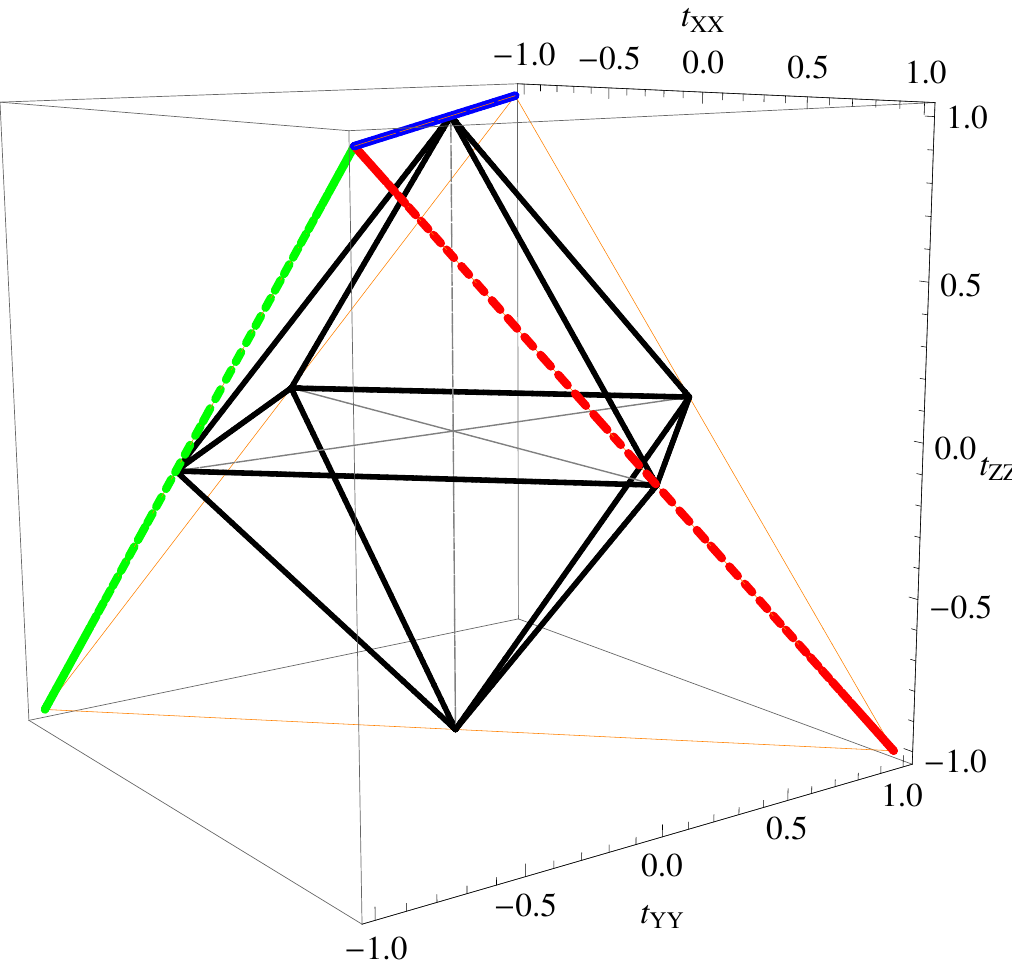}
		\caption{}
		\label{fig:BELL_PhiPlus_Bell_diagonal_maps_EVOLUTION_Ri_1_}
	\end{subfigure}
	\hspace{2em}
	\begin{subfigure}[t]{0.48\textwidth}
		\centering
		\includegraphics[width=\textwidth]{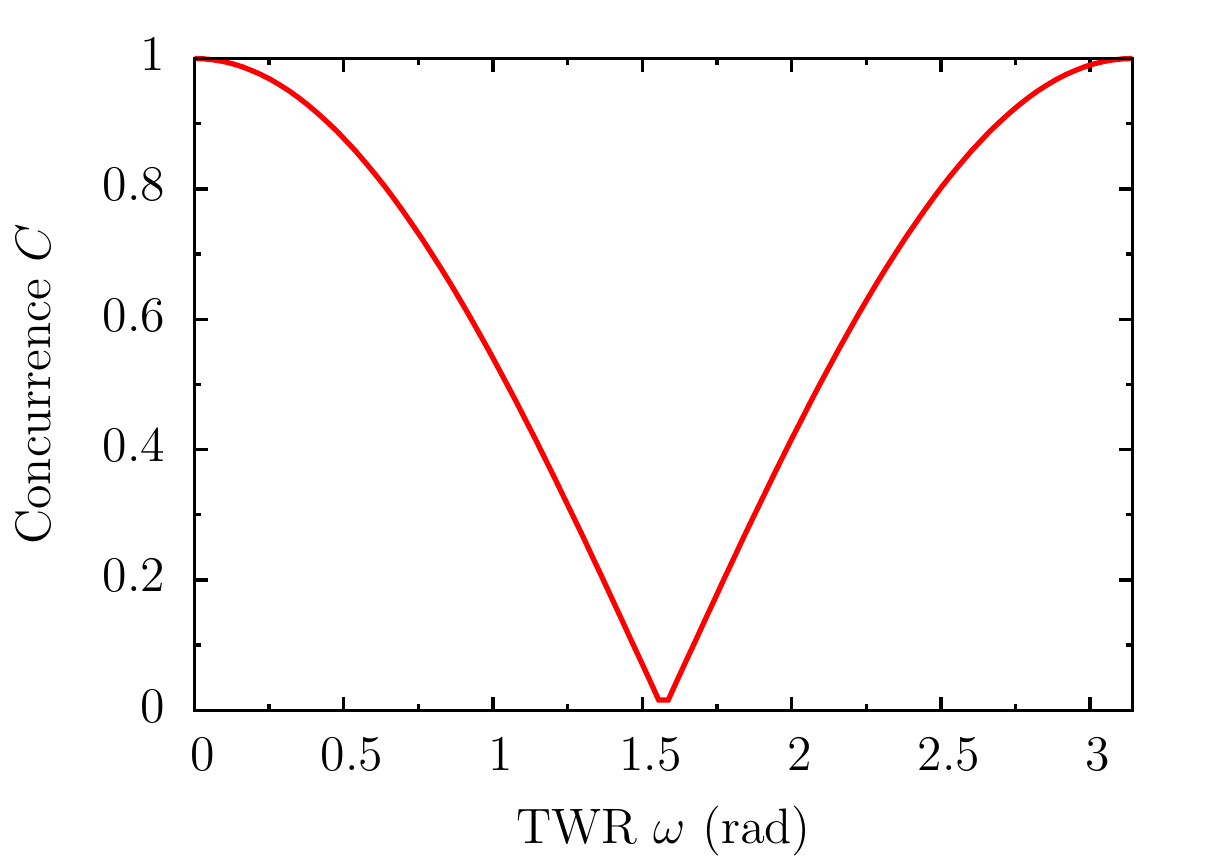}
		\caption{}
		\label{fig:BELL_PhiPlus_Bell_diagonal_maps_CONCURRENCE_Ri_1_}
	\end{subfigure}
	\caption[Spin orbit and concurrence under $R_i \otimes \idd$ with $\rho_{\Sigma}$.]%
{Spin orbit and concurrence under $R_i \otimes \idd$ with $\omega \in [0, \pi]$ generated by $\rho_{\Sigma}$. (a) Initial state $\ket{\Phi_+}$ corresponds to vertex $(1, -1, 1)$, orbit for $R_X \otimes \idd$ is shown red, $R_Y \otimes \idd$ green and $R_Z \otimes \idd$ blue. 
(b) Concurrence has the same shape for all $R_i \otimes \idd$. 
}\label{fig:BELL_PhiPlus_R_i_otimes_1}
\end{figure*}

We start our discussion by considering the Bell states. The state $\ket{\Phi_+}$ is recovered by setting $\lambda = 1$. Fig.~\ref{fig:BELL_PhiPlus_R_i_otimes_1} shows plots of the orbits and the concurrence. 
It becomes clear now that visualization of the orbit provides valuable insight into the behavior of entanglement. For the sake of concreteness, let us focus on $t_{X \otimes \idd}$. Initially the state is at rest, represented by the vertex at $(1, -1, 1)$. As boosts increase, the state moves along the line towards the center of the face (shown red in Fig.~\ref{fig:BELL_PhiPlus_Bell_diagonal_maps_EVOLUTION_Ri_1_}), reaching a separable state represented by $(1, 0, 0)$ at $\omega = \pi / 2$. The latter vector can be written as a convex combination of vectors corresponding to states $\ket{\Phi_+}$ and $\ket{\Psi_+}$,
\begin{align}
\left(1, 0, 0\right) = \frac{1}{2} \left(t_{\Phi_+} + t_{\Psi_+}\right).
\end{align}
As boosts increase further, the state again becomes entangled. Finally, when the Wigner angle is almost $\pi$, the system reaches the vertex $(1, 1, -1)$, that is the boosted observer sees the state $\ket{\Psi_+}$ instead of $\ket{\Phi_+}$.

The behavior of concurrence mimics this pattern. Initially, when the state is maximally entangled the concurrence takes the value one. As the boosts start to increase, this decreases monotonically and reaches zero when $\omega = \pi / 2$. When the boosts become larger, it increases monotonically, approaching one as the rotation becomes close to $\pi$ or equivalently, when boosts approach the speed of light.

The other rotations $R_Y \otimes \idd$ and $R_Z \otimes \idd$ induce similar orbits with vectors given in (\ref{eq:vectorsForR_Y_1AndR_Z_1WithM_Sigma}) and shown as green and blue, respectively, in Fig.~\ref{fig:BELL_PhiPlus_Bell_diagonal_maps_EVOLUTION_Ri_1_}. All three orbits have similar shape, they are related to each other by three-rotations $R(2 \pi n/ 3)$, $n = 1, 2$, where the axis of rotation is the line through the origin $(0, 0, 0)$ and the vertex $(1, -1, 1)$ representing $\ket{\Phi_+}$.

Furthermore, direct calculation shows that the other Bell states exhibit the same behavior under the rotations \mbox{$R_i \otimes \idd$}, $i \in \{ X, Y, Z\}$. For a given state, the rotations generate orbits that are related by three-rotations $R(2 \pi n / 3)$, $n = 1, 2$ around the axis through the origin and the vertex representing the respective state.

The latter two results hold for all nontrivial orbits below; we will therefore refrain from repeating them in the following.

\begin{figure*}[htb]	
	\begin{subfigure}[t]{0.38\textwidth}
		\centering
		\includegraphics[width=\textwidth]{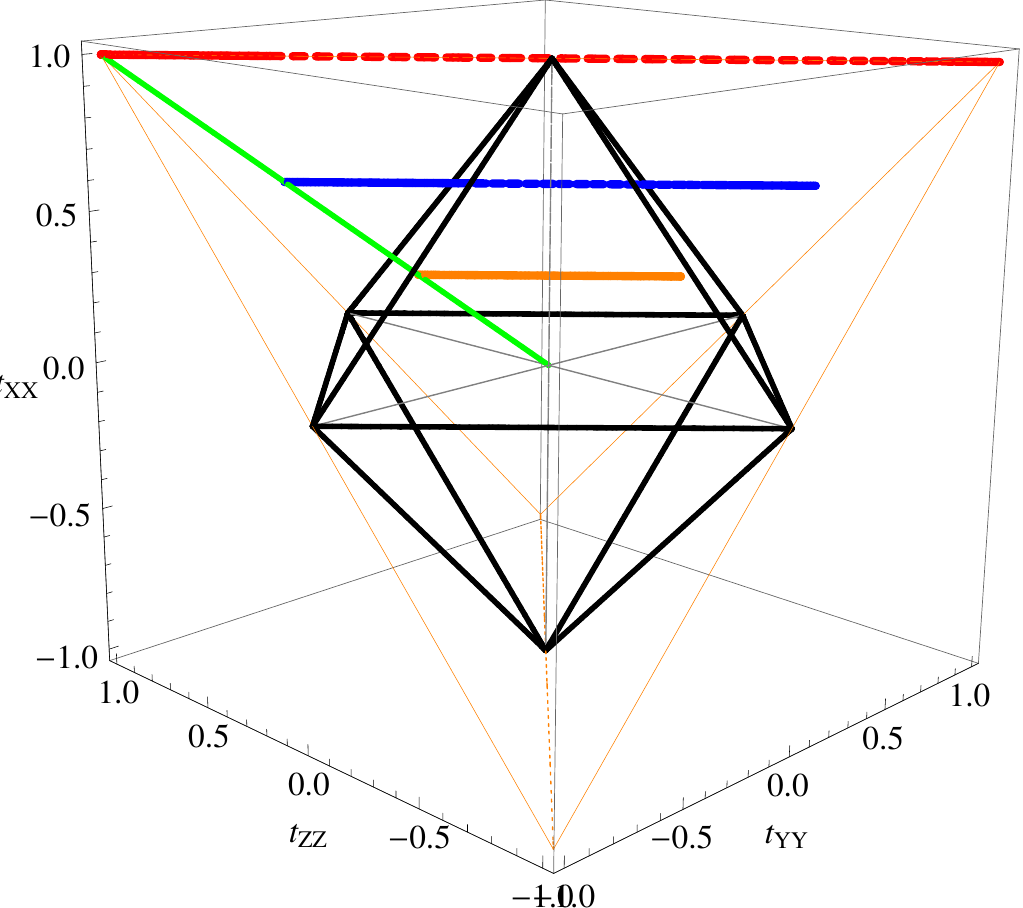}
		\caption{}
		\label{fig:WERNER_PhiPlus_with_PRODUCT_momenta_R_x_1_StateEvolution_}
	\end{subfigure}
	\hspace{2em} 
	\begin{subfigure}[t]{0.48\textwidth}
		\centering
		\includegraphics[width=\textwidth]{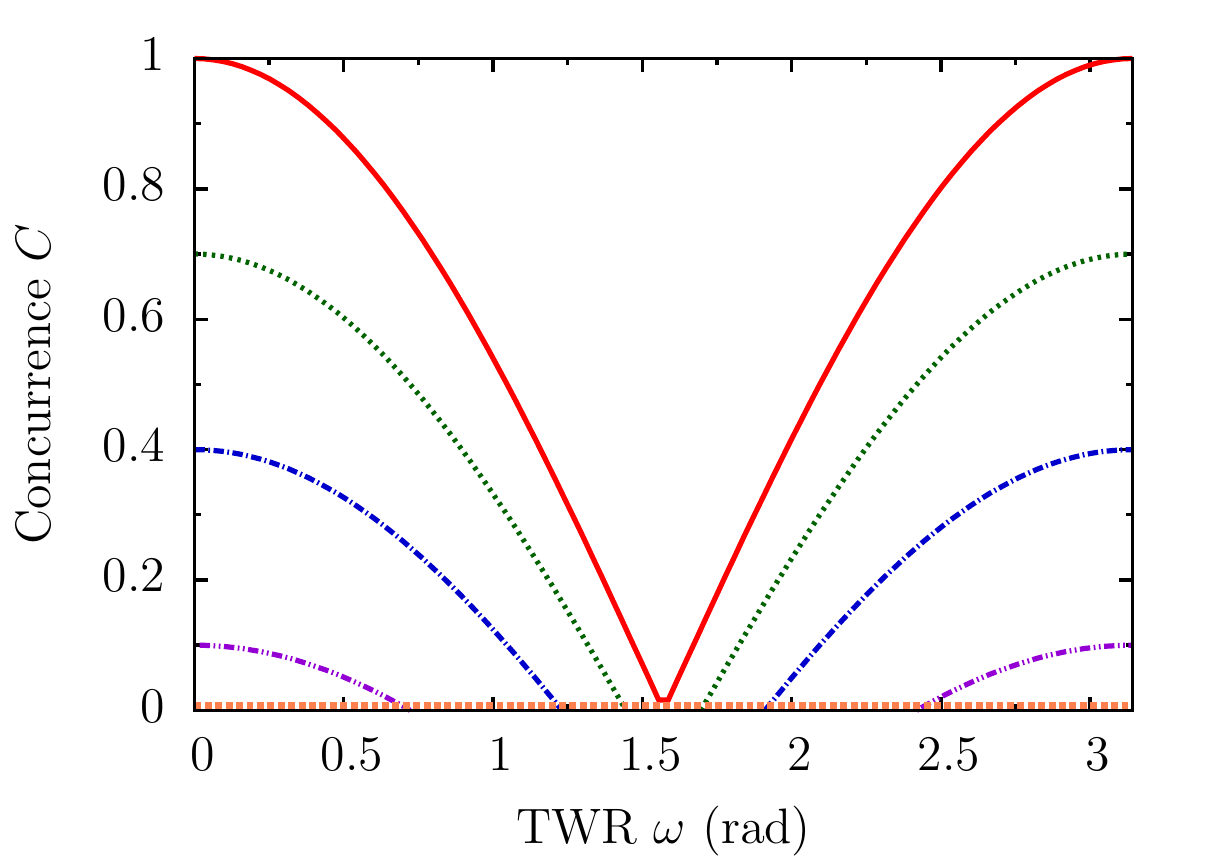}
		\caption{}
		\label{fig:WERNER_PhiPlus_with_PRODUCT_momenta_R_x_1_Concurrence_}
	\end{subfigure}
	\caption[Spin orbit and concurrence under $R_i \otimes \idd$ with $\rho_{\Sigma}$.]%
{Typical spin orbit and concurrence under $R_i \otimes \idd$ with $\omega \in [0, \pi]$ generated by mixed momenta $\rho_{\Sigma}$. 
(a) Initial states $\rho_{W}(\lambda)$ lie on the line connecting the origin to the vertex $(1, -1, 1)$ and correspond to values $\lambda = 1, 3/5, 1/3$ with the respective colors red, blue and orange. 
Note that the figure has been rotated relative to the previous ones. 
(b) Concurrence is shown for $\lambda = 1, 4/5, 3/5, 2/5, 1/3$ with the respective colors red, green, blue, magenta, orange.
}%
	\label{fig:WERNER_PhiPlus_with_PRODUCT_momenta_R_x_1_}
\end{figure*}

Let us next turn to a discussion of the case $0 \leq \lambda < 1$ where the initial state is mixed. Fig.~\ref{fig:WERNER_PhiPlus_with_PRODUCT_momenta_R_x_1_} shows plots of the orbits and concurrences. 
We illustrate spin behavior by plotting three orbits in Fig.~\ref{fig:WERNER_PhiPlus_with_PRODUCT_momenta_R_x_1_StateEvolution_} for three different values of $\lambda$, Fig.~\ref{fig:WERNER_PhiPlus_with_PRODUCT_momenta_R_x_1_Concurrence_} includes the corresponding graphs of the concurrence. Again we see that visualization of the orbits provides valuable insight into the behavior of the state, as well as explaining the characteristics of entanglement.

The initial states with $\lambda < 1$ are mixed and they lie on the (green) line between the vertex $(1, -1, 1)$ and the origin. The initial state with $\lambda = 3/5$ lies just outside the octahedron, still containing some entanglement at $C = 2/5$. When the state is boosted, it moves along the orbit (shown blue) which is parallel to the orbit of the Bell state, becoming separable as it enters the octahedron. To find the value of $\omega$ at this point, we set the concurrence to zero in the first line of Eq.~(\ref{eq:concurrenceForSpinWernerStateWithProductMomenta_R_i_1}), obtaining
\begin{align}
\omega_{\pm} = \arccos\, \pm \left( \frac{1 - \lambda}{2 \lambda} \right). 
\end{align}
For $\lambda = 3/5$ this evaluates to $\omega_+ = 1.23$ and $\omega_- = 1.91$. Thus in the range $\omega \in [1.23, 1.91]$ the spins appear fully separable to the boosted observer. However, as the boosts increase even further, entanglement becomes non-zero again when $\omega$ is larger than $1.91.$ The orbit leaves the octahedron and enters the region of entangled states. As boosts near the speed of light, the Wigner angle approaches $\pi$ and the state is mapped to the point which is a mirror image of the initial state with respect to the plane $\mathcal{P}$ that intersects the origin and the vertices $(1, -1, -1)$ and $(1, 1, 1)$. This is a generalization of the phenomenon we saw in the Bell states where boosts at the speed of light mapped $\ket{\Phi_+}$ to $\ket{\Psi_+}$. In the present case, maximal boosts map the Werner state $\rho_W(\lambda)$ to a another Werner state, which is written as a mixture of $\pouter{\Psi_+}{\Psi_+}$ and the maximally mixed state, 
\begin{align}\label{eq:WernerStateWithPsiPlus}
\rho_{W \Psi_+}(\lambda) = \lambda \pouter{\Psi_+}{\Psi_+} + (1 - \lambda) \frac{1}{4} \idd_4 \quad \mathrm{with}\;\; \lambda \in [0,1].
\end{align}
When $\omega = \pi$, the concurrence is $2/5$, the same value it has in the rest frame.

States that lie initially in the octahedron, for instance when $\lambda = 1/3$, are separable. Boosts map such a state to an orbit which is again parallel to that of the Bell state, with the total orbit being of symmetric shape with respect to the plane $\mathcal{P}$. However, because the whole orbit remains inside the octahedron of separable states, concurrence is zero at all boost values.

\subsubsection{Case $R_i \otimes R_i$}

Rotations of the form $R_i \otimes R_i$ can be again implemented by various geometries. For example, when the boost is in the $z$-direction, then $R_X \otimes R_X$ is realized by the state 
\begin{align}
\frac{1}{4}
\left( \pouter{\ppp_y}{\ppp_y} + \pouter{-\ppp_y}{-\ppp_y} \right) \left( \pouter{\qqq_y}{\qqq_y} + \pouter{-\qqq_y}{-\qqq_y} \right)
\end{align}
where momenta of both particles lie in the $y-z$-plane. From Eqs.~(\ref{eq:stateAsThreeVector}) and (\ref{eq:coefficientsOfStateVector}) we calculate the vector representing the spin under $R_X \otimes R_X$, 
\begin{align}
t_{X \otimes X}(\omega, \lambda) = \lambda \left( 1, -\cos^2 \omega, \cos^2 \omega \right),
\end{align}
which yields for the concurrence
\begin{align}\label{eq:concurrenceForSpinWernerStateWithProductMomenta_R_i_R_i}
C(\omega, \lambda) = \left\{
 \begin{array}{ll}
   -\frac{1}{2} + \lambda + \frac{1}{2} \lambda \cos 2\omega \quad & \text{if}\quad \lambda \in (\lambda_{\text{sep}}, 1] \\
  0\;\;\,\quad\quad\quad\quad\quad\quad\quad\quad\quad & \text{if}\quad \lambda \in [0, \lambda_{\text{sep}}]\\  
 \end{array} \right.
\end{align}
where as above $\lambda_{\text{sep}}$ is the value where the rest frame state becomes separable.

The other realizations $R_Y \otimes R_Y$ and $R_Z \otimes R_Z$ produce similar vectors, 
\begin{align}\label{eq:discreteOrbitPureStateR_i_R_i}
t_{Y \otimes Y}(\omega, \lambda) &= \lambda \left(\cos^2 \omega, -1, \cos^2 \omega\right), \nonumber\\ 
t_{Z \otimes Z}(\omega, \lambda) &= \lambda \left(\cos^2 \omega, -\cos^2 \omega, 1\right).
\end{align}

\begin{figure*}	
	\centering
	\begin{subfigure}[t]{0.38\textwidth}
		\centering
		\includegraphics[width=\textwidth]{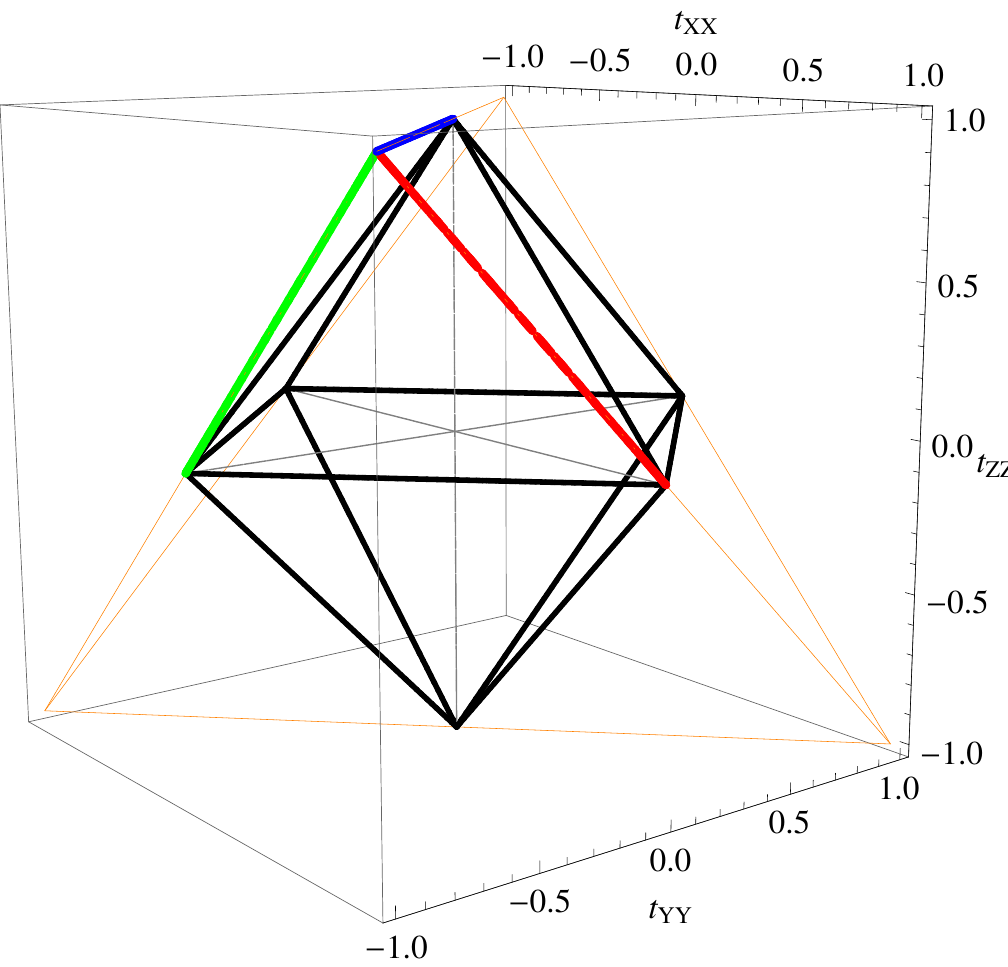}
		\caption{}
		\label{fig:Bell_diagonal_maps_EVOLUTION_Ri_Ri_}
	\end{subfigure}
	\hspace{2em}
	\begin{subfigure}[t]{0.48\textwidth}
		\centering
		\includegraphics[width=\textwidth]{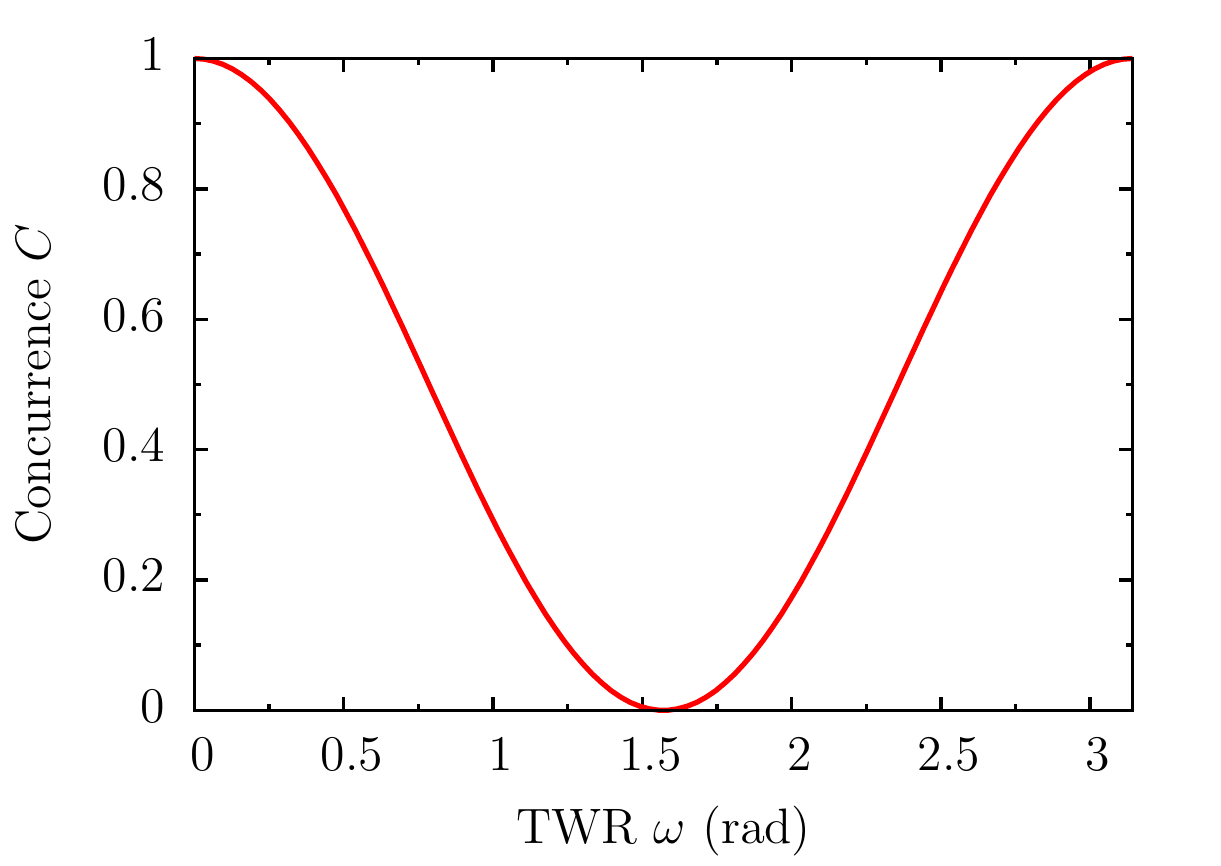}
		\caption{}
		\label{fig:BELL_PhiPlus_Bell_diagonal_maps_CONCURRENCE_Ri_Ri_}%
	\end{subfigure}
	\caption[Spin orbit and concurrence under $R_i \otimes R_i$ with $\rho_{\Sigma}$.]%
{Spin orbit and concurrence under $R_i \otimes R_i$ with $\omega \in [0, \pi]$ generated by momenta $\rho_{\Sigma}$. (a)~Initial state $\ket{\Phi^+}$ corresponds to the vertex at $(1, -1, 1)$, $R_X \otimes R_X$ is shown red, $R_Y \otimes R_Y$ green and $R_Z \otimes R_Z$ blue. (b)~Concurrence has the same shape for all $R_i \otimes R_i$.
}%
	\label{fig:BELL_PhiPlus_R_i_otimes_R_i}
\end{figure*}

We begin the discussion by focussing on the Bell state $\ket{\Phi_+}$, which is the case with $\lambda = 1$. Plots of the orbits and concurrence are shown in Fig.~\ref{fig:BELL_PhiPlus_R_i_otimes_R_i}. 
Let us consider $t_{X \otimes X}$. At the beginning, the effect of boosts is qualitatively similar to the previous pure state case. At rotations smaller than $\pi / 2$, the state $\ket{\Phi_+}$ is again mapped into a mixture of itself and $\ket{\Psi_+}$, moving along the orbit that connects the two states. When $\omega = \pi / 2$, the moving observer sees a separable state. However, at boosts that generate rotations larger than $\pi / 2$, the orbit differs from the previous case as the boosted state moves back along the same path towards the rest frame state. At $\omega = \pi$, we obtain the original rest frame state $\ket{\Phi_+}$.

The concurrence is rather similar to the previous case in that it decreases  monotonically from $1$ to $0$ between $[0, \pi/2]$ and then increases monotonically from $0$ to $1$ between $[\pi / 2, \pi]$, while the precise expression differs slightly from the previous case.

\begin{figure*}[htb]
	\centering
	\begin{subfigure}[t]{0.38\textwidth}
		\centering
		\includegraphics[width=\textwidth]{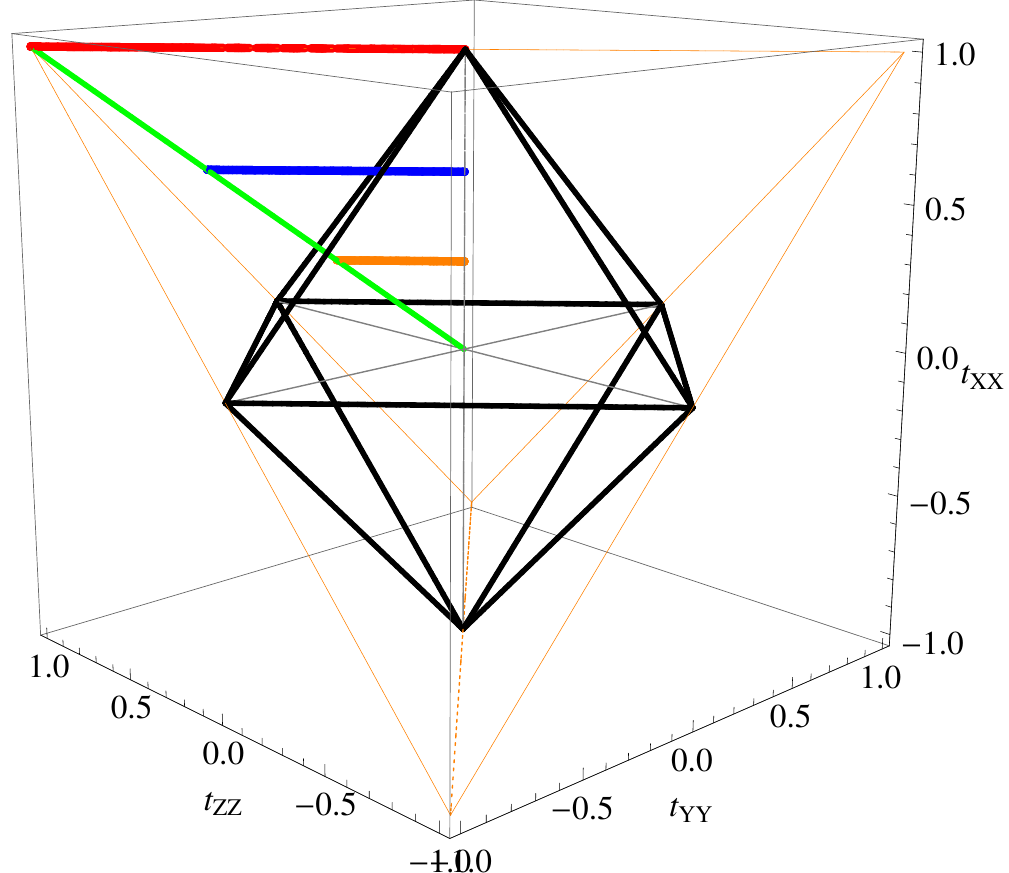}
		\caption{}
		\label{fig:WERNER_PhiPlus_with_PRODUCT_momenta_R_x_R_x_StateEvolution_}
	\end{subfigure}
	\hspace{2em}
	\begin{subfigure}[t]{0.48\textwidth}
		\centering
		\includegraphics[width=\textwidth]{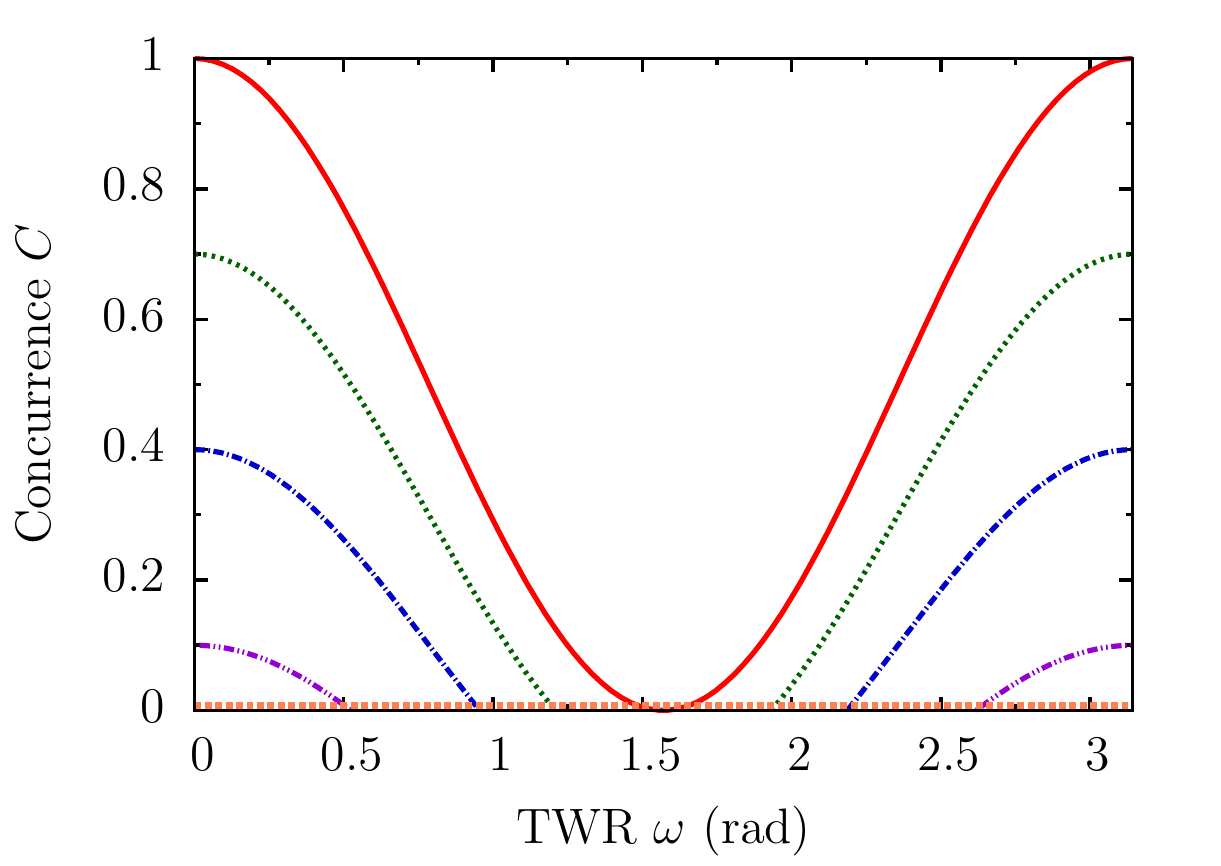}
		\caption{}
		\label{fig:WERNER_PhiPlus_with_PRODUCT_momenta_R_x_R_x_Concurrence_}
	\end{subfigure}
	\caption[Spin orbit and concurrence under $R_i \otimes R_i$ with $\rho_{\Sigma}$.]%
{Typical spin orbit and concurrence under $R_i \otimes R_i$ with $\omega \in [0, \pi]$ generated by mixed momenta $\rho_{\Sigma}$. 
(a) Initial states $\rho_{W}(\lambda)$ lie on the line connecting the origin to the vertex $(1, -1, 1)$ and correspond to values $\lambda = 1, 3/5, 1/3$ with the respective colors  red, blue and orange. 
(b) Concurrence is shown for $\lambda = 1, 4/5, 3/5, 2/5, 1/3$ with the respective colors red, green, blue, magenta, orange.}%
	\label{fig:WERNER_PhiPlus_with_PRODUCT_momenta_R_x_R_x_}
\end{figure*}

Now let us turn to the case of mixed initial states, i.e. $0 \leq \lambda < 1$. Plots of the spin orbits and concurrence are shown in Fig.~\ref{fig:WERNER_PhiPlus_with_PRODUCT_momenta_R_x_R_x_}.
For illustration, the orbits are again shown for three different values of $\lambda$ in Fig.~\ref{fig:WERNER_PhiPlus_with_PRODUCT_momenta_R_x_R_x_StateEvolution_}. In Fig.~\ref{fig:WERNER_PhiPlus_with_PRODUCT_momenta_R_x_R_x_Concurrence_} we plot the corresponding graphs of concurrence.

Many characteristics are similar to the previous case. Orbits of initial states with less than maximal entanglement are parallel to the orbit of the Bell state $\ket{\Phi_+}$. As boost increases, a state that is initially entangled moves towards the octahedron and becomes separable when entering the octahedron. To find the corresponding values of $\omega$, we set the concurrence to zero in Eq.~(\ref{eq:concurrenceForSpinWernerStateWithProductMomenta_R_i_R_i}) and solve for $\omega$, 
\begin{align}
\omega_{k, \pm} = \frac{1}{2} \left( 2 k \pi \pm \arccos \left( \frac{1 - 2 \lambda}{\lambda} \right) \right) , \quad k \in \mathbb{N}.
\end{align}
The solutions relevant in the present case are $\omega_{0, +}$ and $\omega_{1, -}.$ This means $\rho^{\Lambda}_S$ is separable if $\omega \in [\omega_{0, +},\, \omega_{1, -}].$ For instance, a state for which $\lambda = 3/5$, whose orbit is shown blue in Fig.~\ref{fig:WERNER_PhiPlus_with_PRODUCT_momenta_R_x_R_x_StateEvolution_}, has vanishing concurrence if $\omega \in [0.96, 2.19]$. This corresponds to the part of the orbit which resides in the octahedron. In a similar vein, initial states that lie inside the octahedron and are separable follow an orbit for which entanglement remains zero for all boosts. There is a difference from the previous case:\ when boosts approach the speed of light, the state is mapped back to the original state.

\subsubsection{Case $R_i \otimes R_j$}\label{sec:spinWernerStateWithProductMomenta_R_i_R_j}

Rotations around different axis, $R_i \otimes R_j$, can be implemented by momenta that lie in different boost planes. For instance, when boost is in the $z$-direction and momenta are of the form 
\begin{align}
\frac{1}{4} \left( \pouter{\ppp_y}{\ppp_y} + \pouter{-\ppp_y}{-\ppp_y} \right) \left( \pouter{\qqq_x}{\qqq_x} + \pouter{-\qqq_x}{-\qqq_x} \right) 
\end{align}
then spins are rotated by $R_X \otimes R_Y$. We calculate that the three vector corresponding to the state is 
\begin{align}\label{eq:vectorWernerStateUnder_R_X_R_Z}
t_{X \otimes Z}(\omega, \lambda) = \lambda \left( \cos \omega, -\cos^2 \omega, \cos \omega \right),
\end{align}
and the concurrence is given by 
\begin{align}\label{eq:concurrenceForSpinWernerStateWithProductMomenta_R_i_R_j}
C(\omega, \lambda) = 
\left\{
  \begin{array}{ll}
    \frac{1}{8}
      \left( \big|
	    \left| 2 + \lambda + 4\lambda \cos\omega + \lambda \cos2\omega 
	    \right| 
	  \right. 
	\\ 
	  \left.  - 
	    \left| 2 + \lambda - 4\lambda \cos\omega + \lambda \cos2\omega 
	    \right| \big| 
      \right)  
    \\
    + 2 \left( - 2 + \lambda + \lambda \cos2\omega \right)
    \\
    \quad\quad\quad\quad\quad\quad\quad\quad\quad\quad\quad\quad 
    \text{if}\quad \lambda \in (\lambda_{\text{sep}}, 1] \\
    0  \quad\quad\quad\quad\quad\quad\quad\quad\quad\quad\quad\;\;
    \text{if}\quad \lambda \in [0, \lambda_{\text{sep}}]\\
  \end{array} 
\right.
\end{align}
where at $\lambda_{\text{sep}}$ the state becomes separable. The other rotations generate similar vectors 
\begin{align}\label{eq:vectorWernerStateUnder_R_X_R_Z}
t_{X \otimes Y}(\omega, \lambda) &= \lambda (\cos \omega, -\cos \omega, \cos^2 \omega), \nonumber\\
t_{Y \otimes Z}(\omega, \lambda) &= \lambda (\cos^2 \omega, -\cos \omega, \cos \omega).  
\end{align}

\begin{figure*}	
	\centering
	\begin{subfigure}[t]{0.38\textwidth}
		\centering
		\includegraphics[width=\textwidth]{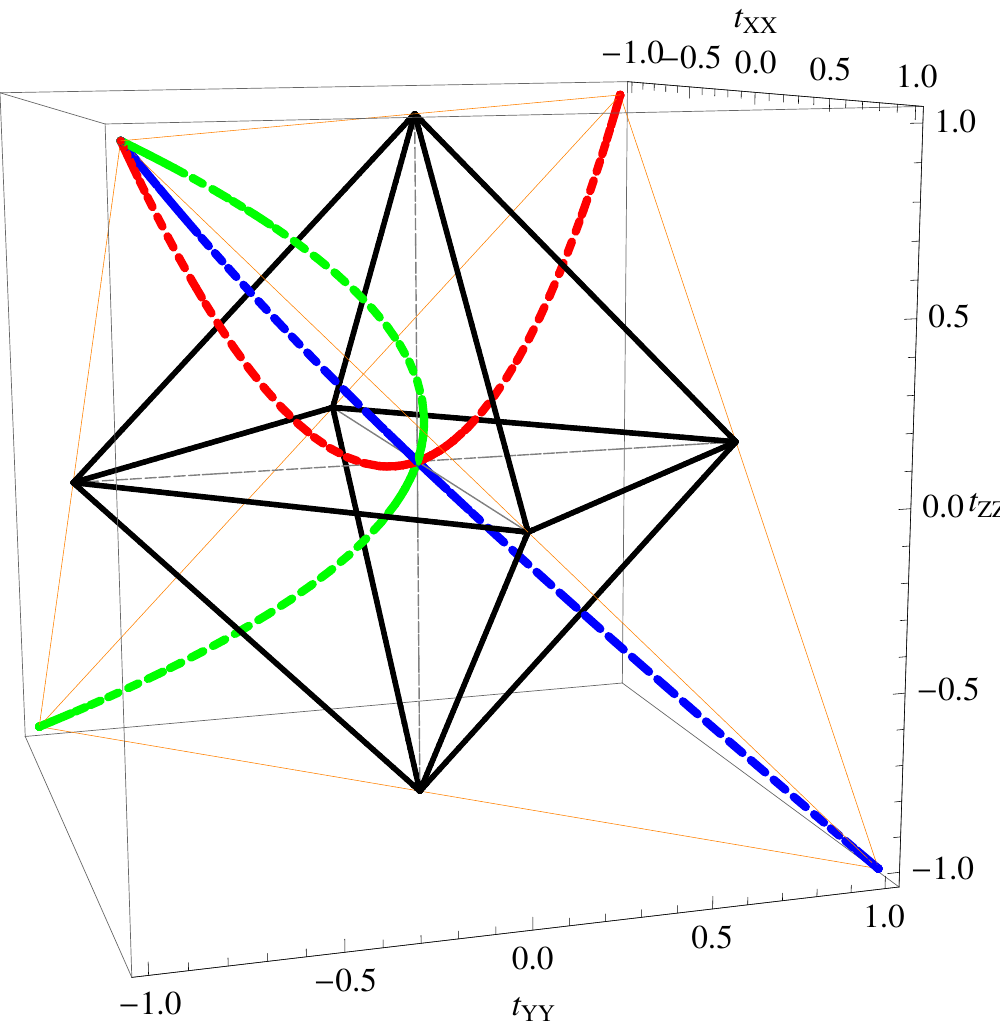}
		\caption{}
		\label{fig:BELL_PhiPlus_Bell_diagonal_maps_EVOLUTION_Ri_Rj_}
	\end{subfigure}
	\hspace{2em}
	\begin{subfigure}[t]{0.48\textwidth}
		\centering
		\includegraphics[width=\textwidth]{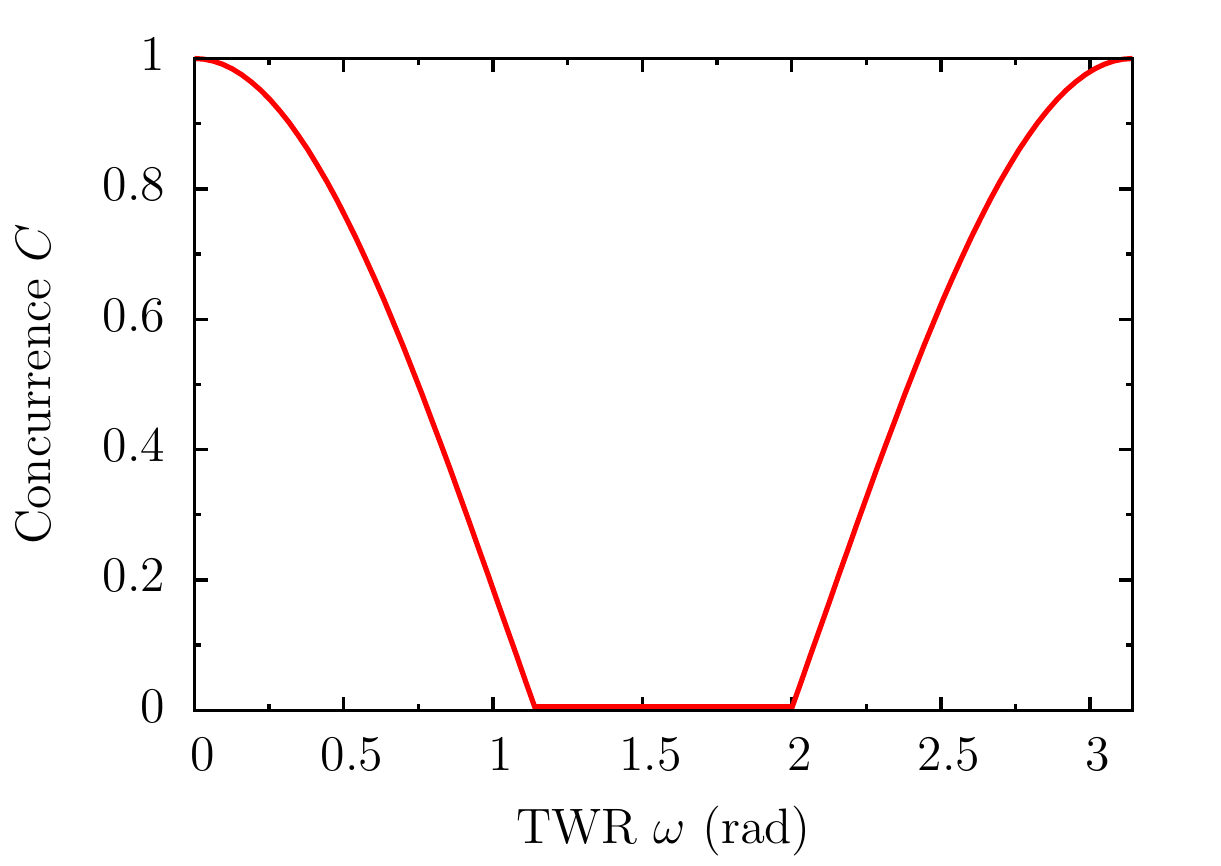}
		\caption{}
		\label{fig:BELL_PhiPlus_Bell_diagonal_maps_CONCURRENCE_Ri_Rj_}
	\end{subfigure}
	\caption[Spin orbit and concurrence under $R_i \otimes R_j$, $i \ne j$ with $\ket{M^{\Sigma}}$.]%
{Spin orbit and concurrence under $R_i \otimes R_j$, $i \ne j$ and $\omega \in [0, \pi]$ generated by momenta $\rho_{\Sigma}$. (a) Initial state $\ket{\Phi^+}$ corresponds to the vertex at $(1, -1, 1)$, $R_X \otimes R_Y$ is shown red, $R_X \otimes R_Z$ green and $R_Y \otimes R_Z$ blue.
(b) Concurrence has the same shape for all $R_i \otimes R_j$.
}%
	\label{fig:BELL_PhiPlus_R_i_otimes_R_j}
\end{figure*}

We begin by considering the Bell state $\ket{\Phi_+}$, the case with $\lambda = 1$. Plots of the spin orbits and concurrences for all the different rotations are shown in Figure~\ref{fig:BELL_PhiPlus_R_i_otimes_R_j}.
The behavior of spin under mixed rotations is quite different from the two previous cases. Let us consider $t_{X \otimes Z}$ as an illustration. Fig.~\ref{fig:BELL_PhiPlus_Bell_diagonal_maps_EVOLUTION_Ri_Rj_} shows that the orbit has the shape of a curve that starts at the vertex $(1, -1, 1)$ which represents the rest state $\ket{\Phi^+}$. It then evolves towards the origin, reaching it at $\omega = \pi / 2$. The second half of the orbit for values $\omega \in [\pi / 2, \pi]$ is symmetric to the first half. The state evolves towards the vertex $(-1, 1, 1)$ which represents the Bell state $\ket{\Phi_-}$, reaching it when the boosts approach the speed of light. The orbit lies in the plane that intersects the initial state $\ket{\Phi_+}$, the origin and the final state $\ket{\Phi_-}$.

It is interesting that the spins become separable when the Wigner angle lies between $[1.14, 2.00]$, see Fig.~\ref{fig:BELL_PhiPlus_Bell_diagonal_maps_CONCURRENCE_Ri_Rj_}. While this might look puzzling if we only knew the behavior of the concurrence, the plot of the orbit clearly shows what is happening. The spin state evolves in the plane that intersects the octahedron of separable states, hitting the face of the octahedron when $\omega = \omega_-$, and then following a path towards the maximally mixed state $\frac{1}{4} \idd_4$ represented by $(0, 0, 0)$. When $\omega = \pi / 2$, the moving observer sees the maximally mixed state. The concurrence of the boosted state becomes non-zero again as $\omega$ becomes greater than $\omega_+$, this corresponds to the point where the spin state leaves the octahedron.

\begin{figure*}	
	\centering
	\begin{subfigure}[t]{0.38\textwidth}
		\centering
		\includegraphics[width=\textwidth]{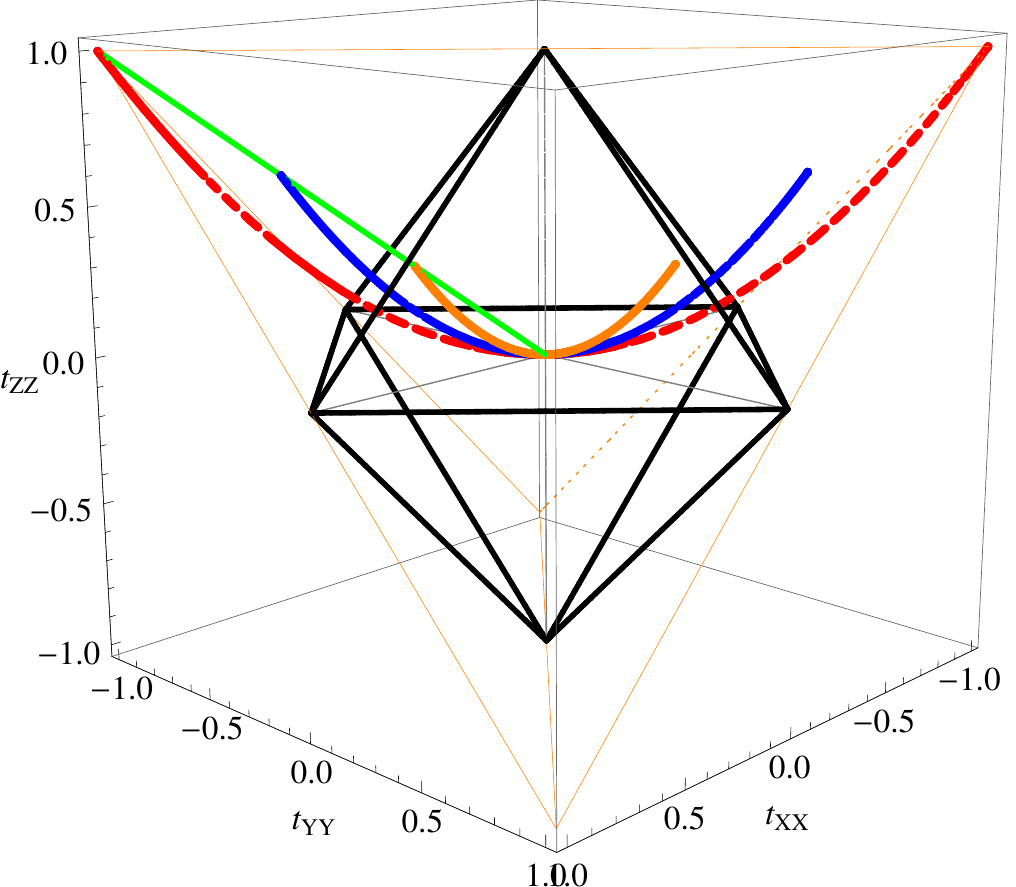}
		\caption{}
		\label{fig:WERNER_PhiPlus_with_PRODUCT_momenta_R_x_R_z_StateEvolution_}
	\end{subfigure}
	\hspace{2em} 
	\begin{subfigure}[t]{0.48\textwidth}
		\centering
		\includegraphics[width=\textwidth]{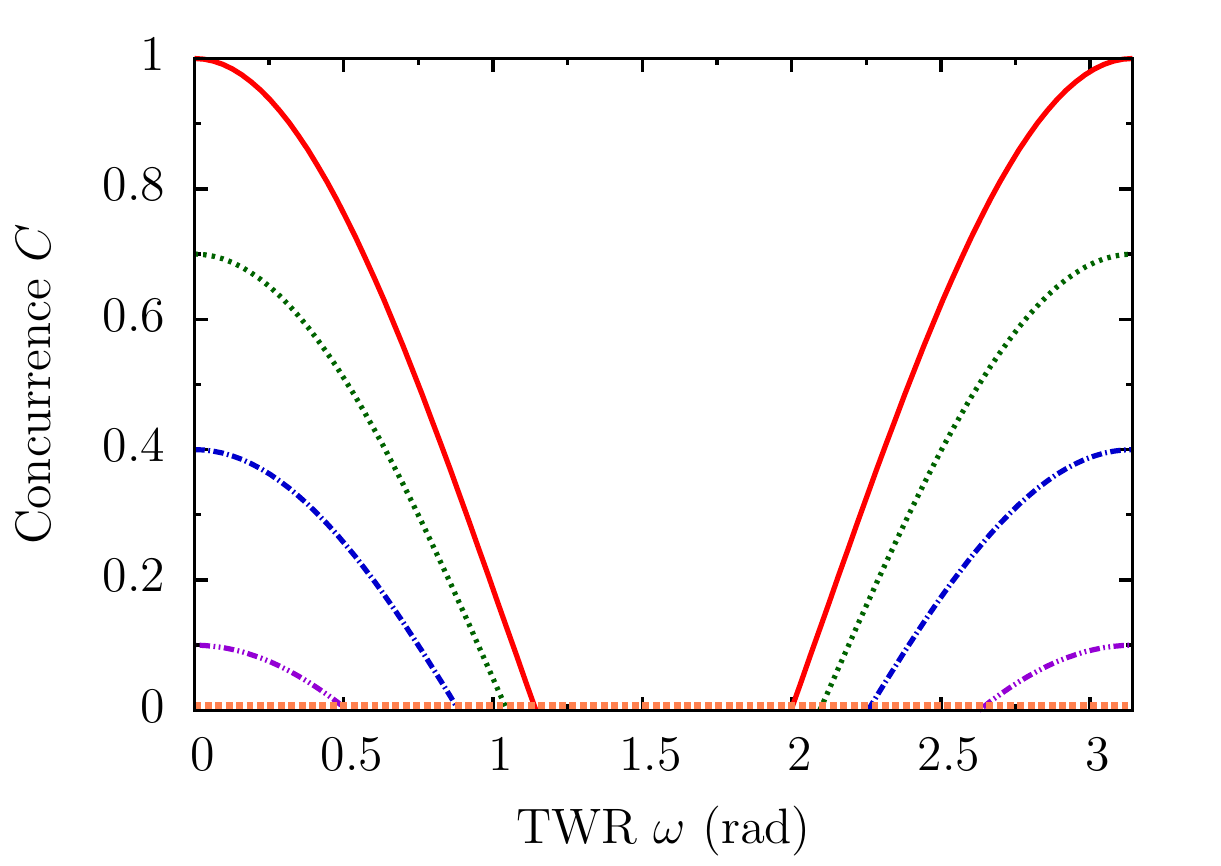}
		\caption{}
		\label{fig:WERNER_PhiPlus_with_PRODUCT_momenta_R_x_R_z_Concurrence_}
	\end{subfigure}
	\caption[Spin orbit and concurrence under $R_i \otimes R_j$, $i \ne j$ with $\rho_{\Sigma}$.]%
{Typical spin orbit and concurrence under $R_i \otimes R_j$, $i \ne j$ with $\omega \in [0, \pi]$ generated by mixed momenta $\rho_{\Sigma}$. 
(a) Initial states $\rho_{W}(\lambda)$ lie on the line connecting the origin to the vertex $(1, -1, 1)$ and correspond to values $\lambda = 1, 3/5, 1/3$ with the respective colors  red, blue and orange. 
(b) Concurrence is shown for $\lambda = 1, 4/5, 3/5, 2/5, 1/3$ with the respective colors red, green, blue, magenta, orange.}%
	\label{fig:WERNER_PhiPlus_with_PRODUCT_momenta_R_x_R_z_}
\end{figure*}

Let us next consider the mixed states, $0 \leq \lambda < 1$. In Fig.~\ref{fig:WERNER_PhiPlus_with_PRODUCT_momenta_R_x_R_z_}, we have plotted the orbits and concurrences, where the orbits are again shown for three different values of $\lambda$. 
We recognize a pattern of behavior that is similar to the previous cases, albeit with a few differences. As before, the states follow an orbit that resides in the octahedron for a range of values around $\pi / 2$. However, the region where the concurrence vanishes is considerably larger than in the previous cases. Also, while we saw above that the orbits of mixed states were parallel to the orbit of the Bell state, here all the orbits pass through the maximally mixed state $\frac{1}{4} \idd_4$. To find the values of $\omega$ for which the concurrence vanishes, we set the concurrence to zero in the first line of Eq.~(\ref{eq:concurrenceForSpinWernerStateWithProductMomenta_R_i_R_j}) and solve for $\omega$,
\begin{align}
\omega_{k, \pm} = k \pi \pm \arccos \left( \frac{\lambda - \sqrt{\lambda + \lambda^2}}{\lambda} \right) ,\quad k = 0, 1,
\end{align}
which entails that the state is separable if $\omega \in [\omega_{1, -},\, \omega_{0, +}].$ For example, when $\lambda = 3/5$, entanglement vanishes in the interval $\omega \in [0.89, 2.25]$.

\subsection{Product momenta $\rho_{\times}$}\label{subsec:productMomentaGingrichAdamiWithWernerState}

In the following sections \ref{subsubsec:MTimesR_iTimes1}--\ref{subsubsec:MTimesR_iTimesR_iAndR_iTimesR_j} we will analyze mixed momenta of the form 
\begin{align}
\rho_{\times} = \frac{1}{8} 
&\left( \pouter{\ppp}{\ppp} + \pouter{-\ppp}{-\ppp} + \pouter{\ppp_{\perp}}{\ppp_{\perp}} + \pouter{-\ppp_{\perp}}{-\ppp_{\perp}} \right) \nonumber\\
&\otimes \left( \pouter{\qqq}{\qqq} + \pouter{-\qqq}{-\qqq} + \pouter{\qqq_{\perp}}{\qqq_{\perp}} + \pouter{-\qqq_{\perp}}{-\qqq_{\perp}} \right),
\end{align}
which are the counterpart of the pure product momenta $\ket{M_{\times}}$.
We will have to analyze only two types of rotation, the $R_i \otimes \idd$ and the combination of $R_i \otimes R_i$ with \mbox{$R_i \otimes R_j$}. 
The latter two are not two distinct cases because a generic expression of $\rho_{\times}$ involves momentum terms that generate both types of rotation. For instance, if the boost is in the $z$-direction and the momenta are constrained to lie in the $x-z$- and $y-z$-planes, we get terms that correspond to the pure momenta $\ket{\pm\ppp_x, \pm\ppp_x}$, $\ket{\pm\ppp_y, \pm\ppp_y}$, $\ket{\pm\ppp_x, \pm\ppp_y}$ and $\ket{\pm\ppp_y, \pm\ppp_x}$, which generate the respective rotations $R_X \otimes R_X$, $R_Y \otimes R_Y$, $R_Y \otimes R_X$ and $R_X \otimes R_Y$. We will also see that the state vectors of both types can be obtained as convex combinations of the vectors we have already calculated above.

\subsubsection{Case $R_i \otimes \idd$}\label{subsubsec:MTimesR_iTimes1}

We begin by considering the case where only the first particle undergoes rotation while the second particle is left alone. If we assume that the boost is in the $z$-direction, then such a scenario is realized when the momentum of the first particle is a mixture of projectors on $\ket{\pm\ppp_{y}}$ and $\ket{\pm\ppp_{x}}$, and momentum $\ket{\ppp_0}$ of the second particle lies at the origin. The resulting state vector is a convex sum of vectors for single particle rotations $t_{X \otimes \idd}$ and $t_{Y \otimes \idd}$ generated by $\rho_{\Sigma}$ in section \ref{sec:spinWernerStateWithProductMomenta_R_i_1}, 
\begin{align}\label{eq:convexCombinationPureGAMomentumXY1}
t_{XY \otimes \idd}(\omega, \lambda) &= \frac{1}{2} \left[ t_{X \otimes \idd}(\omega, \lambda) + t_{Y \otimes \idd}(\omega, \lambda) \right], \nonumber\\
&= \lambda \left( \cos^2 \frac{\omega}{2}, -\cos^2 \frac{\omega}{2}, \cos \omega \right).
\end{align}
The vectors generated by the other rotations can be obtained in the same fashion, 
\begin{align}
t_{XZ \otimes \idd}(\omega, \lambda) &= \lambda \left( \cos^2 \frac{\omega}{2}, -\cos \omega, \cos^2 \frac{\omega}{2} \right), \nonumber\\
t_{YZ \otimes \idd}(\omega, \lambda) &= \lambda \left( \cos \omega, -\cos^2 \frac{\omega}{2}, \cos^2 \frac{\omega}{2} \right).
\end{align}
The corresponding concurrence is 
\begin{align}
C(\omega, \lambda) = 
  \left\{
    \begin{array}{ll} 
      \frac{1}{2} \left( - 1 + \lambda + 2 \lambda \cos\omega \right) \quad & \text{if}\quad \lambda \in (\lambda_{\text{sep}}, 1] \\
      0\;\;\,\quad\quad\quad\quad\quad\quad\quad\quad\quad & \text{if}\quad \lambda \in [0, \lambda_{\text{sep}}]\\  
    \end{array} 
  \right.
\end{align}
where $\lambda_{\text{sep}}$ is the value at which the initial spin state becomes separable.

\begin{figure*}	
	\centering
	\begin{subfigure}[t]{0.38\textwidth}
		\centering
		\includegraphics[width=\textwidth]{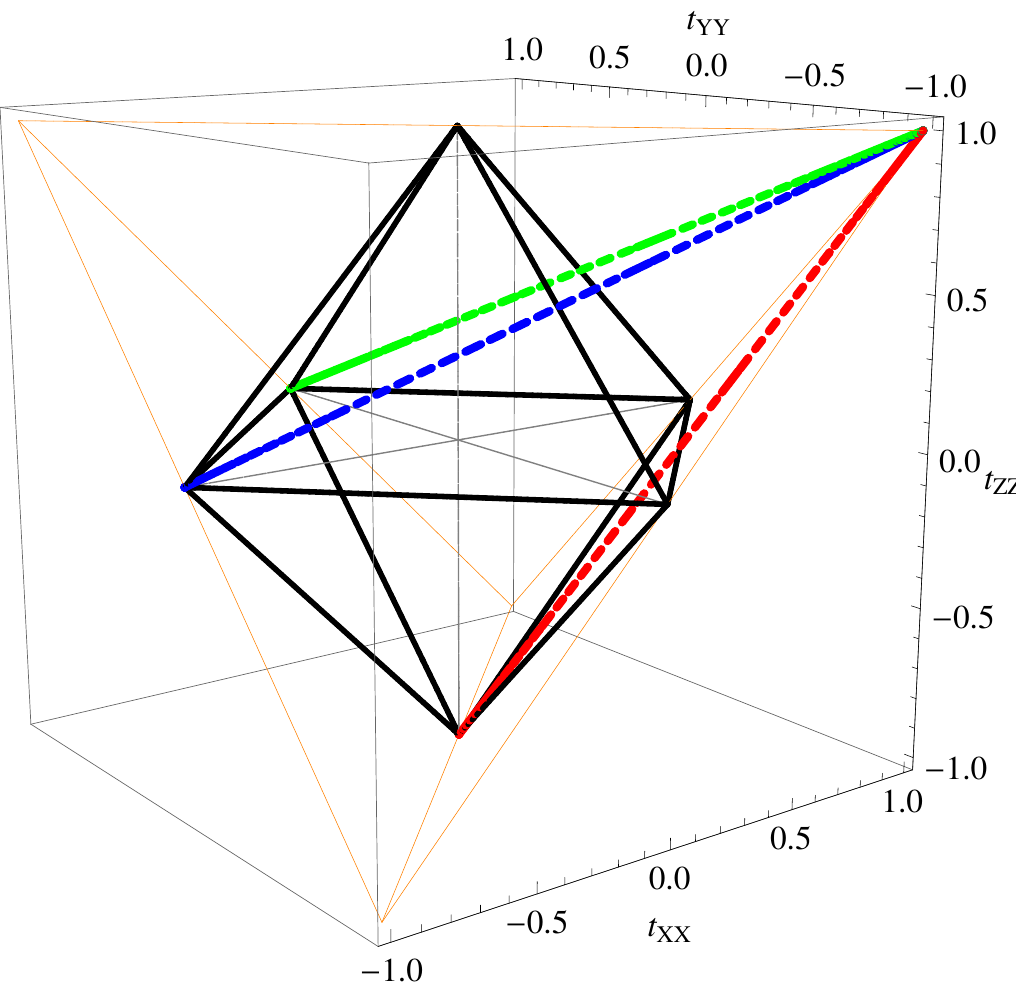}
		\caption{}
		\label{fig:BELL_PhiPlus_with_PRODUCT_momenta_GA_StateEvolution_}
	\end{subfigure}
	\hspace{2em}
	\begin{subfigure}[t]{0.48\textwidth}
		\centering
		\includegraphics[width=\textwidth]{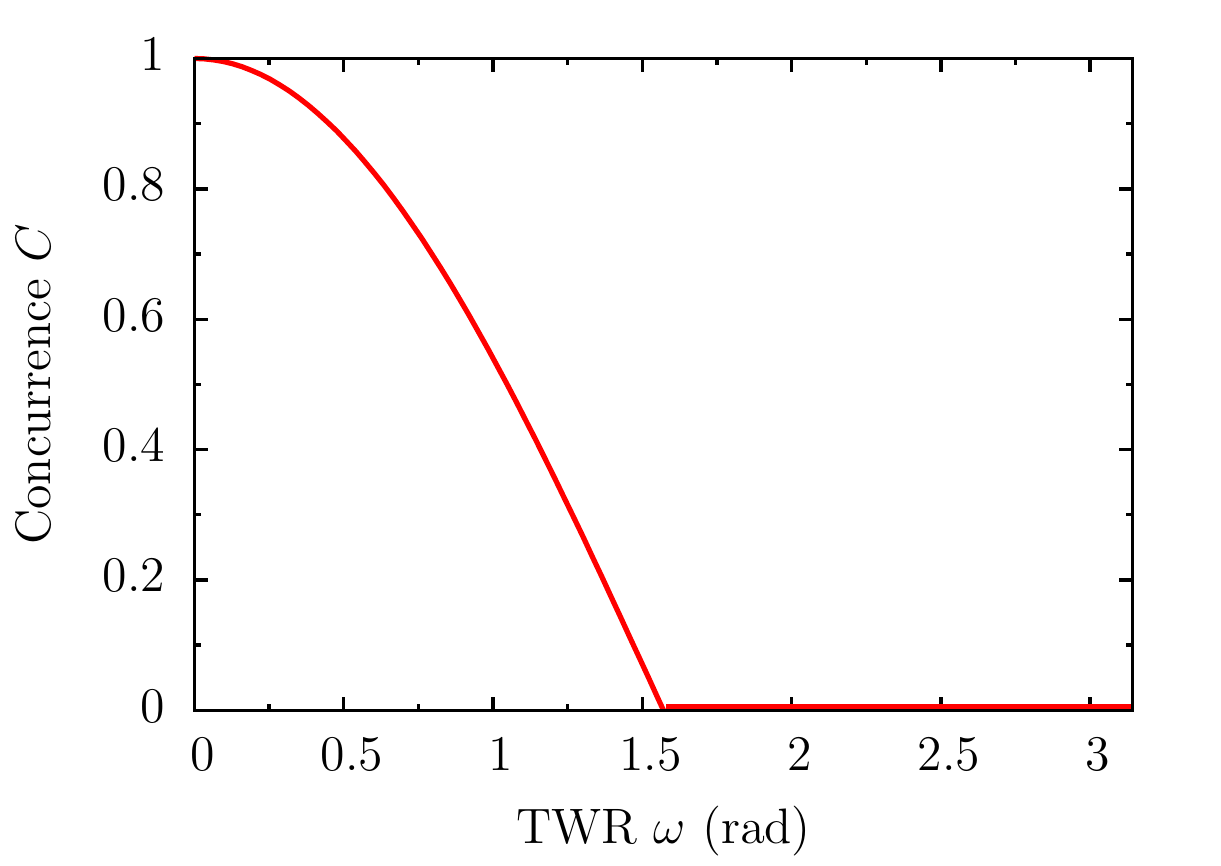}
		\caption{}
		\label{fig:BELL_PhiPlus_with_PRODUCT_momenta_GA_Concurrence_}
	\end{subfigure}
	\caption[Spin orbit and concurrence under $R_i \otimes \idd$ with $\rho_\times$.]{Spin orbit and concurrence under $R_i \otimes \idd$ generated by momenta $\rho_\times$ with $\omega \in [0, \pi]$. 
(a) Initial state $\ket{\Phi^+}$ corresponds to the vertex at $(1, -1, 1)$, the orbit $t_{XY \otimes \idd}$ is shown red, $t_{XZ \otimes \idd}$ green and $t_{YZ \otimes \idd}$ blue.
(b) Spin concurrence has the same shape for all orbits.
}%
	\label{fig:BELL_PhiPlus_with_PRODUCT_momenta_GA_R_i_1}
\end{figure*}

We start by considering the Bell states, i.e. the case $\lambda = 1$. We plot the spin orbits and concurrence for $\ket{\Phi_+}$ in Fig.~\ref{fig:BELL_PhiPlus_with_PRODUCT_momenta_GA_R_i_1}.
Since the orbit of $t_{XY \otimes \idd}$ is a convex sum of vectors for single particle rotations $t_{X \otimes \idd}$ and $t_{Y \otimes \idd}$, it is represented by a line that connects the initial vector $(1, -1, 1)$ for $\ket{\Phi_+}$ and the point $(0, 0, -1)$ that corresponds to the equal mixture of projectors onto $\ket{\Psi_+}$ and $\ket{\Psi_-}$. Accordingly, the concurrence displays the same behavior as that of $t_{X \otimes \idd}$ or $t_{Y \otimes \idd}$ until $\omega = \pi / 2$. However, in contrast to the latter, it vanishes for all values of $\omega$ greater than $\pi / 2$. This is because when boosts induce rotations larger than $\pi / 2$, the state follows a path in the set of separable states on the face of the octahedron until $\omega = \pi$.

\begin{figure*}	
	\centering
	\begin{subfigure}[t]{0.38\textwidth}
		\centering
		\includegraphics[width=\textwidth]{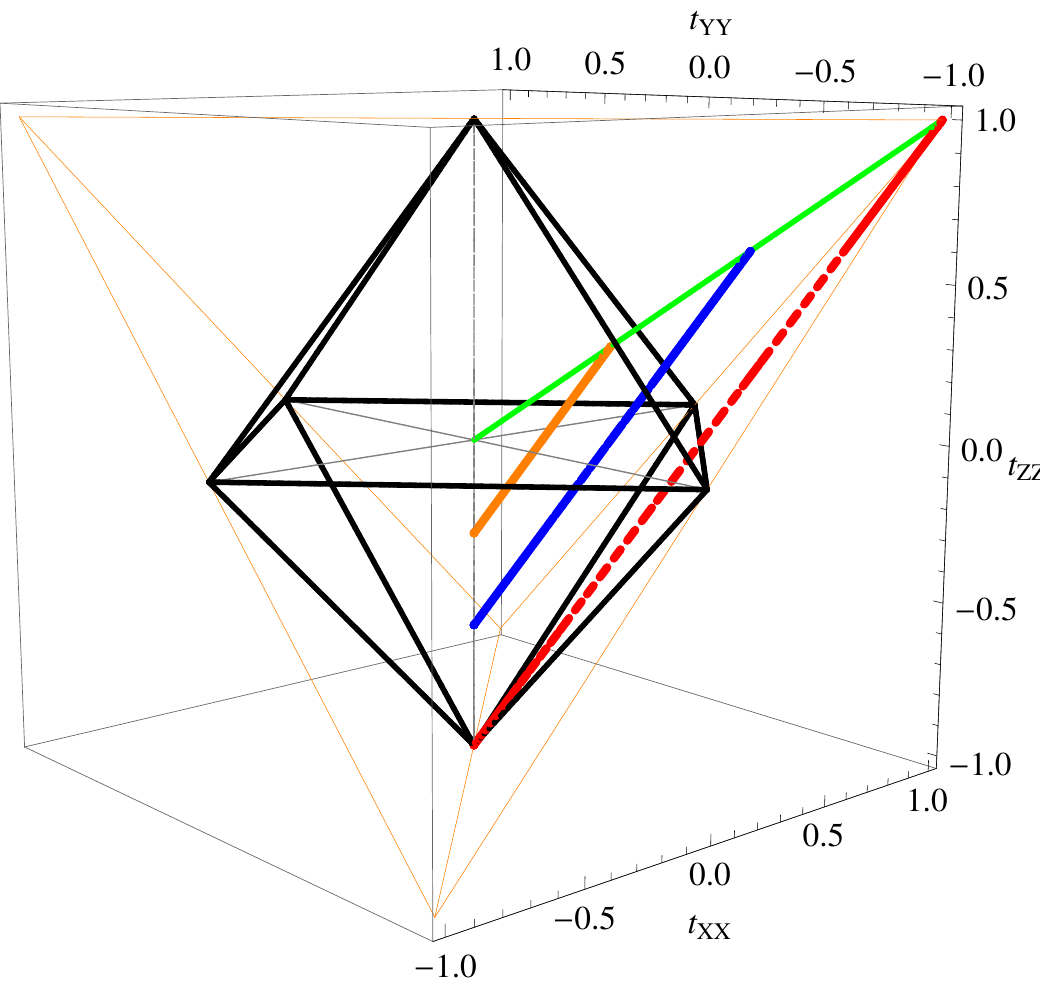}
		\caption{}
		\label{fig:WERNER_PhiPlus_with_PRODUCT_momenta_GA_StateEvolution_R_i_1}
	\end{subfigure}
	\hspace{2em}
	\begin{subfigure}[t]{0.48\textwidth}
		\centering
		\includegraphics[width=\textwidth]{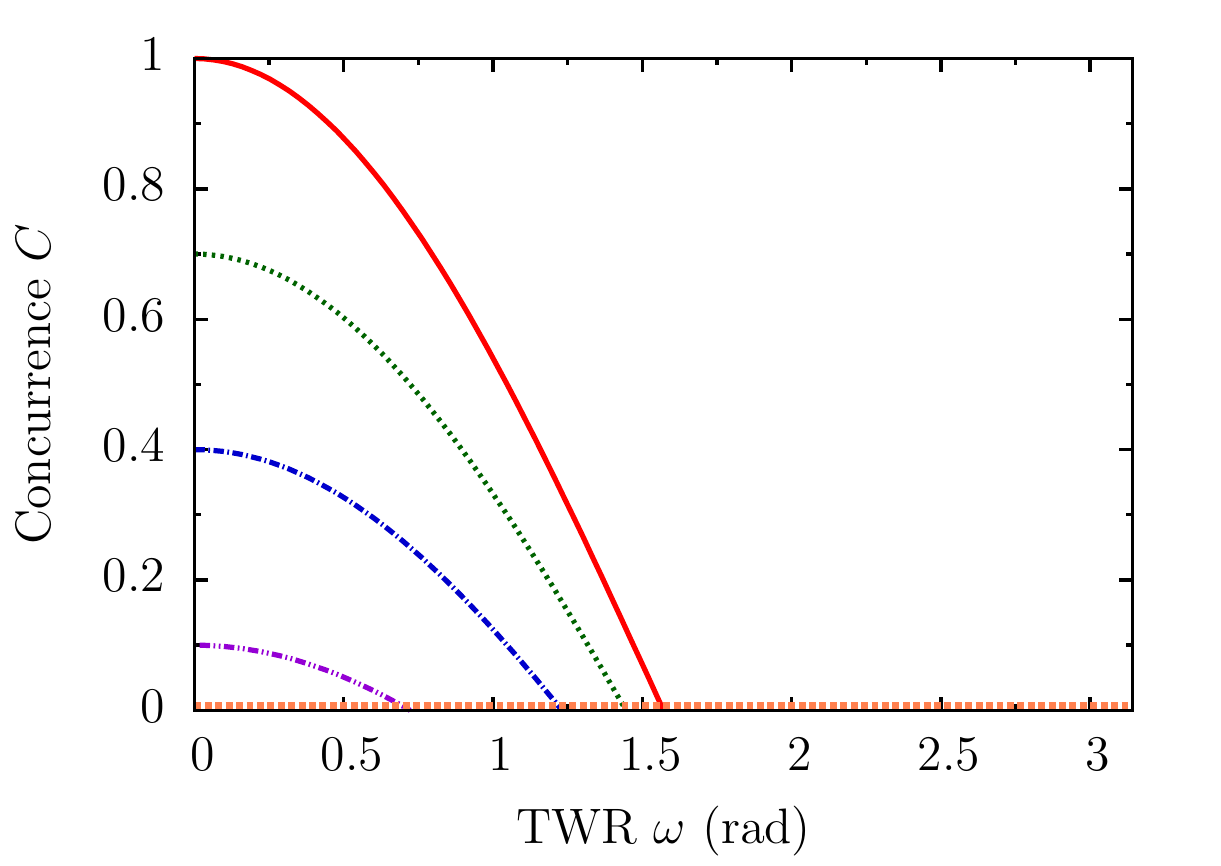}
		\caption{}
		\label{fig:WERNER_PhiPlus_with_PRODUCT_momenta_GA_Concurrence_R_i_1}
	\end{subfigure}
	\caption[Spin orbit and concurrence under $R_i \otimes \idd$ with $\rho_{\times}$.]{Typical spin orbit and concurrence under $R_i \otimes \idd$ generated by momenta $\rho_{\times}$ with $\omega \in [0, \pi]$. 
(a) Initial states $\rho_{W}(\lambda)$ lie on the line connecting the origin to the vertex $(1, -1, 1)$ and correspond to values $\lambda = 1, 3/5, 1/3$ with the respective colors  red, blue and orange. 
(b) Concurrence is shown for $\lambda = 1, 4/5, 3/5, 2/5, 1/3$ with the respective colors red, green, blue, magenta, orange.}%
	\label{fig:WERNER_PhiPlus_with_PRODUCT_momenta_GA_R_i_1}
\end{figure*}

For mixed states, $0 \leq \lambda < 1$, we plot the orbits for three different values of $\lambda$ in Fig.~\ref{fig:WERNER_PhiPlus_with_PRODUCT_momenta_GA_StateEvolution_R_i_1}. The concurrence is shown in Fig.~\ref{fig:WERNER_PhiPlus_with_PRODUCT_momenta_GA_Concurrence_R_i_1}. 
We note that as $\lambda$ decreases, the states start to disentangle at lower values of $\omega$. This is because the orbits remain parallel to the orbit of the Bell state and thus enter the octahedron sooner. Since they are also parallel to the face of the bottom pyramid, the state never escapes the region of separability.

\subsubsection{Case $R_i \otimes R_i$ and $R_i \otimes R_j$}\label{subsubsec:MTimesR_iTimesR_iAndR_iTimesR_j}

For rotations that act on both particles let us consider the scenario where the boost is in the $z$-direction and momenta are constrained to lie in the $x-z$- and $y-z$-planes. The state then consists of terms that correspond to the pure momenta 
$\ket{\pm\ppp_x, \pm\ppp_x}$, $\ket{\pm\ppp_y, \pm\ppp_y}$, $\ket{\pm\ppp_x, \pm\ppp_y}$ and $\ket{\pm\ppp_y, \pm\ppp_x}$, which generate the rotation terms $R_Y \otimes R_Y$, $R_X \otimes R_X$, $R_X \otimes R_Y$ and $R_Y \otimes R_X$. The spin orbit can be calculated by combining the respective vectors,
\begin{align}\label{eq:convexCombinationPureGAMomentumXXYY}
t^{\times}_{X \otimes Y}(\omega, \lambda) &= \frac{1}{4} \left( 
  t^{\Sigma}_{X \otimes X} + 
  t^{\Sigma}_{X \otimes Y} + 
  t^{\Sigma}_{Y \otimes X} + 
  t^{\Sigma}_{Y \otimes Y} \right) \nonumber\\
&= \lambda \left( \cos^4 \frac{\omega}{2}, -\cos^4 \frac{\omega}{2}, \cos^2 \omega \right),
\end{align}
where we have used superscripts to distinguish between the vectors generated by $\rho_{\Sigma}$ and $\rho_{\times}$. The other vectors can be obtained in a similar fashion, 
\begin{align}
t_{X \otimes Z}(\omega, \lambda) &= \lambda \left( \cos^4 \frac{\omega}{2}, -\cos^2 \omega, \cos^4 \frac{\omega}{2} \right),\nonumber\\
t_{Y \otimes Z}(\omega, \lambda) &= \lambda \left( \cos^2 \omega, -\cos^4 \frac{\omega}{2}, \cos^4 \frac{\omega}{2} \right),
\end{align}
where we have omitted superscripts for brevity. The concurrence is given by 
\begin{align}
C(\omega, \lambda) = 
\left\{
  \begin{array}{ll} 
      \frac{1}{16} 
      \left( - 
        \left| 4 \lambda \cos\omega  - \lambda \cos 2\omega  + \lambda - 4 
        \right| 
      \right.
    \\
      \left.
        \quad +\; 4 \lambda \cos\omega + 7 \lambda \cos 2\omega + 9 \lambda - 4 
      \right) 
    \quad\,
    \\
    \quad\quad\quad\quad\quad\quad\quad\quad\quad\quad\quad\quad\text{if}\quad \lambda \in (\lambda_{\text{sep}}, 1],\\
    0  \quad\quad\quad\quad\quad\quad\quad\quad\quad\quad\quad\;\;
    \text{if}\quad \lambda \in [0, \lambda_{\text{sep}}].\\
  \end{array} 
\right.
\end{align}

\begin{figure*}	
	\centering
	\begin{subfigure}[t]{0.38\textwidth}
		\centering
		\includegraphics[width=\textwidth]{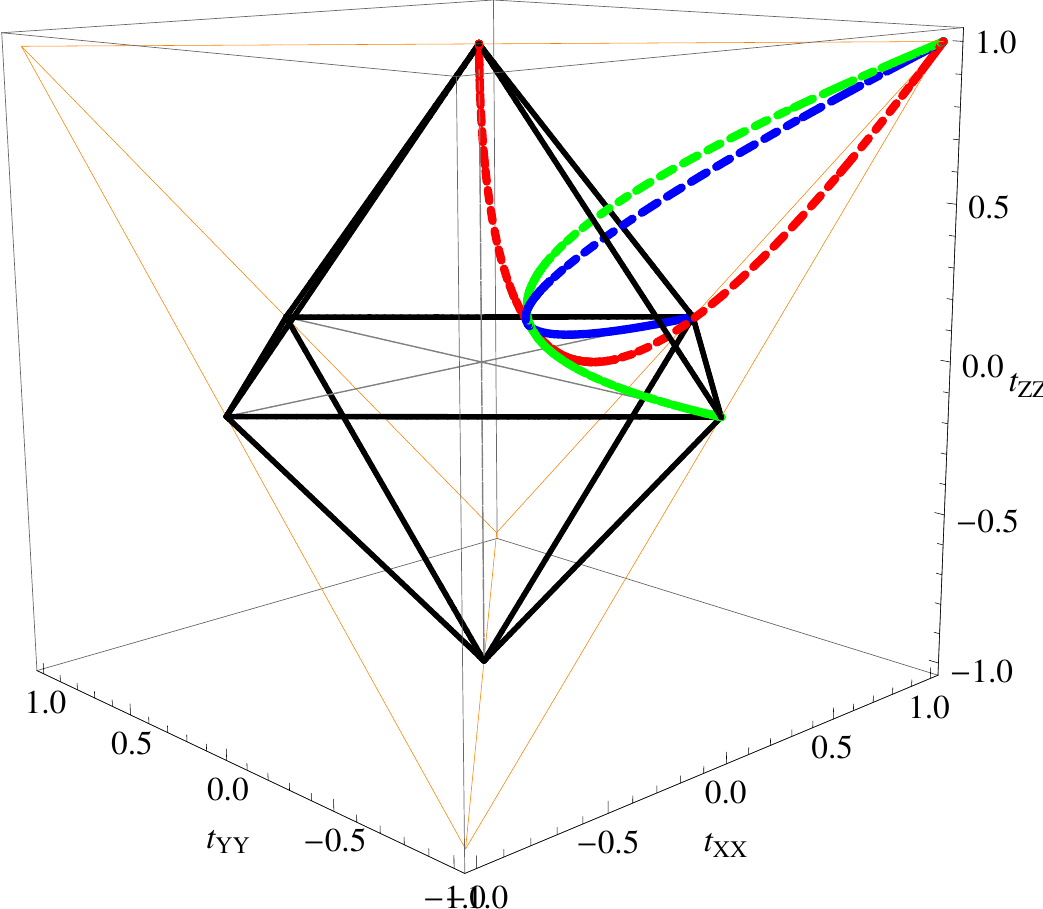}
		\caption{}
\label{fig:BELL_PhiPlus_with_PRODUCT_GA_momenta_R_i_R_i_R_i_R_j_StateEvolution_}
	\end{subfigure}
	\hspace{2em}
	\begin{subfigure}[t]{0.48\textwidth}
		\centering
		\includegraphics[width=\textwidth]{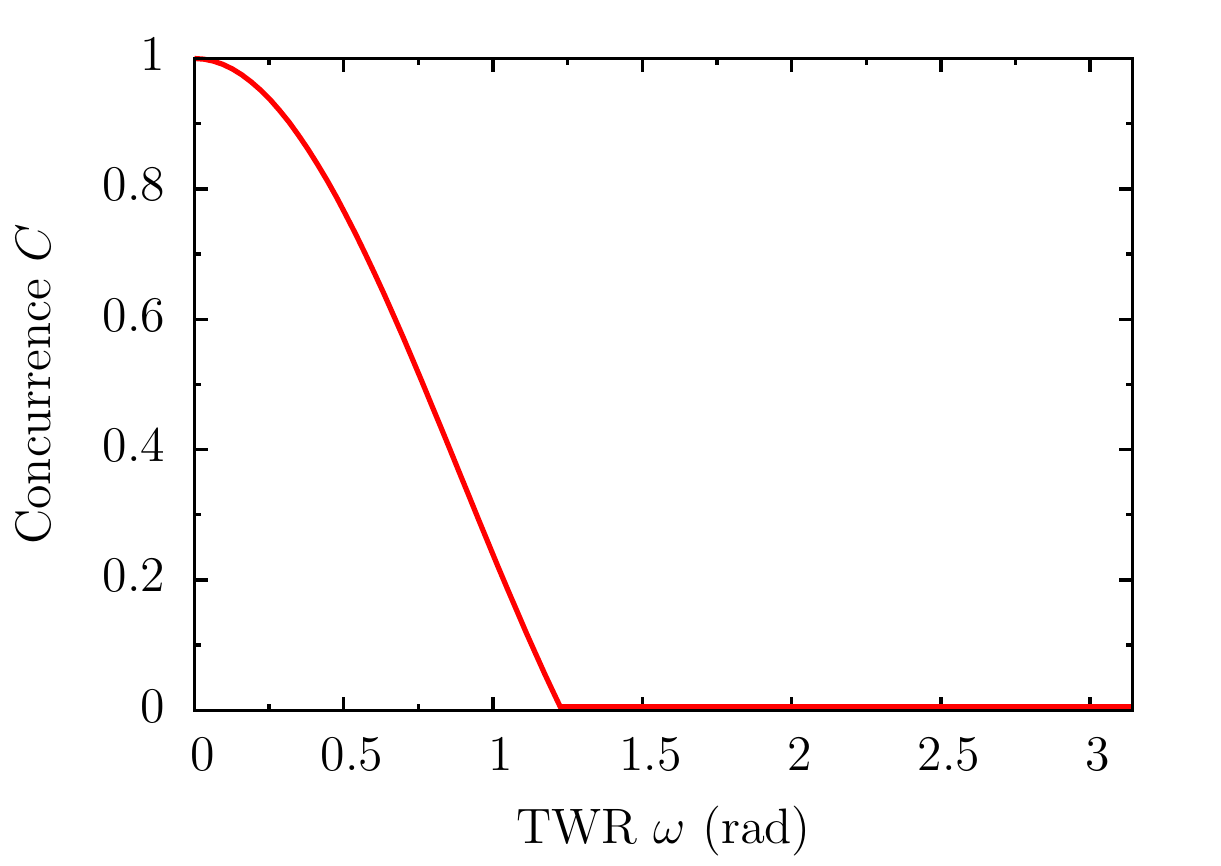}
		\caption{}		\label{fig:BELL_PhiPlus_with_PRODUCT_GA_momenta_R_i_R_i_R_i_R_j_Concurrence_}
	\end{subfigure}
	\caption[Spin orbit and concurrence under $R_i \otimes R_i$ and $R_i \otimes R_j$ with $\rho_\times$.]{Spin orbit and concurrence under $R_i \otimes R_i$ and $R_i \otimes R_j$ generated by momenta $\rho_\times$ with $\omega \in [0, \pi]$. 
(a) Initial state $\ket{\Phi^+}$ corresponds to the vertex at $(1, -1, 1)$, the orbit $t_{X \otimes Y}$ is shown red, $t_{X \otimes Z}$ green and $t_{Y \otimes Z}$ blue.
(b) Spin concurrence has the same shape for all orbits.
}%
	\label{fig:BELL_PhiPlus_with_PRODUCT_GA_momenta_R_i_R_i_R_i_R_j}
\end{figure*}

Considering first the Bell state $\ket{\Phi_+}$, the case with $\lambda = 1$, we plot the spin orbits and concurrence in Fig.~\ref{fig:BELL_PhiPlus_with_PRODUCT_GA_momenta_R_i_R_i_R_i_R_j}.
The orbit exhibits interesting behavior, starting out in a manner similar to $t_{X \otimes Y}$ generated by the symmetric momentum $\rho_{\Sigma}$. However, after entering the octahedron, it changes course and evolves towards the tip of the upper pyramid. When $\omega = \pi$ it reaches the state which corresponds to the equal mixture of projectors onto $\ket{\Phi_+}$ and $\ket{\Phi_-}$. This explains why concurrence vanishes in Fig.~\ref{fig:BELL_PhiPlus_with_PRODUCT_GA_momenta_R_i_R_i_R_i_R_j_Concurrence_} at all boosts that induce rotations larger than $1.23\; \text{rad}$.

It is also instructive to compare the current case with the $R_i \otimes \idd$ in the previous section. While on the face of it the shape of both concurrences is  quite similar, the corresponding orbits follow rather different paths. In analogy to the previous case, the orbit here initially moves downward, while the state disentangles slightly earlier. Soon after entering the octahedron, however, the orbit turns upward and ends at a point which is almost opposite to the one of the final state under $R_{XY} \otimes \idd$ in the previous section. This is another example of how visualization of state change explains differences in the behavior of concurrence which would remain hidden otherwise. It would also explain what happens to entanglement if one changed the parameters that characterize the boost scenario in question.

\begin{figure*}	
	\centering
	\begin{subfigure}[t]{0.38\textwidth}
		\centering
		\includegraphics[width=\textwidth]{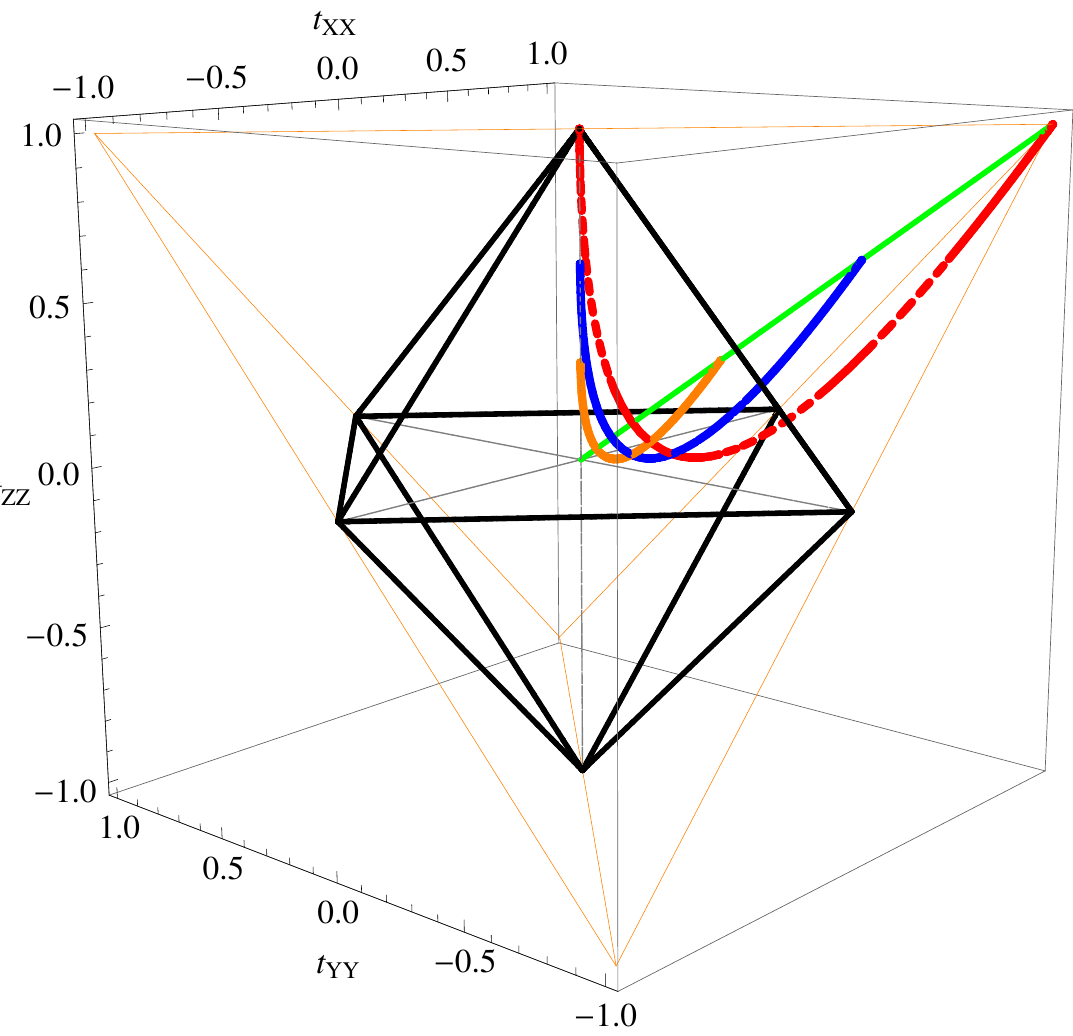}
		\caption{}
		\label{fig:WERNER_PhiPlus_with_PRODUCT_momenta_GA_StateEvolution_}
	\end{subfigure}
	\hspace{2em} 
	\begin{subfigure}[t]{0.48\textwidth}
		\centering
		\includegraphics[width=\textwidth]{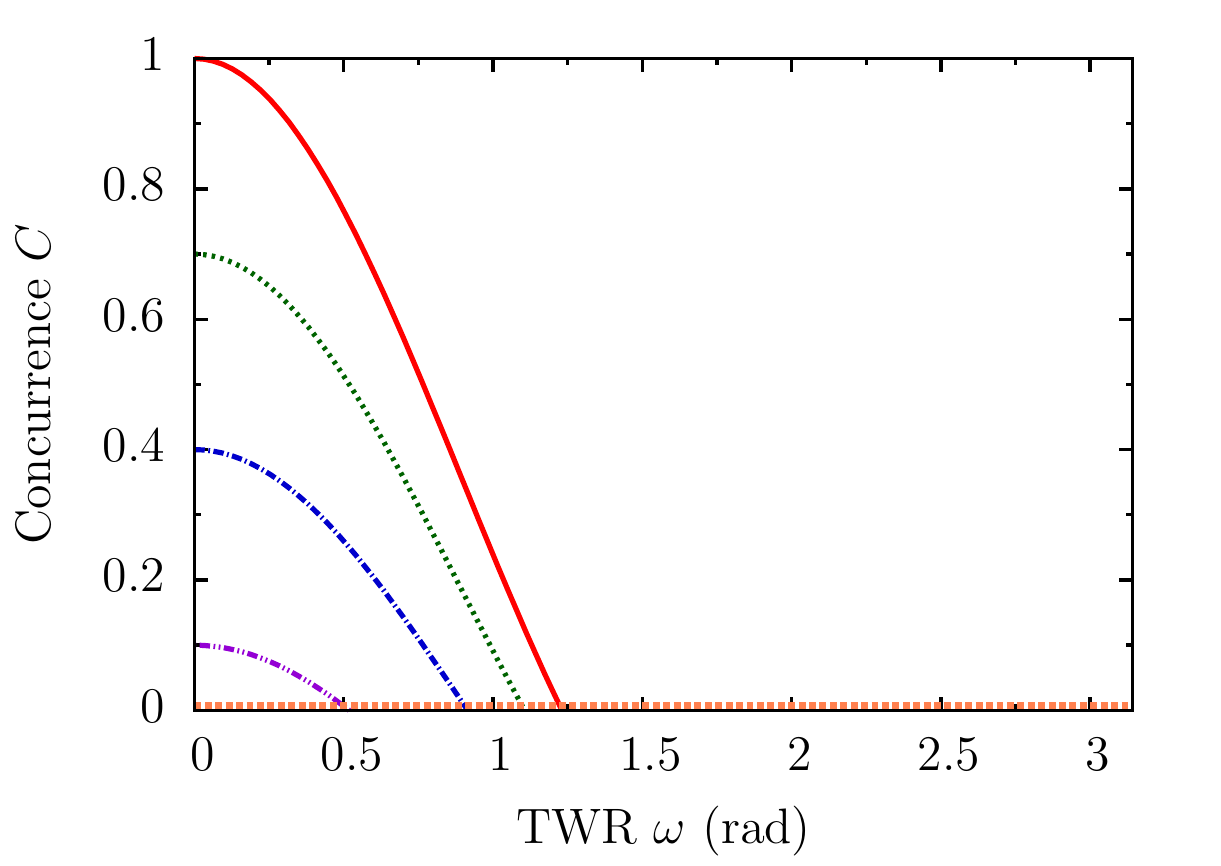}
		\caption{}
		\label{fig:WERNER_PhiPlus_with_PRODUCT_momenta_GA_Concurrence_}
	\end{subfigure}
	\caption[Spin orbit and concurrence under $R_i \otimes R_i$ and $R_i \otimes R_j$ with $\rho_{\times}$.]{Typical spin orbit and concurrence $R_i \otimes R_i$ and $R_i \otimes R_j$ generated by momenta $\rho_{\times}$ with $\omega \in [0, \pi]$. 
(a) Initial states $\rho_{W}(\lambda)$ lie on the line connecting the origin to the vertex $(1, -1, 1)$ and correspond to values $\lambda = 1, 3/5, 1/3$ with the respective colors  red, blue and orange. 
(b) Concurrence is shown for $\lambda = 1, 4/5, 3/5, 2/5, 1/3$ with the respective colors red, green, blue, magenta, orange.}%
	\label{fig:WERNER_PhiPlus_with_PRODUCT_momenta_GA}
\end{figure*}

We now turn to the case of mixed states, $0 \leq \lambda < 1$. Plots of the spin orbit and concurrence are shown in Fig.~\ref{fig:WERNER_PhiPlus_with_PRODUCT_momenta_GA}.
We see that in analogy to the single particle rotations in the previous section, the state begins to disentangle at lower values of $\omega$ as $\lambda$ decreases. Although here the orbits are not strictly parallel to the one of the Bell state, the phenomenon is quite similar. Smaller values of $\lambda$ mean the initial state is closer to the set of separable states and needs less rotation to enter the pyramid and become disentangled. Since all orbits approach the same final state, the entanglement never revives.

It is interesting to note that this momentum state can be employed to model quite accurately the continuous momenta discussed in the seminal paper \cite{gingrich_quantum_2002}, see \cite{palge_surveying_2014} for details.

\section{Entangled momenta}

\subsection{`Entangled' momenta}\label{subsec:entangledMomentaWithWernerState}

In the following sections we will assume that momenta are entangled and take the form of Bell states $\ket{M_{\Phi+}}$ or $\ket{M_{\Psi+}}$ or Bell-like states $\ket{M_{[\Phi+]}}$, $\ket{M_{[\Psi+]}}$. The former are instantiated by rotations of type $R_i \otimes R_i$ whereas the latter occur when rotations are around different axes, $R_i \otimes R_j$, $i \ne j$. As mentioned above, there is no need to consider momenta with other relative phases since they induce the same orbits for the spin state. Also, the type $R_i \otimes \idd$ will be omitted since it is equivalent to the $R_i \otimes \idd$ generated by product momenta $\rho_{\Sigma}$. This is because if the first particle is rotated while the second is left alone, the product momenta $\rho_{\Sigma}$ and any of the entangled momenta are given by the same state. We will also leave out the implementations of concrete rotations since they are analogous to those of product momenta.

As before, for the reason of computational convenience we will use the mixed momenta that correspond to the pure entangled states, 
\begin{align}
\rho_{\Phi+} &= \mathrm{diag}\, \pouter{M_{\Phi+}}{M_{\Phi+}}, \nonumber\\
\rho_{\Psi+} &= \mathrm{diag}\, \pouter{M_{\Psi+}}{M_{\Psi+}}, \nonumber\\
\rho_{[\Phi+]} &= \mathrm{diag}\, \pouter{M_{[\Phi+]}}{M_{[\Phi+]}}, \\
\rho_{[\Psi+]} &= \mathrm{diag}\, \pouter{M_{[\Psi+]}}{M_{[\Psi+]}}. \nonumber
\end{align}
They are clearly not entangled since they contain only the diagonal elements of the projectors on entangled states. We will, however, categorize the resulting spin states as if they had been generated by entangled momenta for the reason highlighted above, namely, that entangled momenta would lead to the same spin states.

\subsubsection{$R_i \otimes R_i$}

The case of two-rotations $R_i \otimes R_i$ for entangled momenta is quite dissimilar from the behavior generated by the product momenta. We begin by calculating the spin orbits generated by $\rho_{\Phi+}$. The three realizations fall into two cases. The $R_X \otimes R_X$ and $R_Z \otimes R_Z$ rotations yield the vectors
\begin{align}
t_{X \otimes X} (\omega, \lambda) &= \lambda \left( 1, -\cos 2\omega, \cos 2\omega \right), \nonumber \\
t_{Z \otimes Z} (\omega, \lambda) &= \lambda \left( \cos 2\omega, -\cos 2\omega, 1 \right),
\end{align}
whereas $R_Y \otimes R_Y$ leaves the state invariant, 
\begin{align}
t_{Y \otimes Y}(\omega, \lambda) = \lambda (1, -1, 1). 
\end{align}
This asymmetry arises from the fact that the momentum state $\rho_{\Phi+}$ consists of terms which induce rotations in the positive direction, $R_i(\omega) \otimes R_i(\omega)$, and in the negative direction, $R_i(-\omega) \otimes R_i(-\omega)$. 
The spin state $\rho_W$ is an eigenstate of such rotations around the $y$-axis, $R_Y(\pm \omega) \otimes R_Y(\pm \omega)$, but not around the other axes, $R_X(\pm \omega) \otimes R_X(\pm \omega)$ and $R_Z(\pm \omega) \otimes R_Z(\pm \omega)$.

The trivial orbit $t_{Y \otimes Y}$ has the concurrence
\begin{align}
C(\lambda) = (-1 + 3\lambda) / 2
\end{align}
whereas the nontrivial orbits $t_{X \otimes X}$ and $t_{Z \otimes Z}$ have 
\begin{align}\label{eq:concurrenceWernerSpinAndR_x_R_x_EntangledMomenta}
C(\omega, \lambda) = 
\left\{
  \begin{array}{ll}
  \frac{1}{2} \left( - 1 + \lambda + 2 \lambda |\! \cos 2\omega | \right) \quad & \text{if}\quad \lambda \in (\lambda_{\text{sep}}, 1] \\
  0\;\;\,\quad\quad\quad\quad\quad\quad\quad\quad\quad & \text{if}\quad \lambda \in [0, \lambda_{\text{sep}}].\\  
  \end{array} 
\right. 
\end{align}

\begin{figure*}	
	\centering
	\begin{subfigure}[t]{0.38\textwidth}
		\centering
		\includegraphics[width=\textwidth]{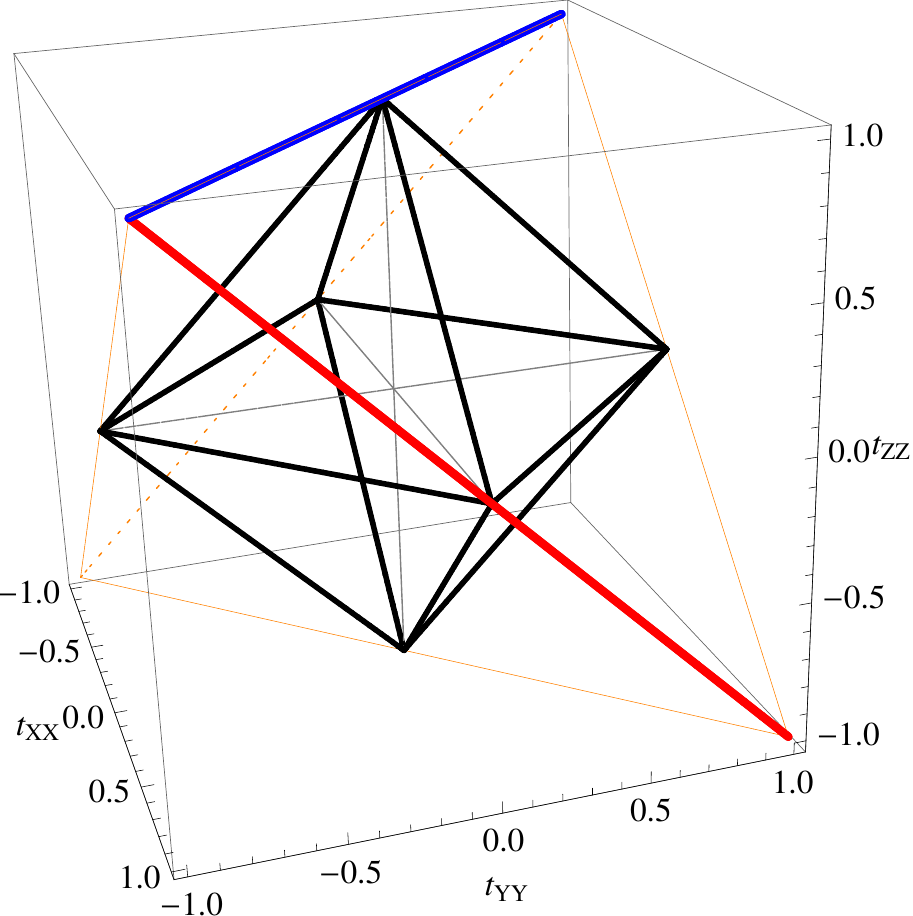}
		\caption{}
		\label{fig:BELL_PhiPlus_with_ENTANGLED_momenta_StateEvolution_R_y_R_y_}
	\end{subfigure}
	\hspace{2em}
	\begin{subfigure}[t]{0.48\textwidth}
		\centering
		\includegraphics[width=\textwidth]{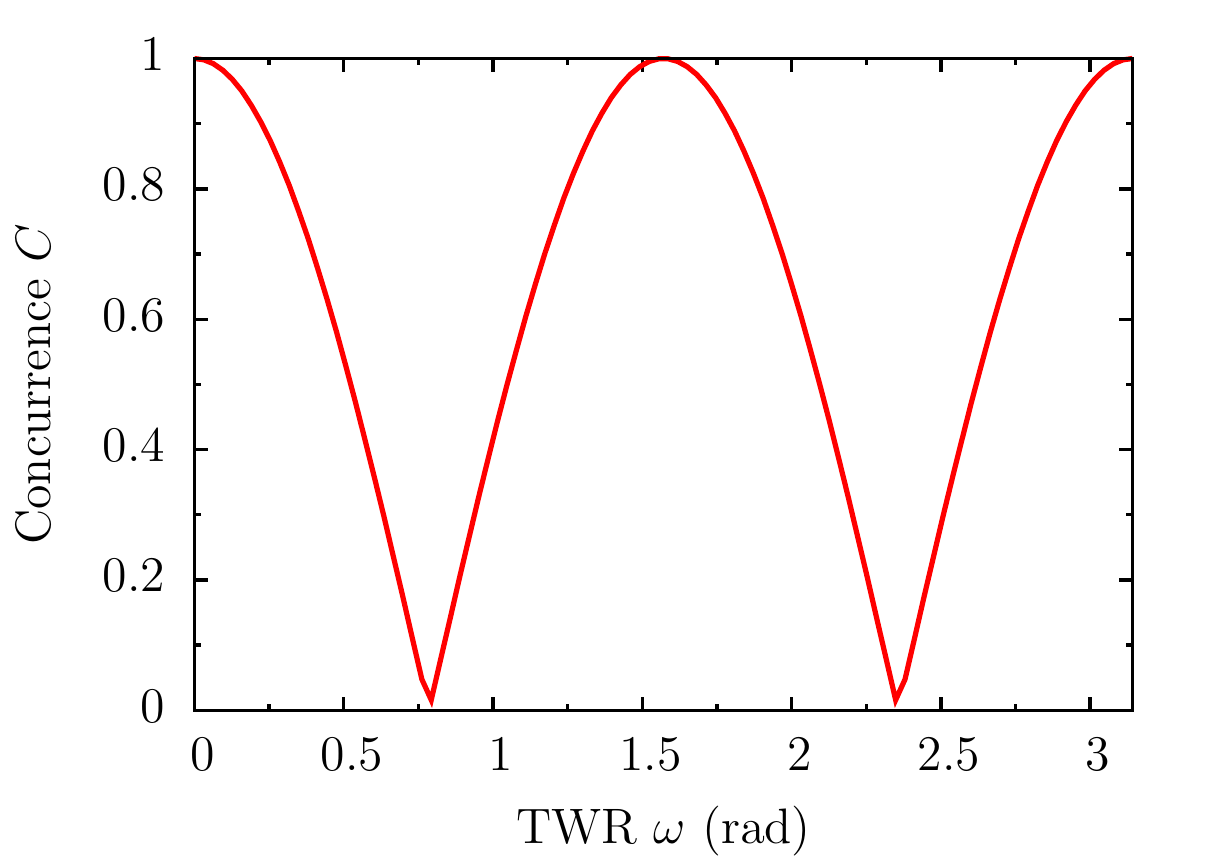}
		\caption{}
	\label{fig:BELL_PhiPlus_with_ENTANGLED_momenta_StateEvolution_R_y_R_y_concurrence}
	\end{subfigure}
	\caption[Spin orbit and concurrence under $R_i \otimes R_i$ with $\rho_{\Phi+}$.]{Spin orbit and concurrence of $\ket{\Phi_+}$ under $R_i \otimes R_i$ with $i \in \{ X, Z\}$ and $\omega \in [0, \pi]$. Entangled momenta are given by $\rho_{\Phi+}$. (a)~Initial state $\ket{\Phi_+}$ corresponds to vertex $(1, -1, 1)$, orbit for $R_X \otimes R_X$ is shown red and $R_Z \otimes R_Z$ is blue. (b)~Concurrence has the same shape for both orbits. 
}%
	\label{fig:BELL_PhiPlus_with_ENTANGLED_momenta_R_y_R_y_}
\end{figure*}

Let us consider first the Bell states, $\lambda = 1$. The non-trivial orbits and corresponding concurrences are shown in Fig.~\ref{fig:BELL_PhiPlus_with_ENTANGLED_momenta_R_y_R_y_}.
It is interesting to compare the orbit in Fig.~\ref{fig:BELL_PhiPlus_with_ENTANGLED_momenta_StateEvolution_R_y_R_y_} with the one obtained earlier in section \ref{sec:spinWernerStateWithProductMomenta_R_i_1} for the single particle rotation $R_i \otimes \idd$ generated by the product momenta $\rho_{\Sigma}$. While they look similar, the one here traverses the same path twice as fast as $\omega$ ranges from $0$ to $\pi$. In analogy to the single particle rotation, the state is sent to $\ket{\Psi_+}$, but in contrast to the single particle rotation, this happens now already at $\omega = \pi / 2$. When the rotation achieves the maximal value $\pi$, the boosted observer sees again the original state $\ket{\Phi_+}$. Accordingly, the concurrence in Fig.~\ref{fig:BELL_PhiPlus_with_ENTANGLED_momenta_StateEvolution_R_y_R_y_concurrence} shows a graph which oscillates twice as fast between its maximal value and zero in the same range of rotation. The other Bell states display similar systematic behavior. Depending on whether or not they are eigenstates of the particular rotation in question, they do or do not show non-trivial orbits and concurrence.

We note that the expression for spin concurrence in Eq.~(\ref{eq:concurrenceWernerSpinAndR_x_R_x_EntangledMomenta}) for the case of pure states, $\lambda = 1$, was first reported in~\cite{jordan_lorentz_2007}. The authors considered a geometry where the momenta of both particles make an angle $\pi / 2$ to the direction of boost, obtaining a change of entanglement shown between $[0, \pi/2]$ in Fig.~\ref{fig:BELL_PhiPlus_with_ENTANGLED_momenta_StateEvolution_R_y_R_y_concurrence}.

\begin{figure*}	
	\centering
	\begin{subfigure}[t]{0.38\textwidth}
		\centering
		\includegraphics[width=\textwidth]{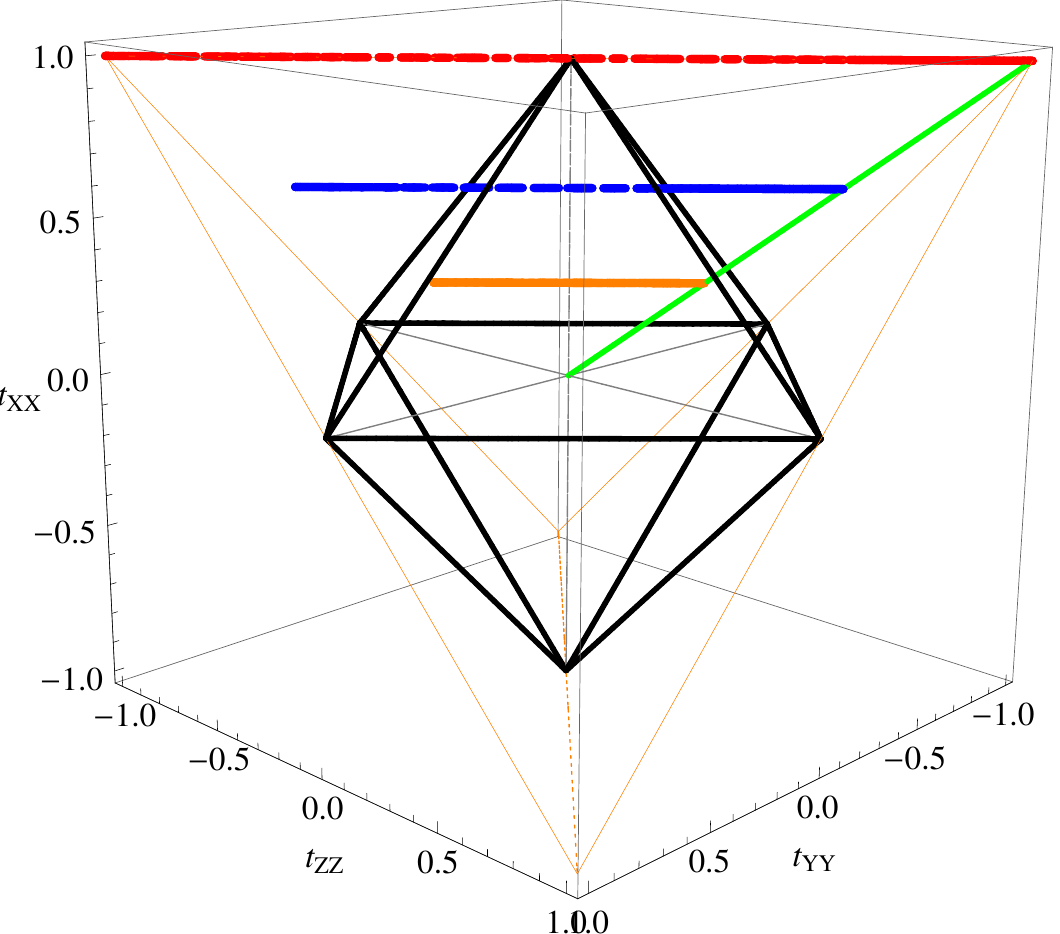}
		\caption{}
		\label{fig:WERNER_PhiPlus_with_ENTANGLED_momenta_R_x_R_x_StateEvolution_}
	\end{subfigure}
	\hspace{2em}
	\begin{subfigure}[t]{0.48\textwidth}
		\centering
		\includegraphics[width=\textwidth]{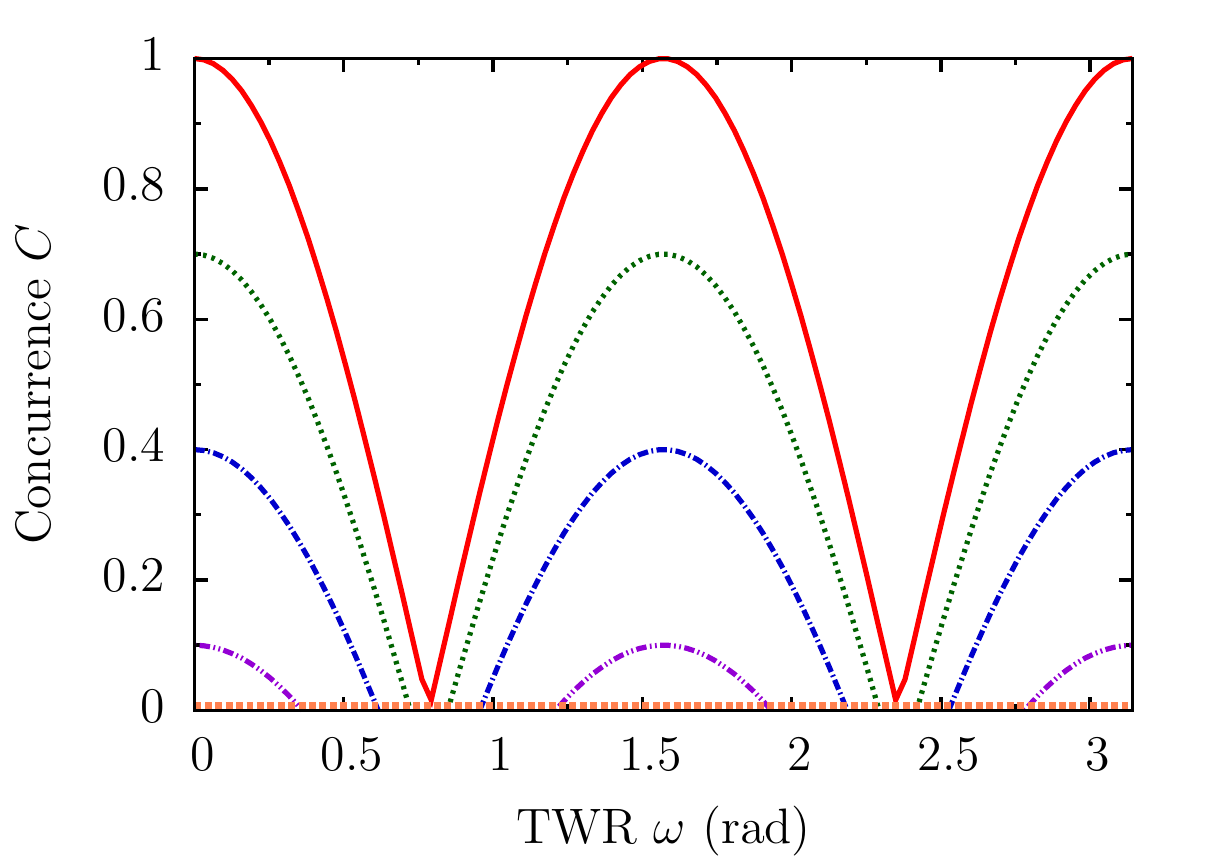}
		\caption{}
		\label{fig:WERNER_PhiPlus_with_ENTANGLED_momenta_R_x_R_x_Concurrence_}
	\end{subfigure}
	\caption[Spin orbit and concurrence under $R_i \otimes R_i$ with $\rho_{\Phi+}$.]%
{Typical spin orbit and concurrence under $R_i \otimes R_i$ with $\omega \in [0, \pi]$ generated by momenta $\rho_{\Phi+}$. 
(a) Initial states $\rho_{W}(\lambda)$ lie on the line connecting the origin to the vertex $(1, -1, 1)$ and correspond to values $\lambda = 1, 3/5, 1/3$ with the respective colors red, blue and orange. 
(b) Concurrence is shown for $\lambda = 1, 4/5, 3/5, 2/5, 1/3$ with the respective colors red, green, blue, magenta, orange.}%
	\label{fig:WERNER_PhiPlus_with_ENTANGLED_momenta_R_x_R_x_}
\end{figure*}

For the general case that includes mixed states, $0 \leq \lambda \leq 1$, we plot the orbits and concurrence in Fig.~\ref{fig:WERNER_PhiPlus_with_ENTANGLED_momenta_R_x_R_x_}.
The orbits of mixed spins are parallel to those of pure spins and share the characteristics described above. The Werner state $\rho_W$ is sent to a counterpart Werner state $\rho_{W \Psi+}$ given in Eq.~(\ref{eq:WernerStateWithPsiPlus}). There are two intervals in Fig.~\ref{fig:WERNER_PhiPlus_with_ENTANGLED_momenta_R_x_R_x_Concurrence_} where the concurrence vanishes since the state moves forward and backward through the octahedron of separable states. The lower the initial degree of entanglement, the larger the part of the orbit in the octahedron, and thus the larger the region of vanishing concurrence.

\subsubsection{$R_i \otimes R_j$}\label{sec:spinWernerStateWithEntangledMomenta_R_i_R_j}

\begin{figure}[htb]
\centering
\includegraphics[width=0.5\textwidth]{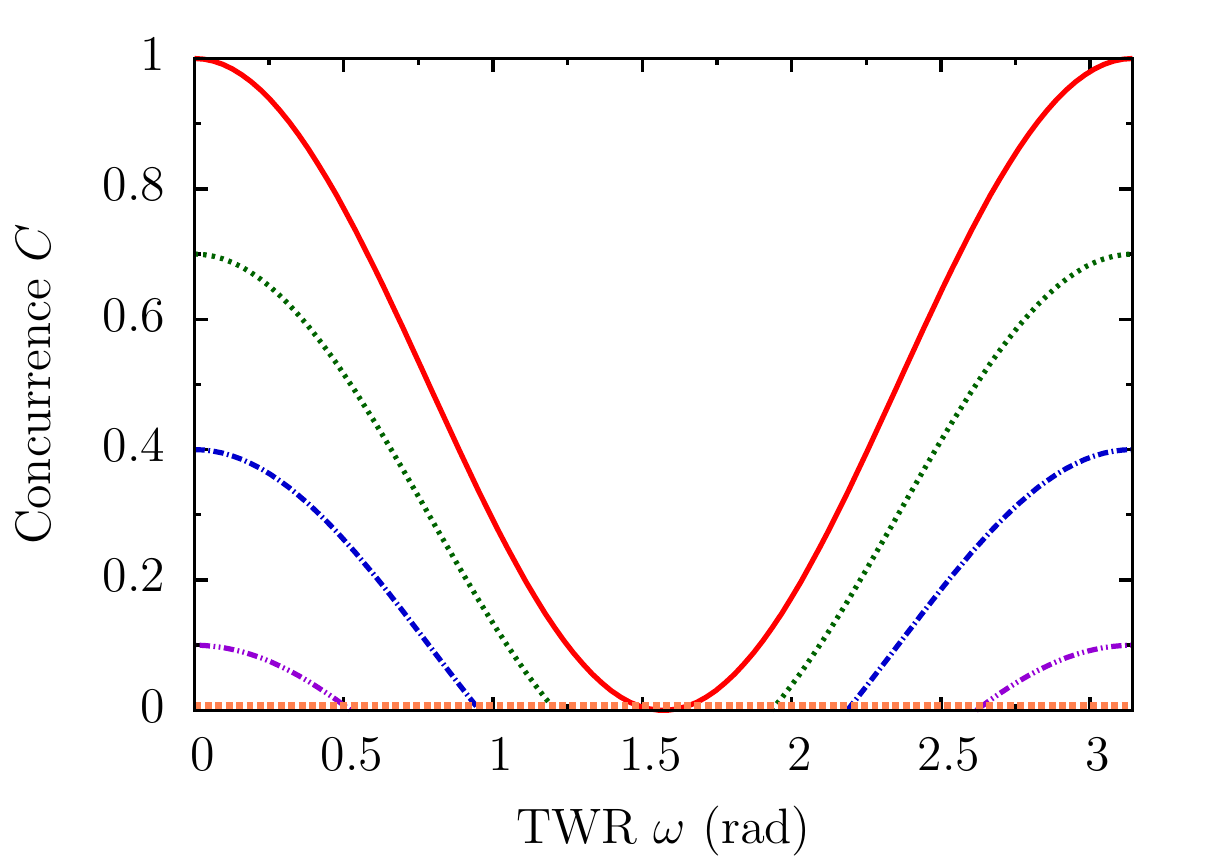}
\caption[Spin concurrence under $R_i \otimes R_j$, $i \ne j$ with $\rho_{[\Phi+]}$ and $\rho_{[\Psi+]}$.]%
{Spin concurrence under $R_i \otimes R_j$, $i \ne j$ with $\omega \in [0, \pi]$ generated by momenta $\rho_{[\Phi+]}$ and $\rho_{[\Psi+]}$.
Concurrence is shown for $\lambda = 1, 4/5, 3/5, 2/5, 1/3$ with the respective colors red, green, blue, magenta, orange.
}\label{fig:WERNER_PhiPlus_with_ENTANGLED_momenta_R_x_R_z_}
\end{figure}

Mixed rotations $R_i \otimes R_j$, $i \ne j$ present the case where the spin states generated by the momenta are not Bell diagonal. For instance, the coefficient matrices $t_{i \otimes j}$, \mbox{$i, j \in \{ X, Y, Z \}$}, generated by the momentum state $\rho_{[\Phi+]}$ for the rest frame spin $\rho_W$ are as follows, 
\begin{align}\label{eq:stateVectorsEntangledMomentaR_i_R_j}
t_{X \otimes Y}(\omega, \lambda) &= 
\lambda 
\begin{pmatrix}
\cos \omega 	&			0	&		0			\\
-\sin^2 \omega	&	-\cos\omega	&		0			\\
0				&			0	&	\cos^2 \omega	\\ 
\end{pmatrix} \,, \nonumber \\
{\footnotesize\phantom{a}} \nonumber \\
t_{X \otimes Z}(\omega, \lambda) &= 
\lambda 
\begin{pmatrix}
\cos \omega 	&			0		&		0			\\
0				&	-\cos^2 \omega	&		0			\\
\sin^2 \omega	&			0		&	\cos \omega		\\ 
\end{pmatrix} \,, \\
{\footnotesize\phantom{a}} \nonumber \\
t_{Y \otimes Z}(\omega, \lambda) &= 
\lambda 
\begin{pmatrix}
\cos^2 \omega 	&			0		&		0			\\
0				&	-\cos \omega	&		0			\\
0				&	-\sin^2 \omega	&	\cos \omega		\\ 
\end{pmatrix} \,. \nonumber
\end{align}
The concurrence is the same for all three states,  
\begin{align}\label{eq:ConcurrenceWernerStateEntangledMomentaRiRj}
C(\omega, \lambda) = 
\left\{
  \begin{array}{ll}
  -\frac{1}{2} + \lambda + \frac{1}{2} \lambda \cos 2\omega \quad & \text{if}\quad \lambda \in (\lambda_{\text{sep}}, 1] \\
   0\;\;\,\quad\quad\quad\quad\quad\quad\quad\quad\quad & \text{if}\quad \lambda \in [0, \lambda_{\text{sep}}] .\\  
  \end{array} 
\right.
\end{align}
Plots of concurrence for different values of $\lambda$  are shown in Fig.~\ref{fig:WERNER_PhiPlus_with_ENTANGLED_momenta_R_x_R_z_}.
Although the states are not diagonal when $\omega \neq 0, \pi$, the structure of $t_{i \otimes j}$ suggests that the orbits are isomorphic to each other. All three matrices contain the same diagonal terms as the vector of $R_i \otimes R_j$ induced by product momenta. In addition, all matrices contain an off-diagonal term $\pm \sin^2\omega$, whose location varies systematically. This allows us to represent the matrices $t_{i \otimes j}$ by a four vector consisting of the diagonal and off-diagonal terms, $(t_{kk}, \pm \sin^2 \omega)$. The three states can be thus seen to be related by a one-one map. They seem to share similar geometric structure as well. The first three components of the four vector represent the vector of $R_i \otimes R_j$, the fourth component varies in the same way (modulo sign) albeit in a different subspace for different rotations.

Now the expression for concurrence (\ref{eq:ConcurrenceWernerStateEntangledMomentaRiRj}) is identical to (\ref{eq:concurrenceForSpinWernerStateWithProductMomenta_R_i_R_i}), i.e. the case of $R_i \otimes R_i$ generated by product momenta. Could this give us clues about the shape of the orbit? Although we cannot say what is the shape of the orbit in the current case, it is definitely different from the one of $R_i \otimes R_i$. 
Taking the case of pure states as an example, this is because while the orbit of the product rotation is cyclic in the sense that it returns to the initial state at $\omega = \pi$, the orbit here starts at $\ket{\Phi_+}$ when $\omega = 0$ and ends at $\ket{\Phi_-}$ with $\omega = \pi$. 
We conclude that more investigation is needed to determine the geometric structure of the orbit but we will not pursue the issue further here since it is not crucial for our purposes.

\section{Summary and discussion}

In this paper we studied spin entanglement of a two particle system. We systematically investigated various boost scenarios involving both product and entangled momenta with the aim of surveying the structure of maps that momenta induce on spins. Momenta were assumed to be discrete and spins in the Werner state. The latter subsume the Bell states when $\lambda = 1$. The results are summarized in Tables~\ref{tab:summaryWernerStatesStudiedMomentumProduct} and~\ref{tab:summaryWernerStatesStudiedMomentumEntangled}.

\renewcommand{\arraystretch}{1.5}
\begin{table*}[tbh!]
\caption{\label{tab:summaryWernerStatesStudiedMomentumProduct}%
Spin orbit and concurrence for $\rho_W(\lambda)$ generated by product momenta $\rho_{\text{EPRB}}$, $\rho_{\Sigma}$ and $\rho_{\times}$. The second column shows a typical orbit.}
\begin{ruledtabular}
\begin{tabular}{lll}
  Momenta &  
  Orbit  &  
  Concurrence 
\\ \hline
  $\rho_{\text{EPRB}}$ &
  trivial or not diagonal  &	
  \!\!\! invariant  
\\ \hline
  $\rho_{\Sigma}$ &
  $\lambda (1, -\cos\omega, \cos\omega)$  &	
  ${
  \begin{array}{ll} \!\!\! \text{max} 
    \left\{ 0, \; \frac{1}{2} 
      \left( - 1 + \lambda + 2 \lambda |\! \cos\omega | 
      \right) 
    \right\} 
    \\   
  \end{array} 
  }$  
\\ \cline{2-3}
  &
  $\lambda (1, -\cos^2 \omega, \cos^2 \omega)$ &	
  ${
  \begin{array}{ll}
  \!\!\!\text{max} \left\{ 0, \; -\frac{1}{2} + \lambda + \frac{1}{2} \lambda \cos 2\omega \right\} 
  \\  
  \end{array} 
  }$  
\\ \cline{2-3}
  &
  $\lambda (\cos\omega, -\cos^2\omega, \cos\omega)$ &	
  ${
  \begin{array}{lll} 
  \!\!\!\text{max}
  \left\{ 0, \; \frac{1}{8}
    \left( \big|
	  \left| 2 + \lambda + 4\lambda \cos\omega
	  \right.  
	\right.
  \right.
  \\
  \quad\quad\quad\;\; 
  \left.
    + \lambda \cos2\omega 
  \right| - 
  \left| 
    2 + \lambda  
  \right. 
  \\
  \quad\quad\quad\;\;
    \left. - 4\lambda \cos\omega + \lambda \cos2\omega 
    \right| \big| 
  \\
  \quad\quad\quad\;\;
  \left.
    \left. + 2 
      \left( 
        - 2 + \lambda + \lambda \cos2\omega 
	  \right)
    \right)
  \right\}
  \\
  \end{array} 
  }$  
\\ \hline
  $\rho_{\times}$ &
  $\lambda (\cos^2\frac{\omega}{2}, -\cos^2\frac{\omega}{2}, \cos\omega)$  &
  ${
    \begin{array}{ll}
    \text{max} \left\{ 0, \frac{1}{2} \left( -1 + \lambda + 2\lambda\cos\omega \right)  \right\} 
    \end{array}
  }$  
\\ \cline{2-3}
  &  	
  $\lambda (\cos^4\frac{\omega}{2}, -\cos^4\frac{\omega}{2}, \cos^2\omega)$  &
  ${
    \begin{array}{ll}
      \text{max} 
      \left\{ 0, \frac{1}{16} 
        \left( - 
          \left| 4 \lambda \cos\omega - \lambda \cos 2\omega  
          \right. 
        \right. 
      \right. 
      \\
      \left.
        \left.
          \quad\quad\quad\quad 
          \left. 
            + \lambda - 4 
          \right| 
          + 4 \lambda \cos\omega + 7\lambda \cos 2\omega 
        \right.
      \right.  
      \\
      \left.
        \left.
          \quad\quad\quad\quad + 9 \lambda - 4  
        \right)  
      \right\} 
    \end{array}
    }$  
\\ 
\end{tabular}
\end{ruledtabular}
\end{table*}

\renewcommand{\arraystretch}{1.5}
\begin{table*}[tbh!]
\caption{\label{tab:summaryWernerStatesStudiedMomentumEntangled}%
Spin orbit and concurrence for $\rho_W(\lambda)$ generated by momenta 
$\rho_{\Phi+}$, 
$\rho_{\Psi+}$,
$\rho_{[\Phi+]}$ and
$\rho_{[\Psi+]}$. 
The second column shows a typical orbit.}
\begin{ruledtabular}
\begin{tabular}{lll}
Rotation  &  Orbit  &  Concurrence 
\\ 
\hline
$R_i \otimes R_i$ &  	
$\lambda (\cos 2\omega, -1, \cos 2\omega)$ &	
${%
  \begin{array}{ll}
    \!\!\!\text{max} \left\{ 0, \; \frac{1}{2} \left( - 1 + \lambda + 2 \lambda |\! \cos 2\omega | \right) \right\}
  \\
  \end{array} 
}$ 
\\  \cline{2-3}
\phantom{$R_i \otimes R_i$} &  	
trivial &	
\!\!\! invariant 
\\ \hline 
$R_i \otimes R_j$ &  	
not diagonal &	
${
  \begin{array}{ll}
    \!\!\!\text{max} \left\{ 0, \; -\frac{1}{2} + \lambda + \frac{1}{2} \lambda \cos 2\omega \right\}
    \\  
  \end{array} 
}$
\\ 
\end{tabular}
\end{ruledtabular}
\end{table*}

We confirm the overall lesson that Lorentz boosts generally cause non-trivial behavior of the spin degree of freedom of a two particle system. However, whether or not, and to what extent, the state and entanglement of spins changes depends substantially on the spin and momentum states involved, as well as on the geometry of the boost scenario. Whereas some states and geometries leave entanglement invariant, others give rise to rapid changes of concurrence. Examples of the former comprise Bell states with product momenta of the form $\rho_{\text{EPRB}}$, and also the case of entangled momenta under rotation of type $R_i \otimes R_i$ where the spin is an eigenstate of rotation. All other types of rotations and momenta were found to bring about entanglement change that ranges from maximal to zero, with the type $R_i \otimes R_i$, where momenta are entangled and spin is not an eigenstate of rotation, causing the fastest decay and rebirth of entanglement.

While the literature on relativistic entanglement commonly analyzes pure entangled states, it is important to consider mixed states as well in order to gain a full understanding. The present work makes a step in this direction by surveying the behavior of the Werner states, whose entanglement ranges between maximal and no entanglement at all. Compared to the pure states, they display less change because their maximal degree of entanglement is bounded by the parameter $\lambda$.

The latter highlights an important conclusion, which applies to both pure and mixed spin states: entanglement between spins cannot increase under Lorentz boosts if there is no spin--momentum entanglement present in the first place. This was first proved for pure states in \cite{gingrich_quantum_2002}. Our investigation shows that the result holds for mixed spin states too. It should be stressed again though that 
it is valid only for systems whose spin--momentum initially factorize. If spin and momentum degrees are initially entangled, the boosted state can be more entangled than the initial one. Although proper study of this interesting case is left for another occasion, as noted in section \ref{sec:theDiscreteModel}, it has been implicit to some extent in the scenarios we examined. The reason is that since Lorentz boosts form a group, we can read all plots in the reverse direction. For instance, the boosted state $\frac{1}{4}\idd$, which is represented by $(0, 0, 0)$ in Fig.~\ref{fig:BELL_PhiPlus_Bell_diagonal_maps_EVOLUTION_Ri_Rj_}, can be transformed to the initial maximally entangled Bell state $\ket{\Phi_+}$ by applying the inverse boost.

We would like to emphasize the usefulness of visualization of spin orbits, which gave further insight into the behavior of entanglement.
The orbit provides a geometric explanation of how boosts affect the state, showing how it traverses the various regions of state space. This means it opens up the possibility of manipulating states to achieve an engineering goal---something that merely plotting the concurrence does not yield.

Perhaps even more valuable are the qualitative insights. We relied on a simple framework that consists of spins at discrete momenta, where the latter are regarded as qubits and boosting means that each spin undergoes a rotation which depends on the boost scenario. The picture can be extended to continuous momenta as well \cite{palge_surveying_2014}. More importantly, it can be used explain the various results reported in the literature, i.e.\ whether or not spin entanglement changes in a particular boost scenario. For example, while \cite{alsing_lorentz_2002} found that Lorentz boosts do not change the degree of entanglement of a maximally entangled Bell state, \cite{gingrich_quantum_2002} show that the boosted observer generally sees a decrease of entanglement. Yet these results are consistent since they employ different momentum states. The system considered in \cite{alsing_lorentz_2002} contains product momenta $\rho_\text{EPRB}$, which induce maps that leave the Bell states invariant as discussed in section~\ref{subsec:productSimpleMomentaPQWithWernerState}. The authors of \cite{gingrich_quantum_2002}, however, assume that momenta are given by origin centered Gaussians. The latter can be modeled accurately by states of the form $\rho_{\times}$, as discussed in \ref{subsubsec:MTimesR_iTimesR_iAndR_iTimesR_j}, and 
they lead to a deterioration of the spin entanglement. 
A similar analysis could be carried out for many two level systems that contain either discrete or continuous momenta.

Finally, in the context of possible implementations of quantum computation or quantum communication protocols in the relativistic setting, the import of our results is twofold. 
On the one hand, relativity may appear as a foe to the quantum information theorist by, for instance, causing unwanted disturbances in an implementation of quantum communication protocol.
Then the results obtained in this paper could be used to engineer states such that the negative effects are diminished.
On the other hand, relativity can be a friend when, in analogy to entanglement, it is regarded as a resource which can be used to generate entangled states or realize, or enhance quantum communication or computation, see e.g.\ \cite{friis_motion_2012, martin-martinez_processing_2013}. The results here could be helpful in finding states and scenarios that help achieve the desired goal. As part of the future work, we envisage working out the implications of foregoing results to quantum information theory in more detail.

\section{Acknowledgments}

This work was financially supported by the United Kingdom EPSRC and the Japan Society for the Promotion of Science (JSPS) Grant-in-Aid for JSPS Fellows No.~25-03757.

\bibliography{paper_PALGE_DUNNINGHAM_2014_TWO_discrete}

\end{document}